THE BOEING COMPANY

# The State of Modeling, Simulation, and Data Utilization within Industry

## An Autonomous Vehicles Perspective

**Fadaie, Joshua G**

# Abstract

The aviation industry has a market driven need to maintain and develop enhanced simulation capabilities for a wide range of application domains. In particular, the future growth and disruptive ability of smart cities, autonomous vehicles and in general, urban mobility, hinges on the development of state of the art simulation tools and the intelligent utilization of data. While aviation based companies have several historical and/or proprietary mission level simulation tools, there is much to learn from the current state of the art within other industries (tangential and competing in scope), as well as in academia.

The purpose of this paper is to address and decompose the simulation capabilities within the key players of the autonomous vehicle and self-driving car industry (Toyota, Waymo, BMW, Microsoft, NVIDIA, Uber, etc.), as well as several notable startups within the high fidelity 3D mapping and simulation domain (Mapper.ai, HERE, Cognata, etc.). While providing an overview of how other companies are using simulation tools and reliable data-sets, this paper will also seek to address several important and related questions, namely: the interaction between simulation and supporting tools/software, the intersection of simulation and the real world, the requirements and utilization of compute infrastructure, the appropriate levels of fidelity within simulation, and how simulation tools are critical to future safety and V&V concerns.

In order for aviation based companies to adequately pursue disruptive mobility within real-world environments, be it in air or on the ground, modeling and simulation tools for autonomous vehicles provide key insights into future development work and are essential technologies.

# Executive Summary

**The aviation industry has a market driven need to maintain and develop enhanced simulation capabilities** for a wide range of application domains. In particular, the **future growth and disruptive ability of smart cities, autonomous vehicles and in general, urban mobility, hinges on the development of state of the art simulation tools and the intelligent utilization of data**. While aviation based companies have several historical and/or proprietary mission level simulation tools, there is much to learn from the current state of the art within other industries (tangential and competing in scope), as well as in academia.

**The purpose of this paper is to address and decompose the simulation capabilities within the key players of the autonomous vehicle and self-driving car industry (Toyota, Waymo, BMW, Microsoft, NVIDIA, Uber, etc.), as well as several notable startups within the high fidelity 3D mapping and simulation domain (Mapper.ai, HERE, Cognata, etc.)**. While providing an overview of how other companies are using simulation tools and reliable data-sets, this paper will also seek to address several important and related questions, namely: the interaction between simulation and supporting tools/software, the intersection of simulation and the real world, the requirements and utilization of compute infrastructure, the appropriate levels of fidelity within simulation, and how simulation tools are critical to future safety and V&V concerns.

More specifically, **this paper will contain a discussion into the following** over several sections:

- An overview of the key players and benefits of simulation
- Briefly discuss an overview of simulation tools and methodologies for programming models
- Rational for why autonomous vehicle based simulation is an excellent guiding example for capabilities needed in modeling urban mobility and smart cities
- Discussion of key players approach to high fidelity modeling, simulation and data utilization, including how they are constructing their simulation, what they are modeling and their overall capabilities
- Discussion of the specific use cases of their simulation as well as its technological and strategic advantages
- An overview of the integrated simulation to real world pipeline
- A look at how scalability, compute infrastructure, and cloud architecture are incorporated
- Safety and validation via simulation

There are several other application domains within the simulation arena that are applicable such as: AR/VR, manufacturing, electronics and IoT, logistics, robotics, and urban management and planning, which are not specifically addressed in this paper.

With that said, in order to properly simulate an autonomous vehicle in modern environments (rural and urban), there is much cross over with the aforementioned domains, such as logistics, robotics, etc. Additionally, in order for aviation based companies to adequately pursue disruptive mobility within real-world environments, be it in air or on the ground, modeling and simulation tools for autonomous vehicles provide key insights into future development work and are essential technologies.

# Table of Contents

# Introduction

Autonomous vehicles (AV) will transform society from social interactions to economics. Until recently the advent of self-driving cars was synonymous with science fiction, or that of a moonshot effort. Now it is increasingly becoming reality with companies such as Uber, Tesla, Waymo and numerous suppliers, startups, and governmental bodies actively participating in the testing of self-driving vehicles within the public spaces; the ultimate goal being to engineer fully autonomous vehicles. Catalyzing the development of self-driving vehicles, and in general autonomous systems is the engineering of new technologies from sensors, robust simulation tools, to scalable, high-throughput compute and storage devices. Additionally, the plethora of available datasets, feasibility of generating additional data, high-fidelity simulation toolsets, and high definition mapping has brought together a formative environment for the creation of self-driving vehicles.

The purpose of this research paper is to focus on key enabling technologies for autonomous systems, primarily simulations toolsets, with an emphasis on mission level simulation environments which test and validate the core functionality/behavior of self-driving vehicles in their environments. The paper will discuss the different methods of constructing simulation toolsets (as evidenced by key players in industry), the various components modeled within the environment, aspects of a robust simulation tool chain, and how simulation will inevitable play a role in safety and verification. Simulation tools alone though are not sufficient to generate self-driving cars and novel vehicle behaviors. It is important to address the inter-play between simulation environments, large scale super-computer clusters which power the rapid analysis of critical scenarios, real-world testing, and the curation of autonomous vehicle data. Proper curation of relevant data from a number of sources directly feeds into both the algorithms and analytics necessary for insight into correct utilization of simulation tools.

The paper is broken into multiple sections discussing key aspects of simulation toolsets. Each section will begin with a brief introduction.

## Autonomous Vehicles Overview

In order to facilitate a better understanding of the simulation environment for autonomous vehicles, a brief overview of the sensors and software on self-driving cars is provided. Self-driving vehicles are characterized by 6 levels of autonomy according to SAE International, where level 0 refers to no automation and level 5 is fully automated with no driver assistance or monitoring [2].

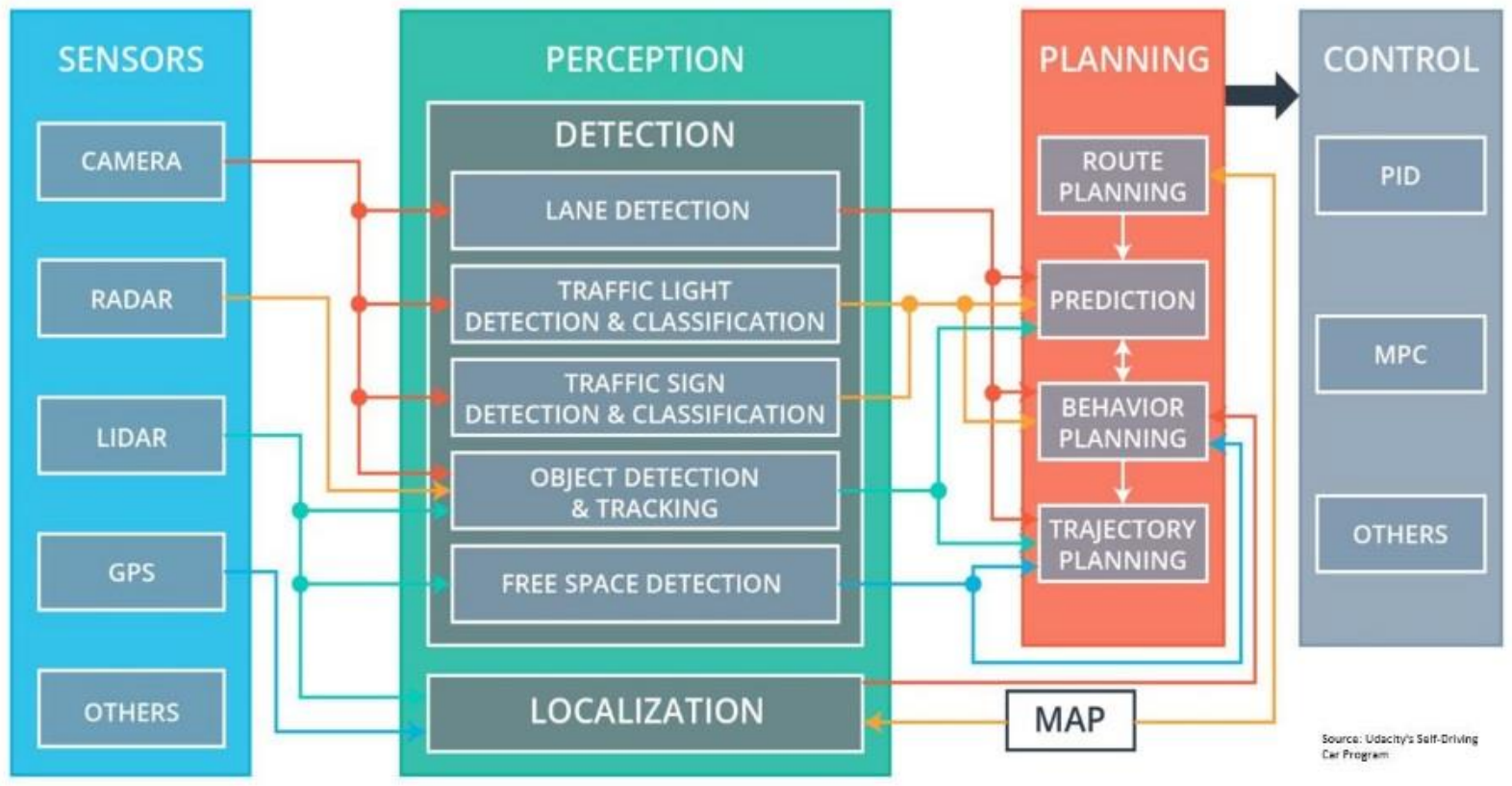

*Figure 1: Autonomous vehicle functionality breakdown. Source: Udacity Self-driving Program.*

Provided above is a breakdown of the software/core components for self-driving vehicles. Different sources will decompose the software stack with subtle variations. Regardless, the main areas of emphasis are in localization, perception, planning which can encompass route planning, behavioral prediction and vehicle trajectory generation, and finally control of the actual vehicle.

More concretely, a number of sensors, further addressed below, enable the vehicle to perceive its environment. High frequency data from cameras, LiDAR, radar and other sensors enable software and machine learning (ML) algorithms to detect relative distances, obstacles, pedestrians, and much more. The software responsible for utilizing sensor data forms the Perception components. Data from LiDAR is additionally used to localize the vehicle with centimeter accuracy, by comparing the generated 3D point cloud with high fidelity 3D maps of the environment stored on the vehicles central computer.

Perception data is then utilized to inform the software and AI/ML components responsible for planning an appropriate route through space. These allowable paths are generated by predicting future movement and behaviors of vehicles, pedestrians, etc. A cost analysis is done within the vehicle to select the most optimal path. From there a trajectory along the selected path is calculated and updated with feedback from inter-vehicle actuators and controls, along with respecting the vehicles dynamical model. Perception, Localization, Planning and Control form the brunt of an autonomous vehicles software stack's responsibility.

Due to the breadth, reliance, and importance of modeling sensors within simulation, it is worth further discussing the technology. Self-driving vehicles have a number of different sensors to provide situational awareness, localization and perception from LiDAR, radar of multiple ranges, GPS, ultrasonic sensors, and multiple cameras. General Motor's autonomous vehicle for example has five LiDARs, 16 cameras and 21 radars enabling it to perceiving the surrounding environment [20]. Each sensor specializes in a particular aspect of the perception problem as shown in Figure 2 and Figure 3.

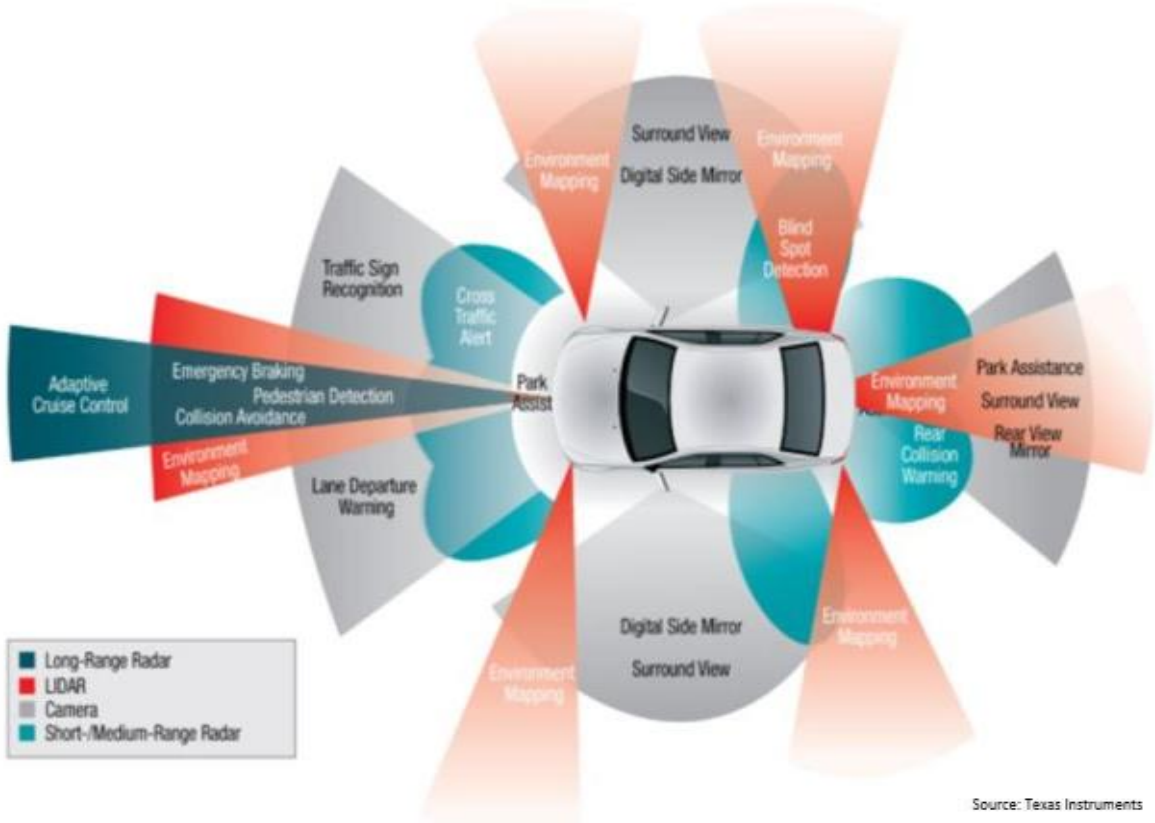

Figure 2: Self-Driving vehicles sensors.

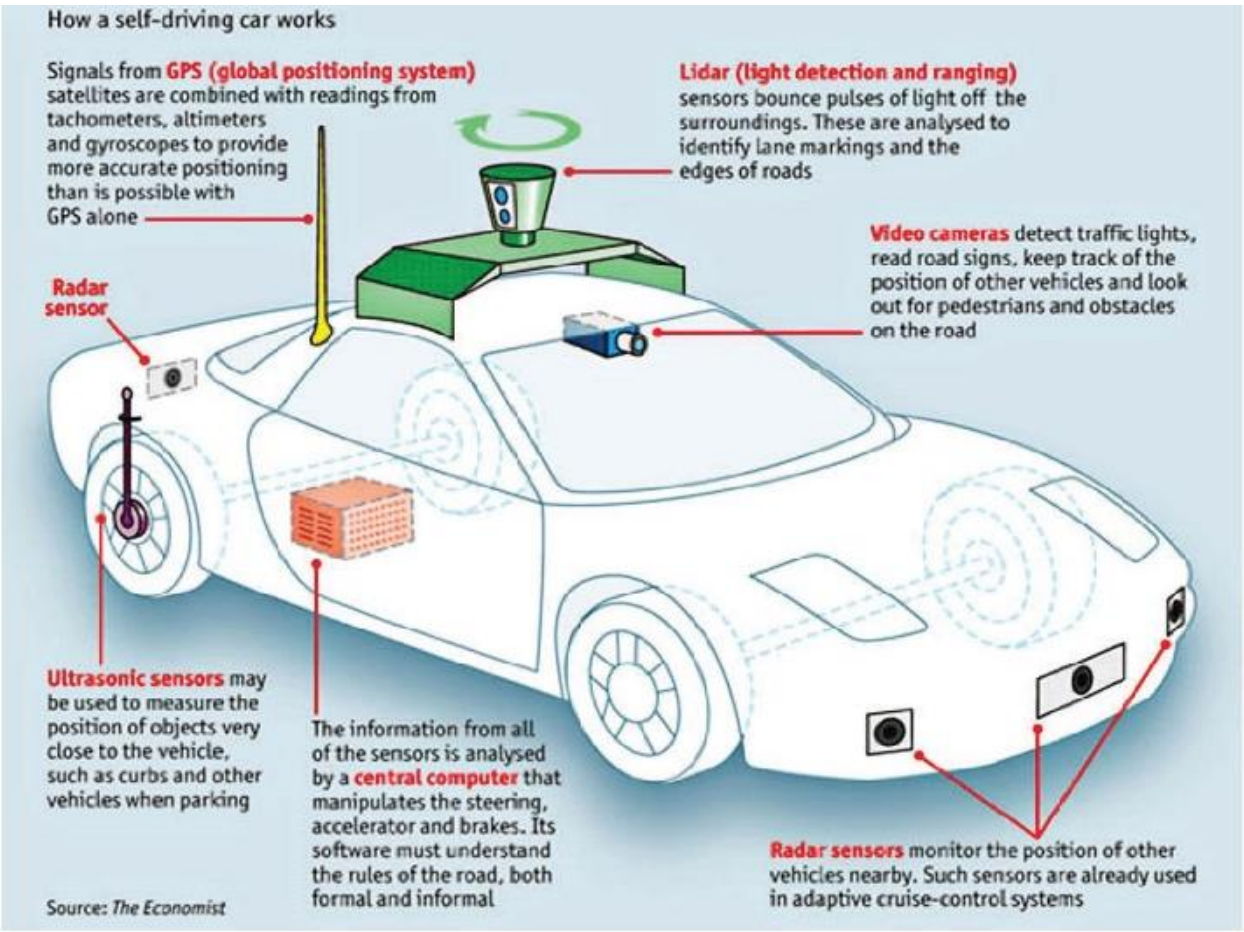

Figure 3: How self-driving cars work [21].

These sensors are being used for mapping, localization, and obstacle perception. LiDAR – Laser Illuminating Detection and Ranging, gathers high resolution geographical information with ranges up to 250 meters [8]. The LiDAR is used to build 3D maps allowing the car to foresee potential hazards by reflecting laser beams off surrounding surfaces. This enables the vehicle to determine the distance and profile of the subject accurately. Radar is used to map and monitor the speed and movement of the surrounding vehicles to avoid potential accidents, detours, traffic delays and any other obstacles by emitting and electromagnetic radio wave [21]. Cameras enable tracking, depth perception and object detection from pedestrians, to traffic lights and traffic signs. Ultrasonic sensors are used to detect and measure objects with close proximity to the vehicle. It is frequently used for automated parking in congested spaces and to sense curb locations [8].

## Autonomous Vehicles Value Proposition

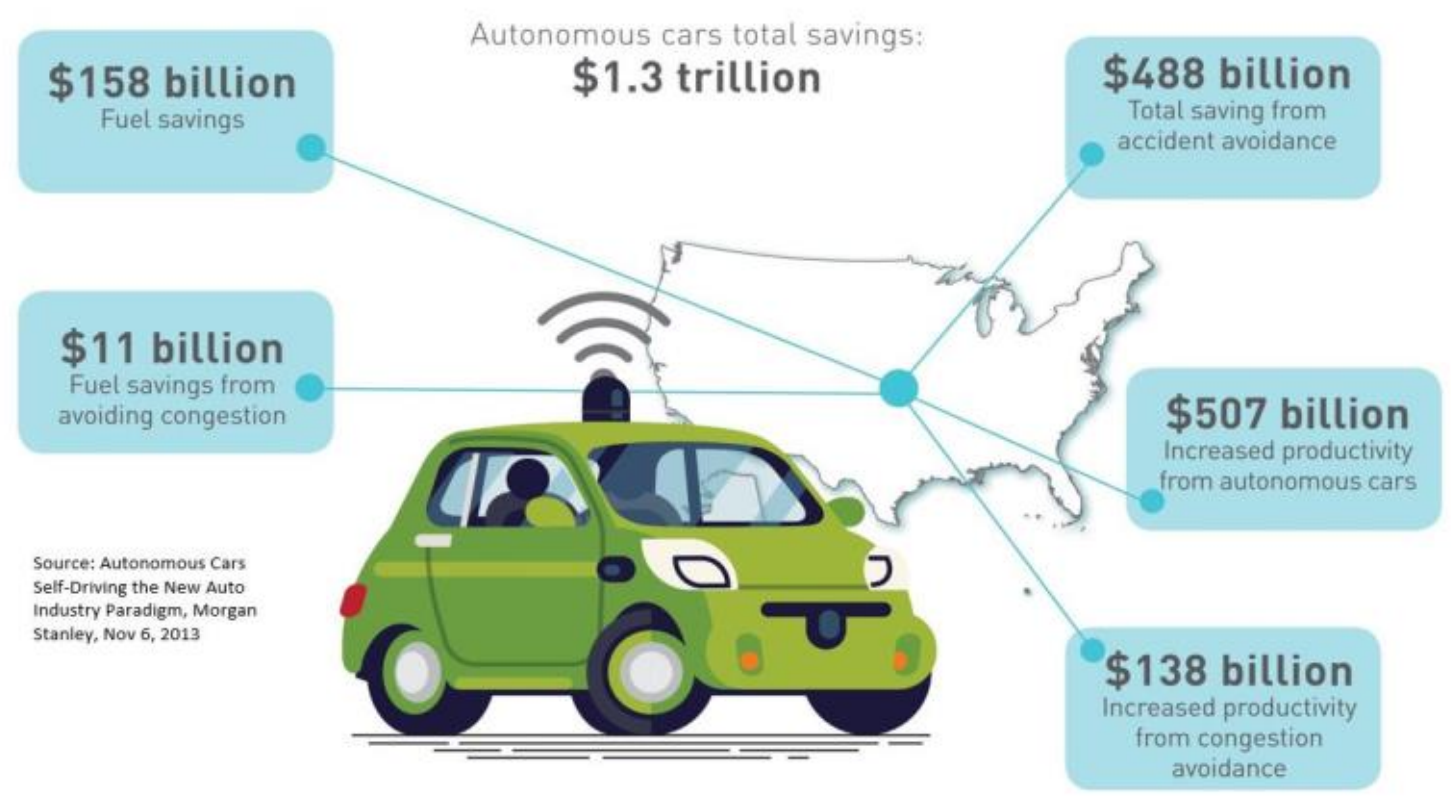

*Figure 4: Savings potential of autonomous vehicles according to Morgan Stanley.*

Futuristic technology aside, autonomous vehicles (AV) aren't without their value from an economic, sociological, and resource management perspective. Autonomous vehicles have the potential to save lives largely caused by human error. In the United States alone, roughly 32,000 people are killed and more than two million injured in crashes every year (Bureau of Transportation Statistics, 2015) [35].

More than 90 percent of crashes are caused by human errors (National Highway Traffic Safety Administration, 2015)—such as driving too fast and misjudging other drivers' behaviors, as well as alcohol impairment, distraction, and fatigue [35]. An autonomous vehicle is never drunk, distracted, or tired; these factors are involved in 41 percent, 10 percent, and 2.5 percent of all fatal crashes, respectively [35].

Aside from increasing safety, self-driving cars could reshape livelihood from a resource/time management perspective, while increasing the availability of transportation through democratization and crowd sharing. Each year, 42 hours of time per person are wasted driving in traffic [46]. Self-driving cars would eliminate much of this loss of time by removing the driver from responsibility, enabling the passenger to be more productive. On the other hand, there are many individuals unable to drive due to impairment and disability. 3 million Americans age 40 and older are blind or have low vision, and 79 percent of seniors age 65 and older live in car-dependent communities [46]. Intelligent cars would increase the mobilization this population.

Autonomous vehicles also lead to an improvement in urban land use by 15-20%, largely through elimination of parking spaces [21]. Self-driving vehicles could facilitate greater emphasis on accessible and shared transportation, increasing utilization from a single person. Instead of spending time looking for a parking spot, an autonomous vehicle could continuously drop off and pick up passengers.

## Increasing Need for Simulation

Simulation has an increasing role in facilitating development of this new transportation system both for land and aerial based autonomous systems. Simulation provides greater insight into all phases of the development lifecycle, across every domain level from mechanical subsystems, structural integrity, electronics, and software development. Two of the main contributors to the need for simulation toolsets, specifically within the autonomous systems domain, is through business value such as development speed, first time quality, and the infeasibility of utilizing real world testing alone.

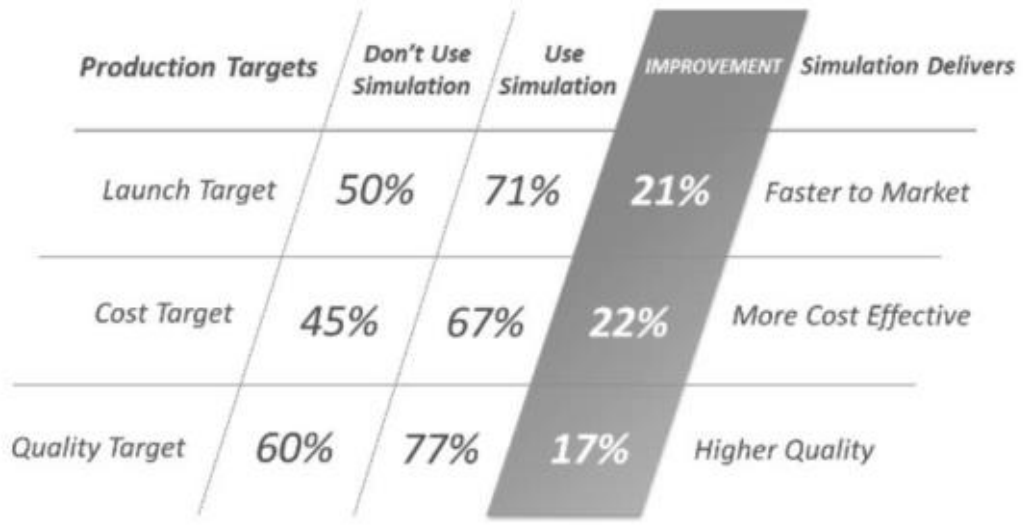

*Figure 5: Research published by Aberdeen advocating the benefits of simulation toolsets from industry surveys [5].*

Simulation capabilities have been shown to substantially enhance the ability to accelerate development throughout the product lifecycle and increase business value to stakeholders. As shown in figure 5, Aberdeen published research from industry surveys comparing companies that do and do not utilize simulation capabilities. Their findings show improved time to market, reduced development costs, and increased first time quality by 21%, 22%, and 17% respectively.

Furthermore, development of autonomous systems and validation of product design is infeasible without an increasing emphasis on simulation.  The RAND Corporation released research indicating the number of miles needed to validate self-driving vehicle performance against that of human drivers. A summarization of their finding is provided in Figure 6. Through statistical analysis, they concluded that best case scenario millions of miles are needed for comparison to a human. Worst case is in the billions of miles.

More concretely, a human driver has roughly 1.09 fatalities for every 100 million miles driven. In order to conclude with 95% confidence that autonomous vehicles decrease failure rates by 20%, 11 billion miles need to be driven, which would take approximately 400 years through fleet driving alone. The number of years was calculated by utilizing a fleet size of 100 autonomous vehicles driving non-stop every day each year at 25 miles per hour.

The research showcases not only the impressive benefit of simulation testing but the necessity in order to generate enough evidence for vehicle performance. To conclude, RAND advocated the need for alternative methods of testing in conjunction with real world testing. They suggested possible methods as “accelerated testing, virtual testing and simulations, mathematical modeling and analysis” [35].

| Statistical Question | How many miles (years[a]) would autonomous vehicles have to be driven... | Benchmark Failure Rate<br>(A) 1.09 fatalities per 100 million miles? | (B) 77 reported injuries per 100 million miles? | (C) 190 reported crashes per 100 million miles? |
|---|---|---|---|---|
| | (1) without failure to demonstrate with 95% confidence that their failure rate is at most... | 275 million miles (12.5 years) | 3.9 million miles (2 months) | 1.6 million miles (1 month) |
| | (2) to demonstrate with 95% confidence their failure rate to within 20% of the true rate of... | 8.8 billion miles (400 years) | 125 million miles (5.7 years) | 51 million miles (2.3 years) |
| | (3) to demonstrate with 95% confidence and 80% power that their failure rate is 20% better than the human driver failure rate of... | 11 billion miles (500 years) | 161 million miles (7.3 years) | 65 million miles (3 years) |

[a] We assess the time it would take to compete the requisite miles with a fleet of 100 autonomous vehicles (larger than any known existing fleet) driving 24 hours a day, 365 days a year, at an average speed of 25 miles per hour.

*Figure 6: Statistical analysis of miles driven for comparison of autonomous system performance to human drivers [35].*

## Key Players Overview

Provided below is a summary of the key industry players mentioned throughout this paper. These companies come from a number of different backgrounds and specializations from commercial technology, automotive vehicle suppliers, manufacturers, simulation experts, startups, etc. Each of these companies has provided capabilities either directly or indirectly contributing to the development of high fidelity simulation toolsets for autonomous vehicles. While a breadth of industry contributors were analyzed, some of the other major autonomous vehicle suppliers or well-known startups in the market such as Tesla, Baidu, or Drive.ai were not mentioned explicitly due to a commonality in approach similar to other companies discussed, or emphasis on data generation through fleet management over simulation (such as Tesla).

| **Technology Companies (Software, Hardware, etc.)** | |
|---|---|
| Waymo (Google) | Waymo performs a comprehensive mix of simulation (Waymo 'Carcraft'), closed course testing (in their 'Castle' facility), and real world testing. Their simulation environment and vehicle development is mature by comparison to some of the other industry members. Their simulation environment does not cover the perception problem (i.e. it does not use realistic graphics). Waymo has 4 generations of self-driving vehicles, 9 years self-driving in more than 25 U.S. cities, 5 million real-world miles on public roads, and 5 billion self-driven miles simulated [46]. |
| Uber | Uber provides a sophisticated simulation environment for autonomous vehicles. Similar to Waymo, their simulation environment does not provide photo-realistic graphics [45]. Additionally, Uber provides open source visualization toolsets for analyzing mobility and geography [11]. |
| NVIDIA | NVIDIA provides expertise in simulation, hardware, large scale infrastructures, and data analysis/utilization. NVIDIA's autonomous vehicle portfolio of technology and services include DRIVE AGX on-board computers, DRIVE Constellation driving simulation software, and Project Maglev. NVIDIA provides technology for autonomous systems level 2-5. They have support from Daimler Benz, Bosch, Continental, Audi, Toyota, etc. [33]. |
| **Open Source Simulations and/or Providers** | |
| CARLA | CARLA (Car Learning to Act) is an open source simulator for self-driving cars developed in partnership between Intel Labs, Toyota Research Institute and |
| | the Computer Vision Center, Barcelona. CARLA is a service oriented, high fidelity, realistic graphics environment built on top of Unreal Engine. The simulation environment allows testing from perception to vehicle control [15]. |
| Microsoft AirSim | AirSim is an open source simulator for autonomous systems created by Microsoft [17]. The platform is intended for experimentation with AI, deep learning, computer vision and reinforcement learning. AirSim provides realistic vehicle dynamics and environments for drones, cars, and more through Unreal Engine and Unity game engines [31]. |
| **HD Mapping Companies** | |
| HERE | HERE specializes in constructing high fidelity mapping of an environment using sensor equipped autonomous vehicles and crowd-sourced resources. HERE works with industry partners such as Audi, BMW, Daimler, etc. to provide their autonomous vehicles with continuously updated HD Maps for localization. HERE HD Live Map initially covers North America and Western Europe with ongoing global coverage roll-out [24]. |
| Mapper.ai | Mapper.ai is building high fidelity maps utilizing autonomous vehicle partner companies and freelance mappers. The maps are updated real-time, precise and customizable to customer needs  [19]. |
| **Automotive Companies** | |
| BMW | BMW utilizes a mix of simulation, hardware in the loop testing, high fidelity simulators, closed course and real world testing to develop there fleet of autonomous vehicles, which in 2017 comprised of 40 7-Series vehicles [6]. The BMW Group is building a new Driving Simulation Centre with an estimated cost of 100 million euros [7]. |
| Ford | Ford validates their self-driving vehicles through a three step process of simulation, closed-course testing and real world testing. Ford has similar simulation capabilities to GM, Waymo and Uber. Ford partnered with Argo AI to develop the “brains and senses” of their AV [18]. |
| GM | GM is developing autonomous vehicles, notably CRUISE AV which has five LiDARs, 16 cameras and 21 radars to handle perception. They are testing in Phoenix and San-Francisco [20]. GM has several high fidelity simulators which have been used to test vehicles such as the Cadillac Super Cruise, CT6 Sedan, etc. These 360-degree simulators enable roll, pitch and yaw of a vehicle to simulate movement, and monitor biometrics and facial expressions [47]. |
| **Simulation Toolset Providers** | |
| DSpace | DSpace creates systems to develop and test electronic control units and mechatronics for a wide range of industries: automotive, aerospace, defense, industrial automation, etc. DSpace provides the Automotive Simulation Models tool suite featuring Simulink (MathWorks product) models for simulated testing of passenger cars and trucks, as well as their components. [16]. |
| ANSYS | ANSYS provides modeling and simulation tools for mechanical analysis, fluid analysis, thermal analysis, systems analysis, functional safety, display systems and more. They provide a full suite of tools aimed at development of vehicles in high-fidelity environments. ANSYS is moving autonomous driving and active safety systems development to the digital twin [8]. They provide an |
| | autonomous vehicles open simulation platform for the development/production lifecycle [3]. |
| RightHook | RightHook is a startup providing simulation toolsets for autonomous vehicles. RightHook has a number of tools available: Traffic simulation, RightMap for 3D physics based environments using HD Maps data, hardware in the loop, RightWorldHD for software testing of virtual drivers, and scenario generation [38]. |
| Cognata | Cognata is a startup specializing in testing and evaluation of self-driving cars through high fidelity, realistic graphics simulation. Cognata uses computer vision and deep learning algorithms to automatically generate a simulated environment [12]. Cognata has partnered with NVIDIA, and Microsoft to build a cloud-based simulation platform [39]. |

## Rational for Autonomous Vehicle Simulation as Use Case

The state of modeling, simulation and data utilization in autonomous systems, namely from the perspective of self-driving vehicles, is a valuable use case and ultimately a good indication of state of the art. The primary reason is that autonomous systems and their associated simulations are a defining feature of future smart cities. Further, the simulation toolsets and capabilities encompassed within self-driving vehicles (land and air based) touch on many of the core technologies for smart cities such as advanced sensor systems, integration with machine learning, appropriate tools to visualize and analyze the urban space and associated data, inter-vehicle and environment communication, to name a few. A more detailed explanation is summarized below:

- Simulation toolsets for autonomous systems, in general, include the desired fidelity for modeling and analyzing smart cities, as well as encompass the full-pipeline from electronics and materials simulation to mission systems
- The lines between the autonomous vehicle (ground based) industry and aviation industry is being blurred, especially with the advent of future consumer transport systems (i.e. autonomous air taxis). Additionally the complexity of the problem shares many of the same features to the aviation industry from routing and planning, logistics, perception, intra and extra networking, communications, controllability, observability, and safety.
- The sensor technologies used have much overlap to the aviation industry.
- Future competitors within the autonomous vehicles (air, ground and everything in between) are striving to gain leverage in the looming market (from Waymo, GM, Ford, Uber, etc.). With that in mind, it is imperative that aviation industry understand, and leverage the technologies developed within these industries, particularly in regards to simulation environments, and supplemental toolsets and techniques.
- The AV simulation environment, particularly for mission level simulations, contain much of the details and models necessary for analyzing future smart cities, in general both urban and rural environments settings. These simulations tools are capable of simulating visually realistic, high fidelity 3D models of an urban environment incorporating both static (street signs, stop lights, building, structures, etc), dynamic landscapes (pedestrians, other vehicles, etc.), and physical phenomena (weather, physics, etc.).

For these and a number of other reasons explored within the document, autonomous systems and self-driving vehicles was taken as an example in understanding state of the art simulation toolsets.

## Insufficiencies of Alternative Simulation Domains

There are a number of other simulation domains - robotics, the game industry, manufacturing and smart factories, urban planning and management, and simulation devices: augmented reality (AR), virtual reality (VR) that could have been insightful use cases. However, many of these alternative simulations are highly domain specific, and in other cases lack the level of fidelity, detail, controllability, and environmental conditions needed to model smart cities.

For instance open source simulators and consumer gaming are an interesting use case, but often lack the robustness, extensibility and infrastructure necessary to model such complex phenomena inherent in the autonomous systems environment. “Open-source racing simulators such as TORCS do not present the complexity of urban driving: they lack pedestrians, intersections, cross traffic, traffic rules, and other complications that distinguish urban driving from track racing. And commercial games that simulate urban environments at high fidelity, such as Grand Theft Auto V do not support detailed benchmarking of driving policies: they have little customization and control over the environment, limited scripting and scenario specification, severely limited sensor suite specification, no detailed feedback upon violation of traffic rules, and other limitations due to their closed-source commercial nature and fundamentally different objectives during their development” [15].

Additionally, there is strong overlap between simulating autonomous systems and other domains. Much of the sensors, actuators, control systems, and software tested via simulation are inherent aspects of robotics, manufacturing and smart factories. Further with the level of fidelity in AV simulations (i.e. through utilization of HD maps for self-driving car localization and simulation construction), a comprehensive analysis of urban planning, management, movement and communication could be performed directly or with some modifications.

# Simulation Overview

In general simulation tools enable the virtual replication of physical phenomena for analysis and development. This section will provide a brief overview into the purpose of simulation tools and various software/modeling methodologies utilized, as well as techniques used for programming systems within simulated environments.

According to ANSYS, (a major developer and supplier of simulation toolsets) simulation provides three broad benefits: [5]

1. Faster time-to-market: Simulation is conducted in a virtual environment and is significantly faster than physical prototyping and testing, expediting a new product’s time-to-market.

2. Reduced cost: Being virtual, simulation is far less expensive than physical prototyping and testing, and can cut costs by an order of magnitude.
3. Enhanced product quality: Simulation provides deep insights into the underlying physics involved in the construction and operation of a product, helping solve quality issues upfront before physical development of the product and throughout the development lifecycle.

In general, simulation tools enable developers and stakeholders to ask many "what if questions" within their analyzed domain. Simulation enables users to modify key parameters within their system and its modeled ecosystem, run the scenario, and then collect data for post processing and analysis. Further a simulation toolset enables the user to more strictly control the environment and conditions under which a system functions. This can be helpful for reproducing conditions that are difficult to experience or uncommon in the real-world.

Systems and the scenarios under which systems operate within simulation can be programmed through a number of different methodologies from sophisticated user interfaces, object oriented programming, agent-based design concepts, and model-based design as examples. The latter three will be briefly discussed. Examples and explanations of such techniques are given throughout the paper.

## Object Oriented Simulation Programming

Object oriented programming is a traditional development methodology used for interacting with, and constructing the simulated environment and underlying models. Object oriented design and programming is commonly used within software engineering to organize the hierarchical interactions and associations between various subcomponents. Such design typically groups systems and their underlying sub-systems and components into a class based architecture, where each class has a number of attributes, and methods that can be performed. In the case of autonomous systems, these classes could be abstract classes such as 'Vehicle' which has abstract turn, accelerate, decelerate, etc. methods, or concrete classes such as GM's Cruise AV which might derive from the Vehicle class and have methods that provide vehicle specific implementations for the accelerate, decelerate, and turn methods.

## Agent Based Simulation Programming

An agent-based paradigm can be considered an extension to object-orientation, whereby an agent represents an object having control of its execution/actions based on its understanding of own-ship and environment state, as well as internal memory [1]. Instead of being expressed in terms of attributes and logic-based methods, a software agent is primarily described in terms of its intended actions. This can be described as stating agents' responses instead of identifying classes, methods and properties [1]. "Agent Based Modeling and Simulation (ABMS) refers to a category of computational models involving dynamic actions, reactions and intercommunication protocols among the agents in a shared environment in order to evaluate their design and performance and derive insights on their emerging behavior and properties" [1].

## Model Based Design

Model based design abstracts way some of the specificities of object oriented development and in general document based software development, in favor of code/module reusability, development speed and enhanced conceptualization of the system/sub-system. A diagram of model-based design is provided below. In this example from ANSYS, the SCADE model based design framework is being utilized to construct the state machine logic and interactions for cruise control. Through model based design systems are not constructed via document level software programming (line by line programmed code), but at the module/model level where each model has a distinct set of inputs and outputs and relations to other models within the system.

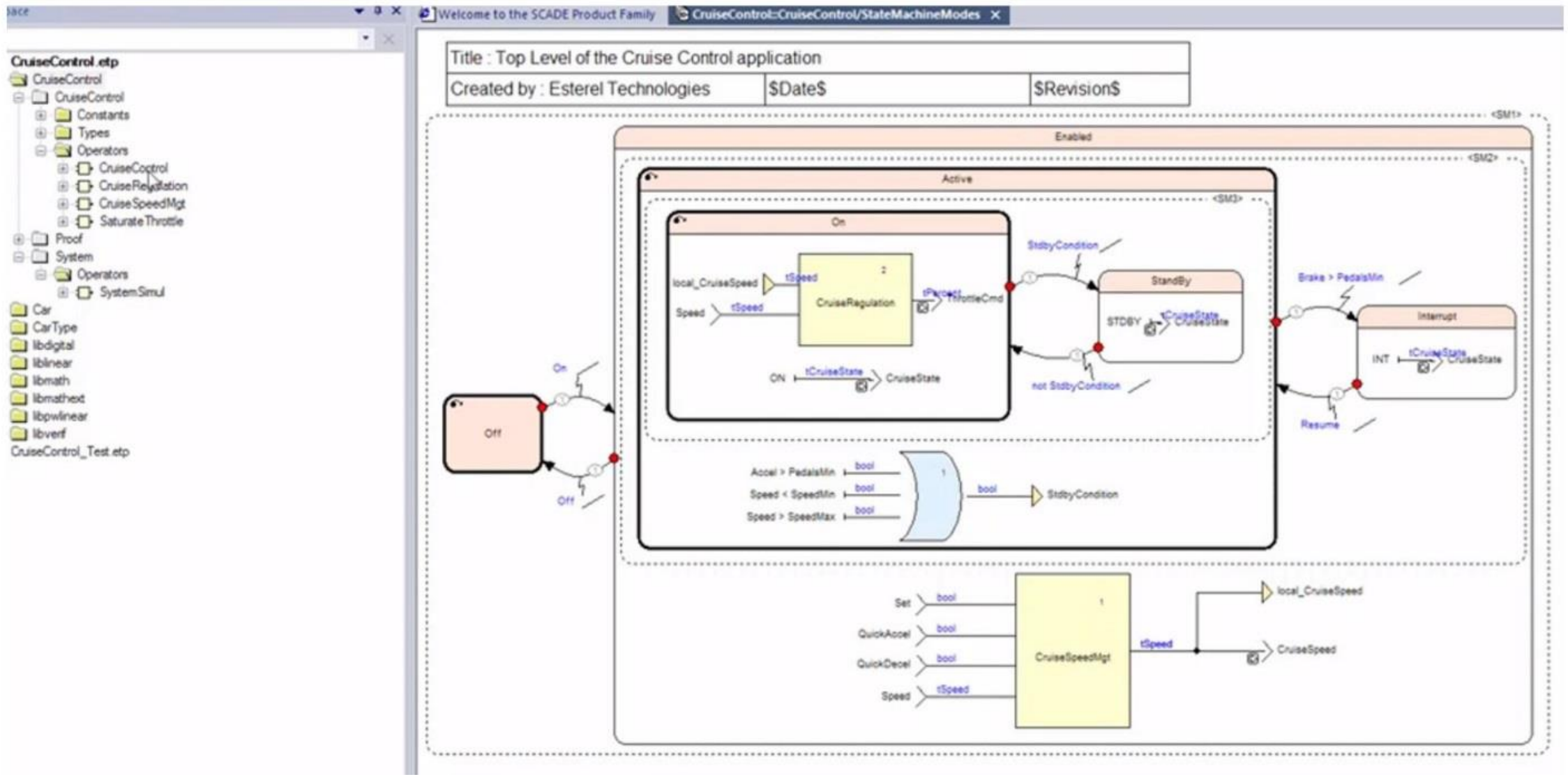

*Figure 7: Example of Model-Based Design for cruise control application [3].*

# Autonomous Vehicle Simulations

There are a number of simulation toolsets needed in order to design, test and validate the construction and implementation details of autonomous systems. The following subsections discuss the various types of simulation toolsets from defining the physical structure of the vehicle and subcomponents to validating software infrastructure. While multi-physics, sensor, functional safety and vehicle dynamics related simulations are discussed, the majority of this subsections will discuss mission/scenario level simulation. Mission level simulation enables testing of the software stack in a realistic virtual environment.

The diagram in Figure 8 provides such an overview of the various simulation toolsets. While the diagram is not all encompassing, it provides insight into the numerous simulations tools that need to be represented throughout the full development lifecycle of an autonomous vehicle. The pipeline of simulation tools will be discussed. Additionally, important characteristics of mission level simulation tools are addressed throughout this section, from visualization/visual fidelity, controllability, extensibility, simulation interfacing, parameterization of the scenario, usability, reusability, throughput, experience replay, and software behavior visualization.

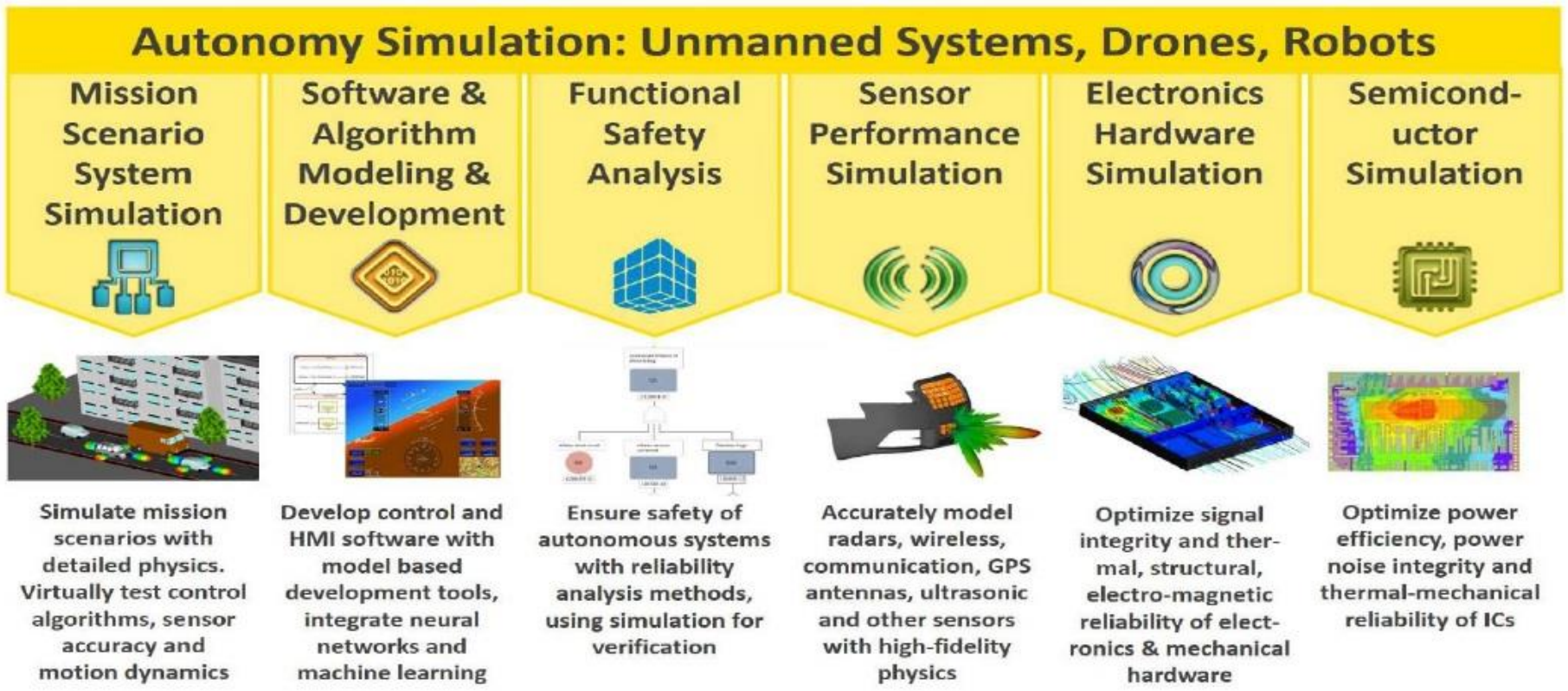

Furthermore, discussion on HD Mapping will be provided. HD Maps are foundational to many industry leaders construction of autonomous systems simulations.

*Figure 8: Different layers of autonomous systems simulation toolsets [42].*

## Mission Level Simulation Environment

In general, the simulation environment discussed will be mission level simulation which regards the actual behaviors, prediction, and control of the vehicle in a complex environment. Mission level simulation toolsets test (or have the capability to test) the complete pipeline of vehicle components; components which utilize the inputs and outputs to control actuators, the vehicle dynamics model, sensors and sensor processing, software control algorithms, which could be provided as software in the loop (SIL) or embedded on the actual hardware running in the loop (HIL). Further, mission level simulations model the 3D scenarios under which an autonomous vehicle will be operating - from the motion of moving objects such as pedestrians, vehicles, cyclists, animals, etc. to the various terrains in which a vehicle maneuvers. A summary of models contributing to mission level simulation tools is provided in Figure 9.

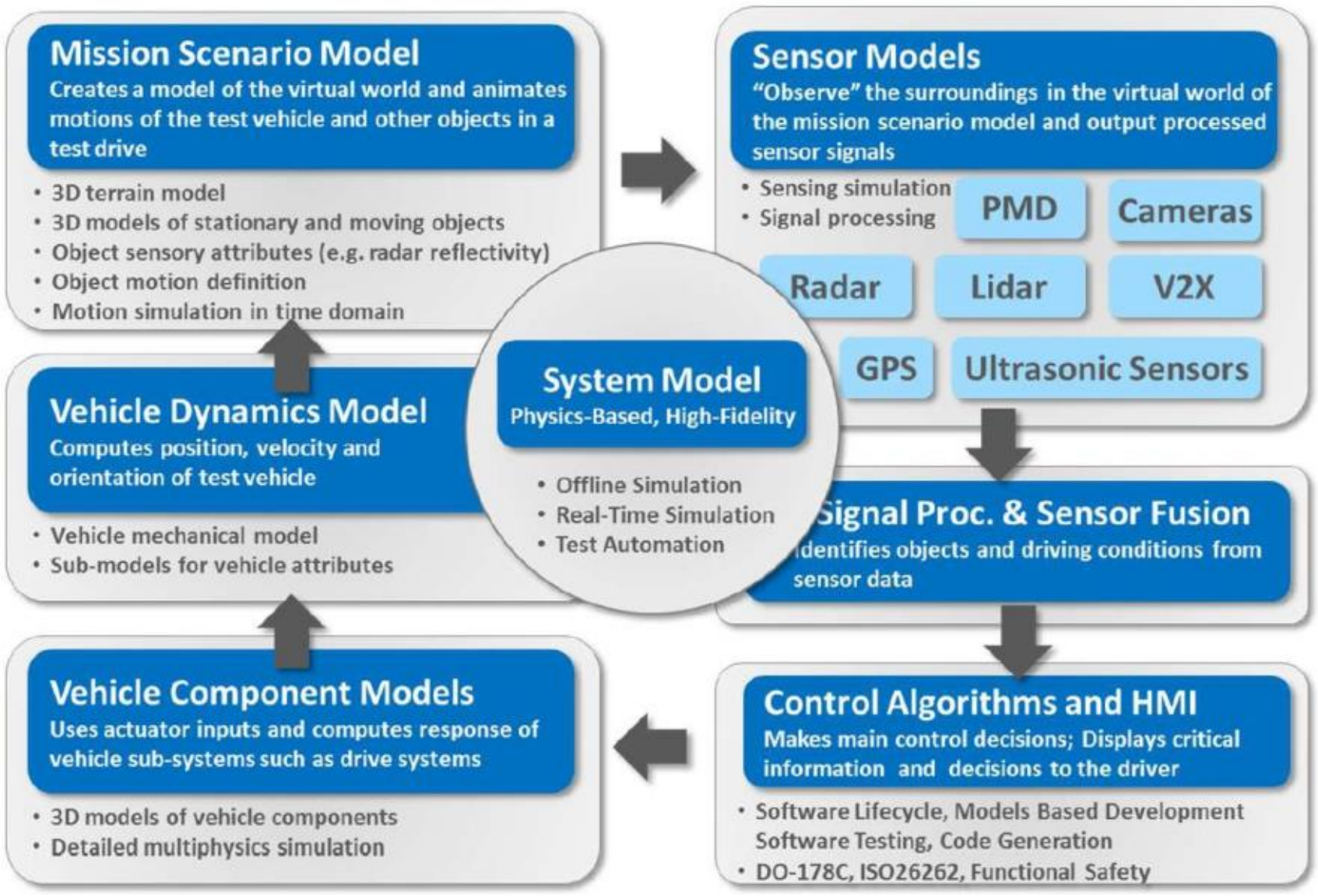

*Figure 9: Overview of Mission Level Simulation toolsets. [42].*

## Construction of Simulation

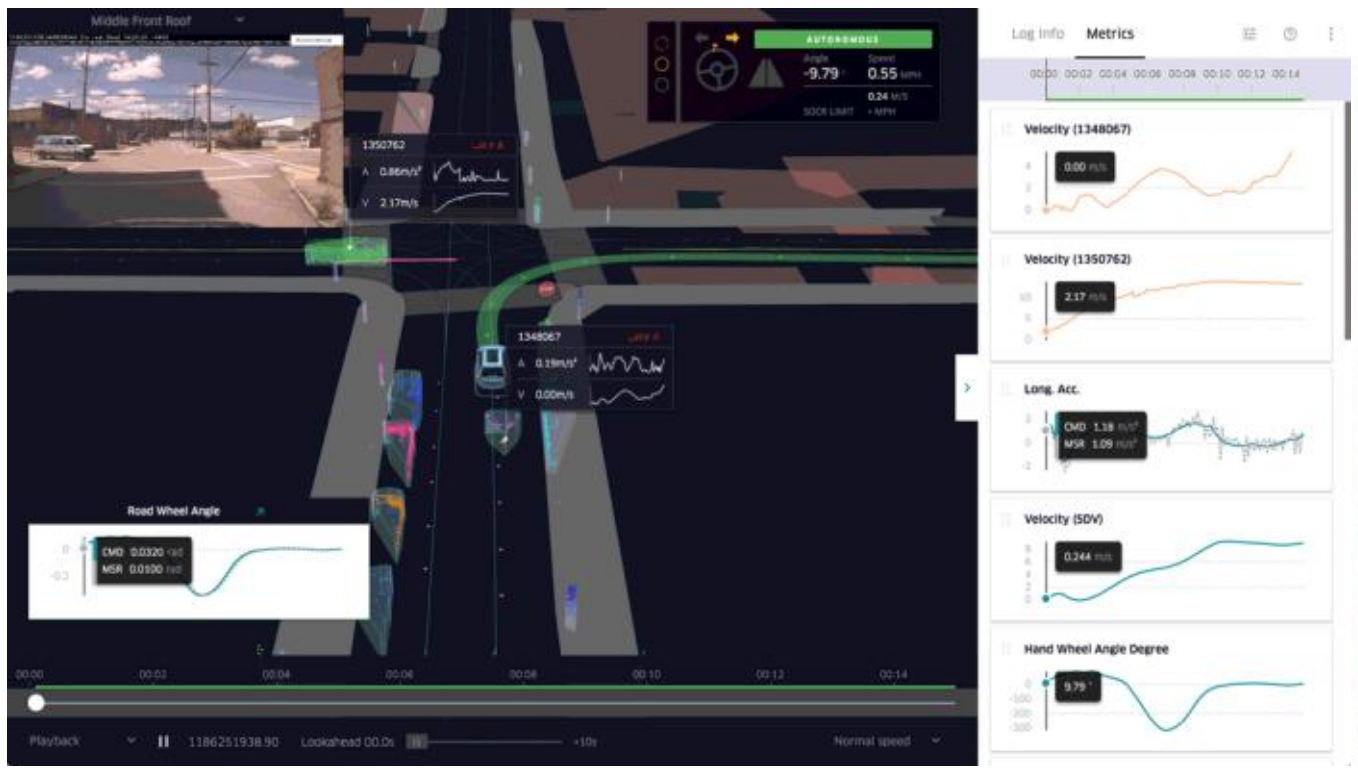

*Figure 10: Uber's autonomous driving programs simulation toolset [10].*

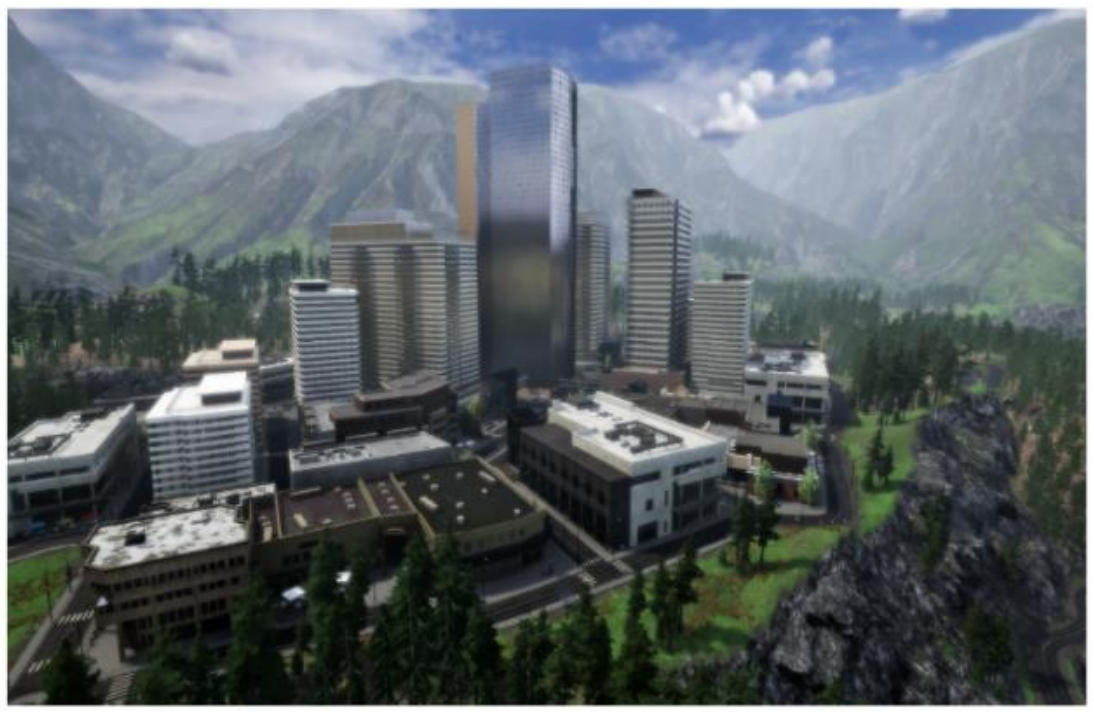

*Figure 11: Microsoft AirSim Windridge City environment in Unity [17].*

The construction of a state of the art autonomous vehicle simulation environment is not a straight forward endeavor. There are several approaches toward constructing a simulation environment from utilizing HD Maps of a real space as the base foundation, solely relying on game engines, or a combination thereof. Further, the level of detail and graphics/visualization of the environment differ greatly between competing companies. Take Waymo and NVIDIA as an example. Compared to NVIDIA, Waymo does not utilize a photorealistic environment. However, there does not appear to be a 'correct' method determining the level/amount of visualization.

This section will expose some of the techniques utilized by autonomous vehicle companies, open source consumer environments, and startups alike. The details of the models implemented within these environments will be discussed in later sections.

There are other perspectives on creating photorealistic and high fidelity simulated environments, which are not explored within the paper. Such approaches are using AI to generate the simulated environments automatically from HD Maps and other data sources. Companies like startup Cognata are exploring these approaches.

### *HD Mapping*

HD Maps are a crucial feature for the development of autonomous road vehicles. Not only are HD maps intended to provide real time feedback of a vehicles environment, enable precise localization of the vehicles while navigating through a space, but they are also heavily used in the construction of modern simulation environments for companies such as Waymo [46], Uber [10], and NVIDIA [22]. This section discusses construction of simulation via High Fidelity Maps of the environment. A more detailed explanation of how HD maps are created, and the types of information contained are within the section discussing High Fidelity Mapping.

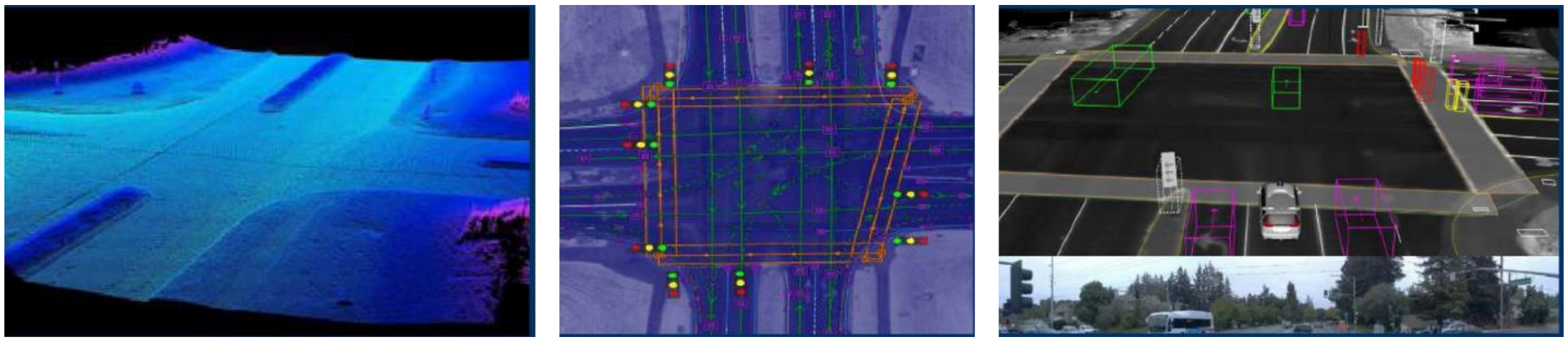

*Figure 12: Waymo's method for constructing their simulated environment. (Left) The base foundation for simulation is generated by mapping a real environment via LiDAR and other sensors data feeds. (Middle) Salient information such as location of street signs, traffic lights, distances between road lanes, etc. is then added to the map foundation. (Right) From here vehicles, pedestrians, and other dynamic features are added to the environment (vehicles are green and purple boxes, pedestrians yellow, cyclists in red) [46].*

An overview of Waymo's methodology for constructing their AV simulation environment is provided in Figure 12. The base foundation for their simulation is generated from data gathered via LiDAR and other sensors, constructing a map of the real environment. Waymo, HERE, Uber and a number of other companies have vehicles configured with the appropriate sensors and computer's necessary to map out an entire environment. For instance, before Waymo can develop a vehicle capable of driving through a Phoenix city suburb, Waymo has safety operators manually drive the test vehicles to map out the operational domain precisely. This provides centimeter level accuracy between the real and simulated world.

Once the base foundation of the environment is generated, salient information such as location of street signs, traffic lights, distances between road lanes, direction of traffic flow, etc. are then added to the map. From here vehicles, pedestrians, and other dynamic features can be simulated within the virtual environment (vehicles are green and purple boxes, pedestrians yellow, and cyclists in red as shown in Figure 12). The dynamic features of the environment could be generated from real world vehicle logs collected from driving through an urban space, or purely through user designed simulation.

It is important to note that the simulation environment produced by companies such as Uber, and Waymo, as shown in the images above, do not provide sufficient levels of detail for handling the Perception problem. In the case of Uber, HTML5 and JavaScript are utilized to construct the visualization aspect of their environment, instead of realistic game engines [10]. More concretely, Waymo's and Uber's simulation environment do not test the AV software's ability to detect pedestrians, vehicles, and cyclists, nor does it test the ability of a vehicle to localize itself in space. This part of the design and development challenge is handled externally from their mission level simulation. Waymo and Uber's simulation environment provide the ability to test path planning, behavioral prediction, trajectory planning and actuation of vehicle's controls.

However, while the perception problem is not directly modeled within this mission level simulation environment, it does not mitigate the capability and relevancy of the tool. "The power of the virtual worlds of Carcraft is not that they are a beautiful, perfect, photorealistic renderings of the real world. The power is that they mirror the real world in the ways that are significant to the self-driving car and allow it to get billions more miles than physical testing would allow. For the driving software running the simulation, it is not *like* making decisions out there in the real world. It is *the same* as making decisions out there in the real world" [29].

### *Photorealistic Simulation (Including and Not Including HD Maps)*

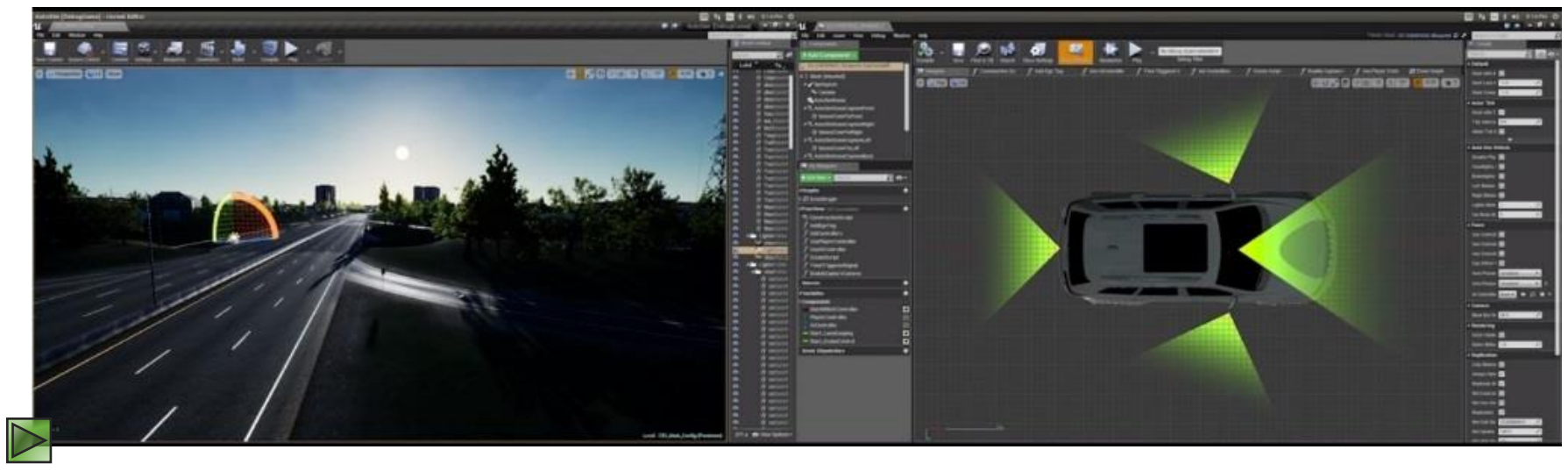

*Figure 13: NVIDIA DRIVE Sim video combining photorealism and HD Mapping (video provided via YouTube).*

Compared to Uber and Waymo, other companies take a different approach by including enough details in the simulated environment to consider the Perception problem. Some of these companies such as Microsoft's open source AirSim platform, and Toyota Research's CARLA simulation solely rely on game engines such as Unity and Unreal Engine which can provide photorealistic rendering. AirSim and CARLA, do not utilize HD Maps as a foundation for simulated arenas. These simulation tools leverage company designed models, as well as crowd-sourced asset libraries to construct the simulated world from the ground up.

The group responsible for CARLA used these asset libraries to construct their environment. The environment is "composed of 3D models of static objects such as buildings, vegetation, traffic signs, and infrastructure, as well as dynamic objects such as vehicles and pedestrians. All models are carefully designed to reconcile visual quality and rendering speed: we use low-weight geometric models and textures, but maintain visual realism by carefully crafting the materials and making use of variable level of detail. All 3D models share a common scale, and their sizes reflect those of real objects. At the time of writing, our asset library includes 40 different buildings, 16 animated vehicle models, and 50 animated pedestrian models. We used these assets to build urban environments via the following steps: (a) laying out roads and sidewalks; (b) manually placing houses, vegetation, terrain, and traffic infrastructure; and (c) specifying locations where dynamic objects can appear (spawn)" [15].

On the other hand companies such as NVIDIA, as shown in Figure 13 provide a toolset leveraging a combination of both techniques: HD mapping to model an environment precisely coupled with photorealistic visualization of a space.

## Environment Models

In order to construct a simulation that appropriately mimics the environmental conditions, a number of models need to be incorporated. The breadth, fidelity, and diversity of these models is certainly dependent on the type of problem you intend to solve with the simulation toolset. For example if the user is only interested in testing software functionality regarding path planning then less emphasis needs to be placed on the graphical representation and diversity of the pedestrians and vehicles in the environment and more emphasis on the behaviors/dynamics of the objects in order to appropriately capture real life movements. On the other hand if the user desires to test the full software stack, including perception related algorithms then the simulation environment needs to include realistic diversity in the visual representation of vehicles, pedestrians, traffic signs, etc.

This section aims to describe and highlight the particular models necessary in both photo-realistic and non-photorealistic environments. When necessary, models that apply to particular types of simulation (photorealistic and non-photorealistic) will be mentioned.

Provided in Figure 14 is an overview of the primary models that need to be included in most AV simulations. The compartmentalization of the various model types as provided by NVIDIA is not the only method for organizing such details. However, in general it is fairly representative of the features that exist in most companies' frameworks from Uber, GM, Waymo, etc.

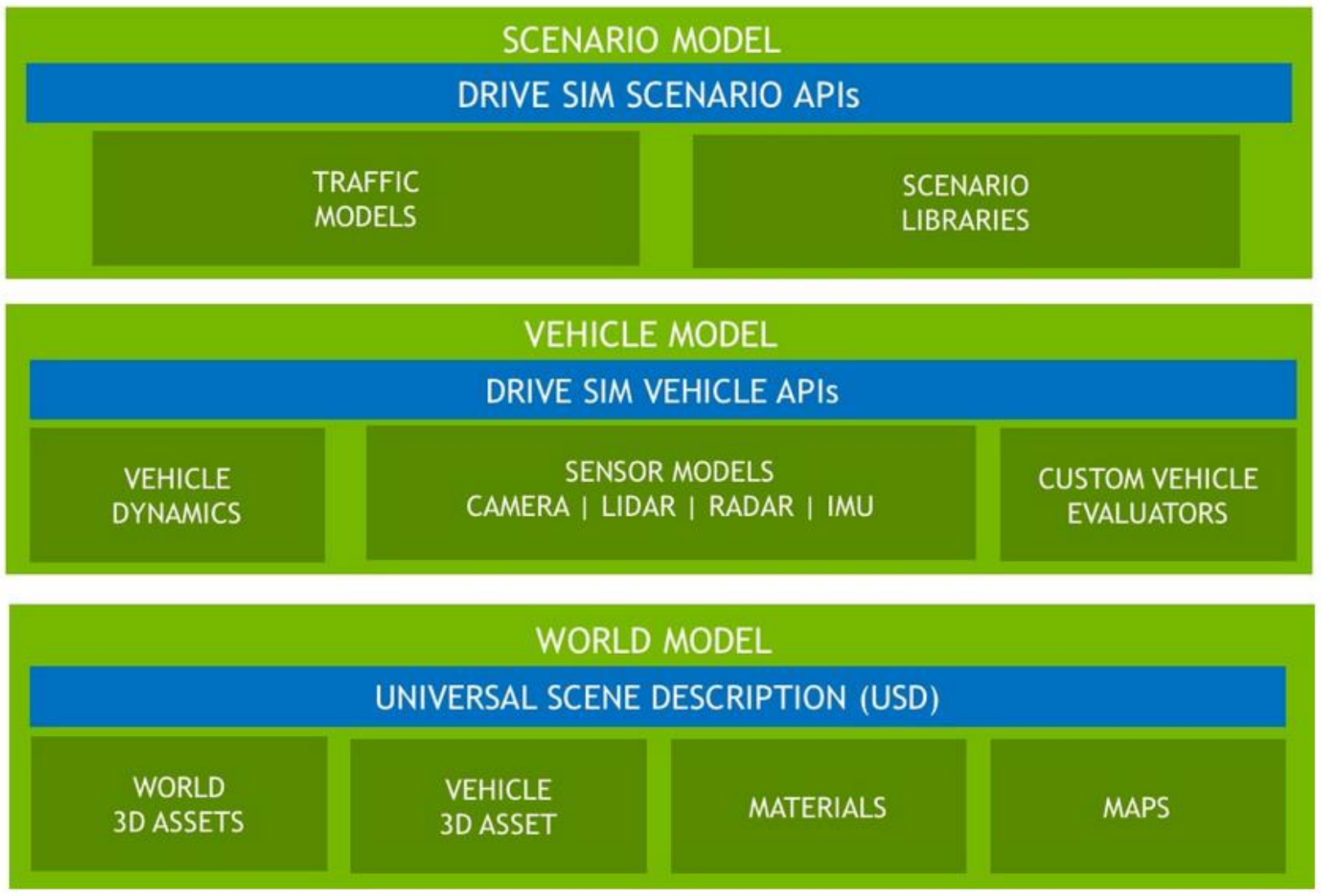

*Figure 14: NVIDIA Drive Constellation's Drive SIM simulation toolset. Models contained within the simulation [22].*

At a high level, the world model provides an accurate representation of the area under which one intends to drive. For instance, if a vehicle is driving through a simulated version of San Francisco, the virtual environment needs to mimic the 3D assets present within that environment from terrain (such as elevation, ground surface types), traffic signs, traffic lights, roads, intersections, highways, buildings, trees, and the various dynamic objects such as animals, pedestrians, and vehicles (which can exhibit much diversity and individuality). In many cases, the virtual model of the world is highly representative/intended to be a direct copy of the real world environment. A vehicle can drive from a clear sunny morning in Mountain View, California, through dense fog in San Francisco without ever leaving the simulation platform [22]. Further, depending on the visual fidelity, textures, lighting, atmospheric conditions, weather, etc. need to be included as well.

The vehicle models include the dynamical characteristics and configuration of the autonomous system. Vehicle dynamics could include the various degrees of freedom, motion control, center of mass and inertial properties, suspension, braking, acceleration/deceleration, powertrain, etc. The virtual car must also be able to behave in simulation as it would in the real world. Actions like braking, accelerating onto a highway or driving over a bumpy road should exhibit the same vehicle dynamics as if they were actually happening to the vehicle [22]. For the sensors, this could include the various sensors connected to the vehicle such as IMU, camera, LiDAR, radar, GPS, ultrasonic sensors, etc. A self-driving car might have various combinations of these sensor systems at any time. Additionally, the models for each sensor enable the configuration of their location and orientation in relation to the external environment and vehicle itself [38]. Note, in the case of less photo-realistic simulations tools, the level of fidelity needed for the sensors is lower. In less visually realistic environments, the sensor model might just provide a probability distribution for object detection and/or simple range values/field of view for each sensor type.

The scenario models might control the higher level features of the environment such as direction and density of traffic flow, the location and area in which the simulation takes place, the number and types of vehicles within the particular environment simulated, etc. Scenario models describe the environment under which an autonomous system acts by choosing the appropriate lower level models (vehicle models, sensor models, world models, object dynamics, etc.) that are desired within the virtual test.

The following sub-sections shed some additional light into the various models that need to be present within the simulated environment. Note: Not all models are discussed in further detail.

### *Static Environmental Elements*

The environment is composed of many 3D models for static objects such as buildings, vegetation, traffic signs, infrastructure, road types, lane markings, and the equipment/items carried by the pedestrians. Figure 15 highlights the static as well as dynamic models and attributes within an environment.

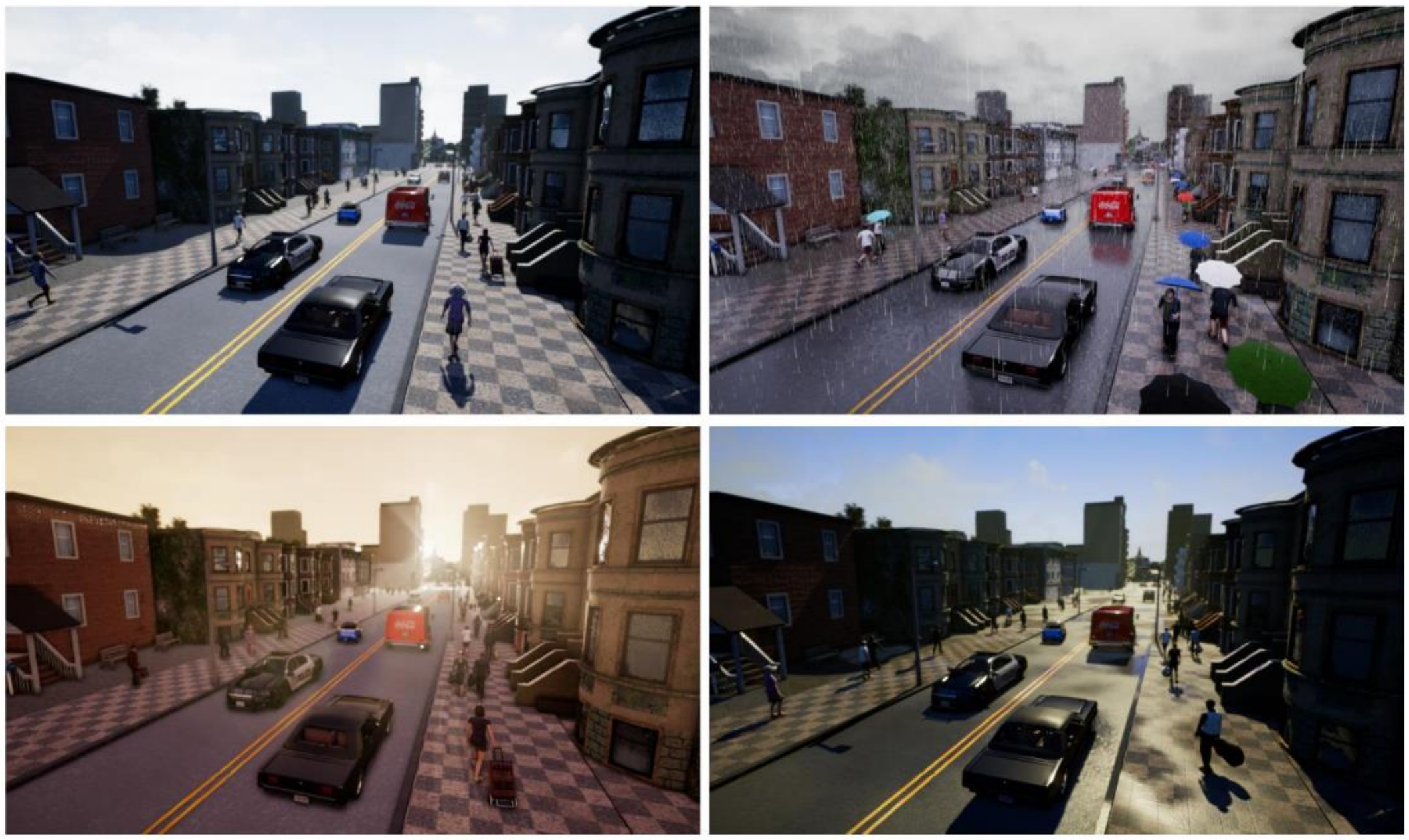

*Figure 15: A street in one of the towns modeled within CARLA simulator, shown from a third-person view in four weather conditions. Clockwise from top left: clear day, daytime rain, daytime shortly after rain, and clear sunset [15].*

### *Pedestrians and Vehicles, and their Configuration*

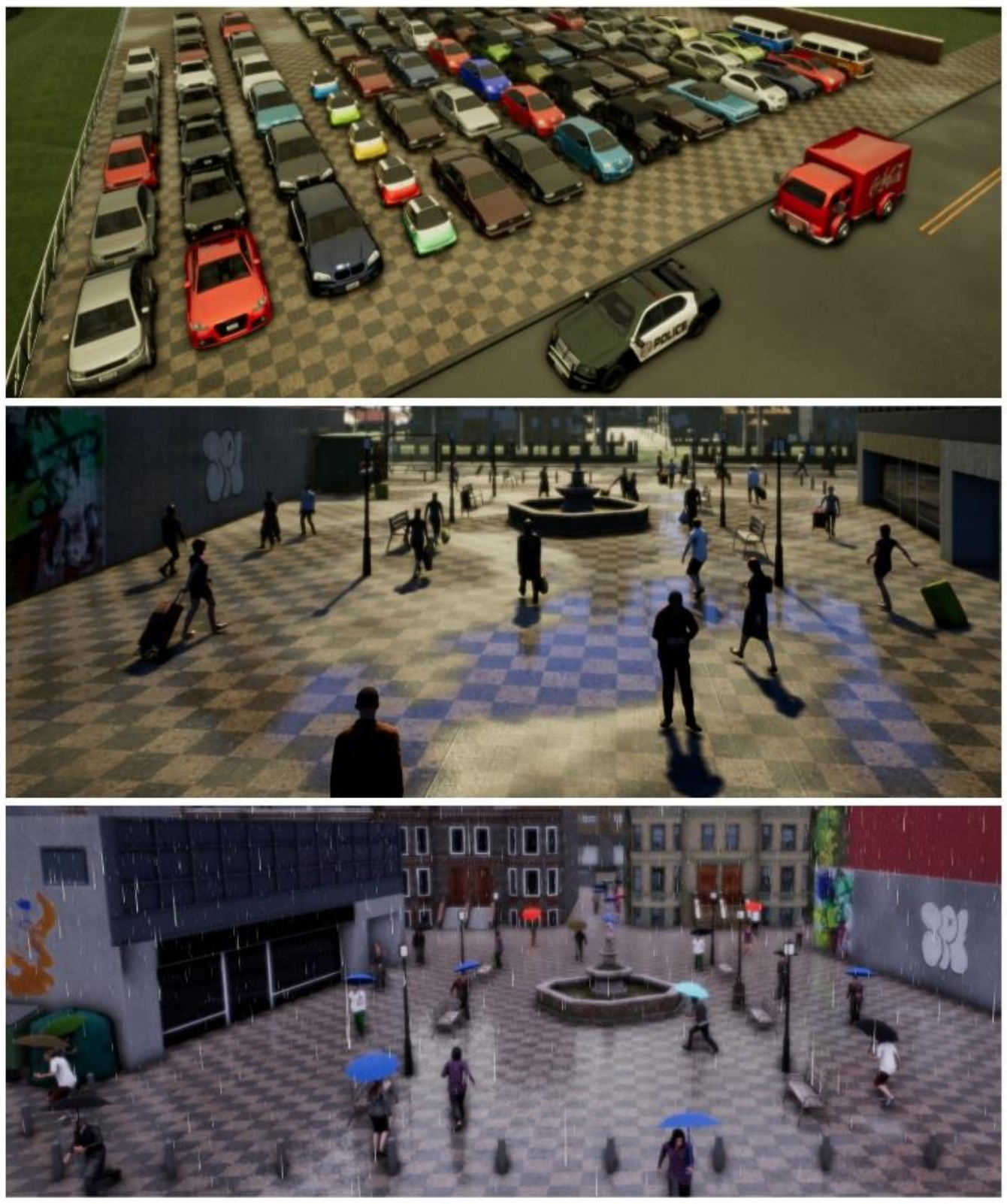

*Figure 16: Diversity of cars and pedestrians currently available in CARLA [15].*

A virtual environment needs to model the appropriate level of diversity, as representative of the real world. As shown in Figure 16, there are a variety of pedestrian and vehicle models - factoring in geometries, shapes, colors, supporting items, etc. Further, each pedestrian is not only wearing unique clothing, but are diversified by the physical items on their person, from umbrellas, to luggages, bags and phones. CARLA, the simulator supported in part by Toyota Research, has an asset library including 40 different buildings, 16 animated vehicle models, and 50 animated pedestrian models [15]. Capturing various sensor combinations and the underlying physics of sensor models is likewise important, particularly if perception is modeled in the simulation toolset.

Regardless of the level of visual fidelity, much of the physical elements represented in Figure 15, need to be represented in the virtual environment. Even if the vehicles and pedestrians are modeled as simple bounding boxes, each might have different dynamics models. For instance a large delivery truck will have different inertial properties and powertrain compared to a speedy luxury vehicle. Further, pedestrian will also exhibit different models of motion depending on simulated age, size, health, etc.

In CARLA for example, kinematic parameters were adjusted for realism. They also implemented a basic controller that governs non-player vehicle behavior: lane following, respecting traffic lights, speed limits, and decision making at intersections [15].

### *Weather and Atmospheric Conditions*

Weather, atmospheric conditions, and lighting are all important aspects of a simulation environment. The vehicle needs to be exposed to various physical phenomena including gravity, air-density, air pressure and magnetic field [40]. The different properties of virtual lighting can add realism to a scene: emitted light, ambient light, diffuse and specular. These light properties can produce the appearance of shininess, object reflections, and stronger illumination of certain aspects in a scene. Additionally, as shown in Figure 15, different weather conditions can be considered such as snow, rain, fog, sunshine, etc.

### *V2X Communication*

IoT backed autonomous vehicles will have various channels of communication connected between different 'intelligent' entities within the environment. Figure 17 provides an overview of the interconnected nature between an autonomous vehicles and various 'smart and connected' technologies, people, networks, vehicles, and city infrastructure.

A simulation might incorporate the various types of communication available to specific platforms based on their hardware/software capabilities, and likewise model "physical restrictions like the broadcasting radius of a certain robot" [32].

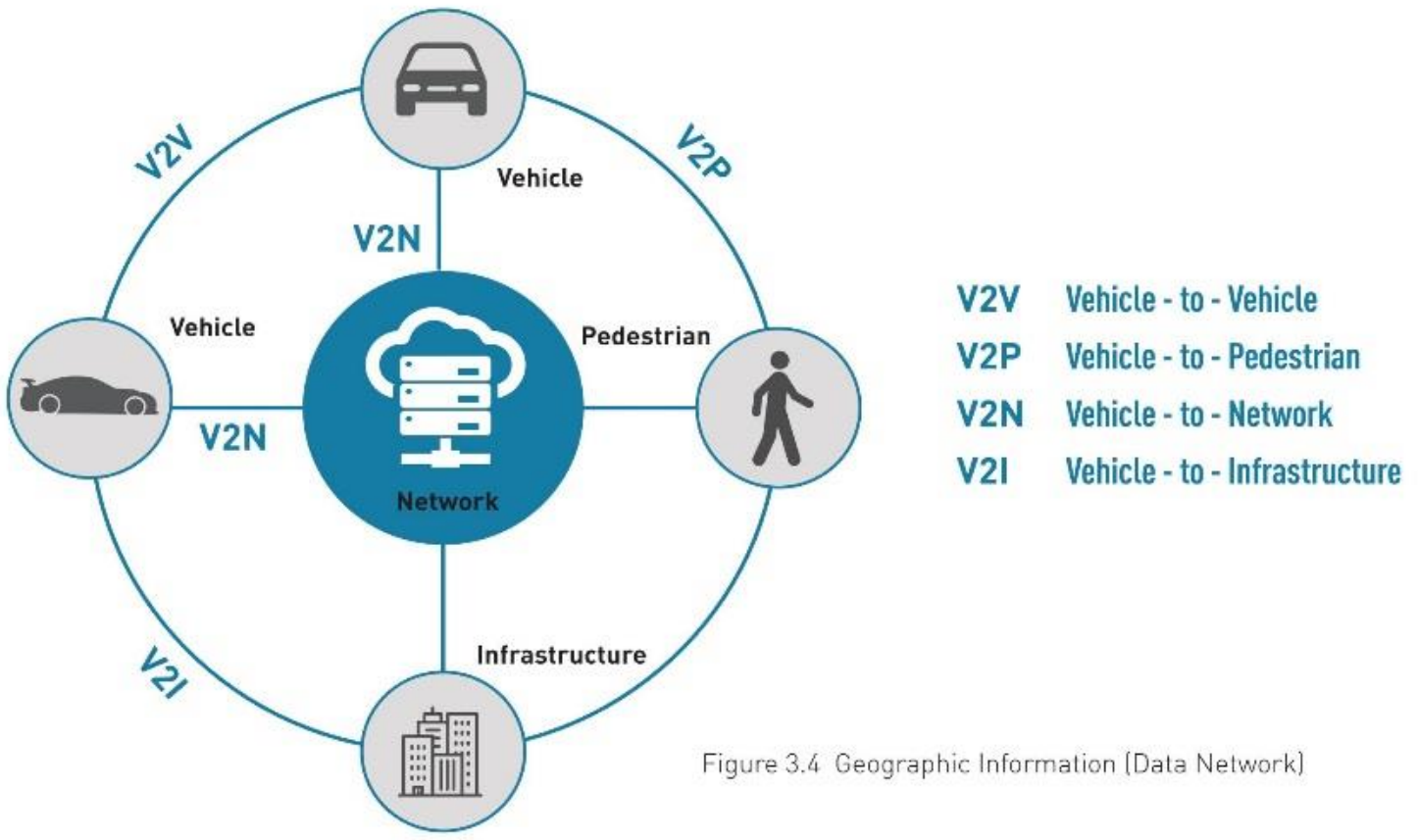

*Figure 17: Vehicle to 'X' communication network. In a connected Internet of Things enabled smart city a vehicle can communicate with a number of objects: other vehicles, sub-components within own-ship, pedestrians, external networks, and infrastructure which could include things like smart traffic lights or smart signs [21].*

"The idea is to employ a small radio transmitter and receiver on each vehicle that broadcasts information about location, speed and direction to other vehicles within several yards. This will help provide warnings to guide the driver about when it is safe to change lanes, speed and merge thereby helping electronic safety systems work safely.

Once V2V is successfully established the next step would be to develop the Vehicle to Infrastructure (V2I). The idea behind V2I is an integrated data network between the vehicle and the roadside infrastructure such as traffic signals, roadway sensors, pedestrians (V2P), etc. It is predicted that the first V2I systems will be developed and employed by 2020." [21]

### *Other: Sensor Noise, Failure Modes, Collision*

Sensor noise (model imperfections), failure modes, and collisions are other considerations which did not fit in one of the major sub-sections, but are still desirable to model.

- Collision detection: the simulators ability to model collisions and boundary effects between various objects.

- Failure simulation: the simulators ability to model failure modes. This can occur, in some form, within most models of the simulation ecosystem (e.g. not sending information to a sensor, or sending corrupt data, ignoring actuator commands, etc.).
- Sensor noise (and in general model noise): The real world is liable to imperfections due to signal noise, scratches on surfaces, component degradation over time, and many other factors.

## Realistic Models - Physical Mimicry

Models are not just developed at the whim of a software, hardware or mechanical engineer, but are backed in most cases by strong mathematical rigor.

An NVIDIA representative commented, “A simulated test environment is more than just a virtual car on a virtual road. It takes model building as intensive as those for movies, and as detailed and accurate as the blueprints for the city roads and highways the car will eventually drive on” [22].

By way of example, it is necessary to model things such as vehicle dynamics, physics, sensors, etc. with enough rigor to ensure sufficient similarity between real world and simulated models. The examples provided below are taken from AirSim, which models both quadcopters and self-driving cars.

### *Motion Control Realism*

As shown in Figure 18, the vehicle dynamics of a virtual quad copter was compared to the physical implementation for various maneuvers within AirSim. The time-series plots and graphics show a similar performance for circular and square maneuvers. For both the simulation and the real-world flights, they collected location of the vehicle in local NED coordinates along with timestamps. Real world testing utilized the Pixhawk v2 flight controller mounted on a Flamewheel quadrotor frame. The sensor measurements were recorded on the Pixhawk device itself. The simulated quadrotor model was configured using the measured physical parameters. Simulated sensor models were based on the sensor data sheets. [40]

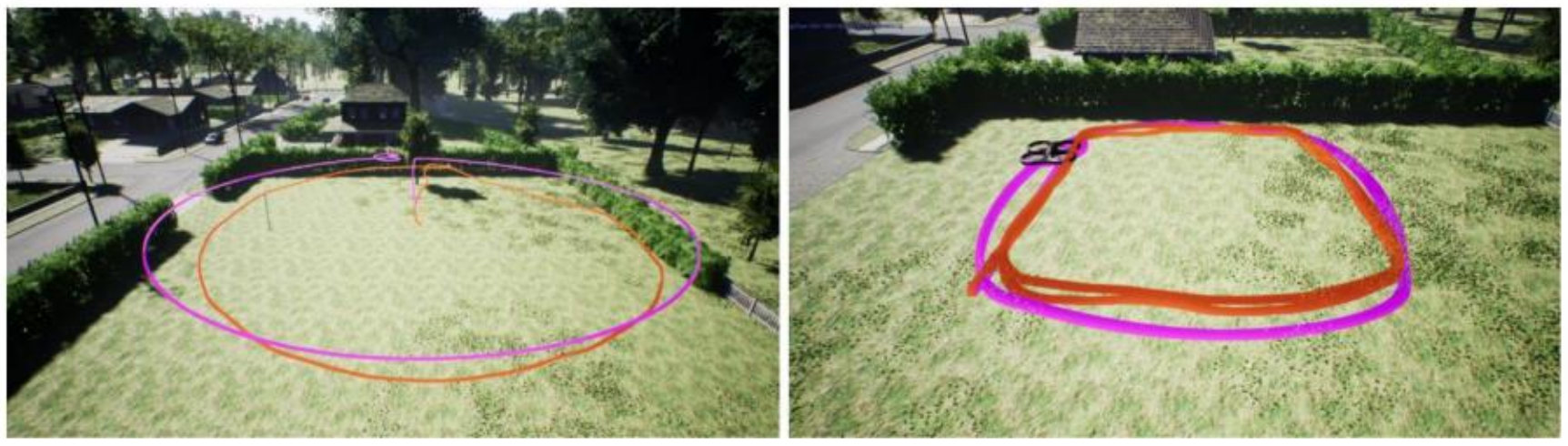

(a) Circle maneuver (b) Square maneuver

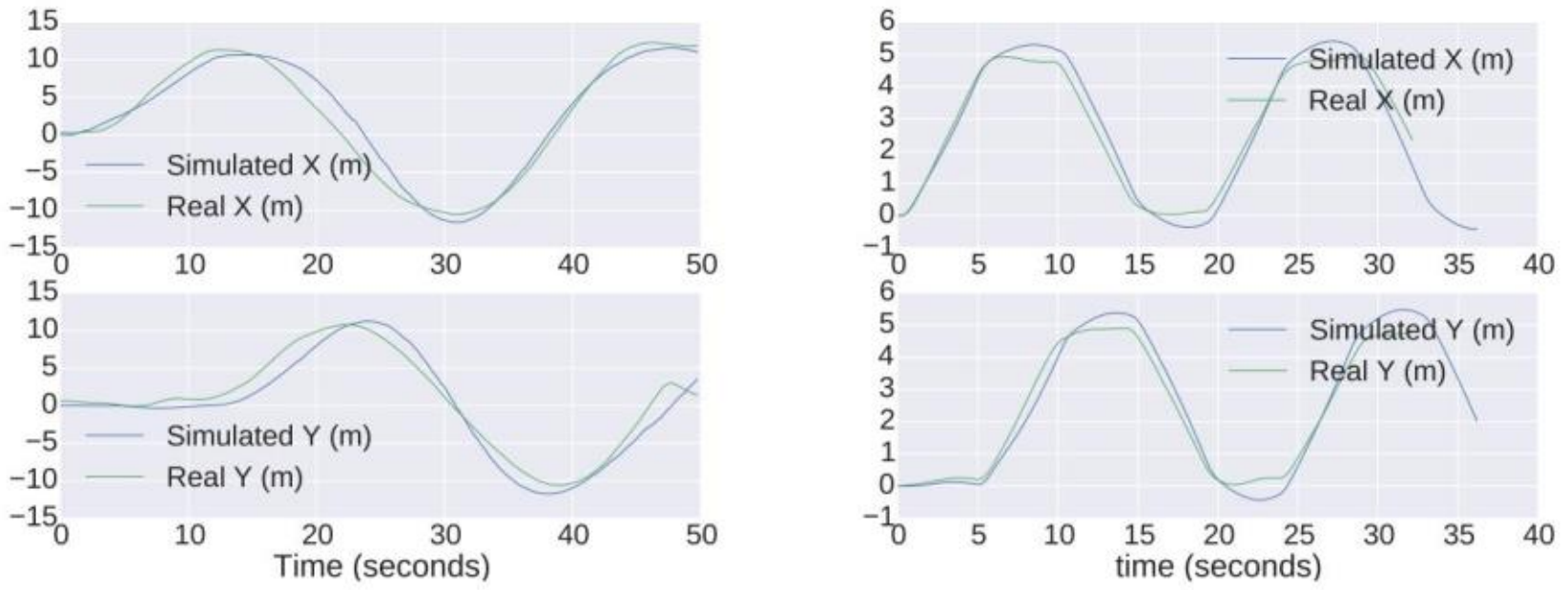

(c) Space-Time Plot for Circle (d) Space-Time Plot for Square

*Figure 18: Evaluating the differences between the simulated and the real-world flight. In top figures, the purple and the red lines depict the track from simulation and the real-world flights respectively [40].*

### Physics Realism

In the case of AirSim, realism was added to the environment by modeling gravity, magnetic fields using the titled dipole model which resembles Earth as a perfect dipole sphere, air pressure and density. The physics engine computed the next kinematic state for each vehicle given the forces and torques acting on the body. The kinematic state of the body is expressed using 6 quantities: position, orientation, linear velocity, linear acceleration, angular velocity and angular acceleration [40].

### Sensor Realism

In AirSim, various sensors were modeled, such as the accelerometer, gyroscope, barometer, magnetometer and GPS. Several sensors were investigated and compared to their physical implementations, namely the barometer (MEAS MS5611-01BA), the magnetometer (Honeywell HMC5883) and the IMU (InvenSense MPU 6000) [40]. Figure 19, shows the results from comparing the real world and simulated barometer (measures atmospheric pressure) and magnetometer (measures magnetic fields). Results from the IMU are not shown.

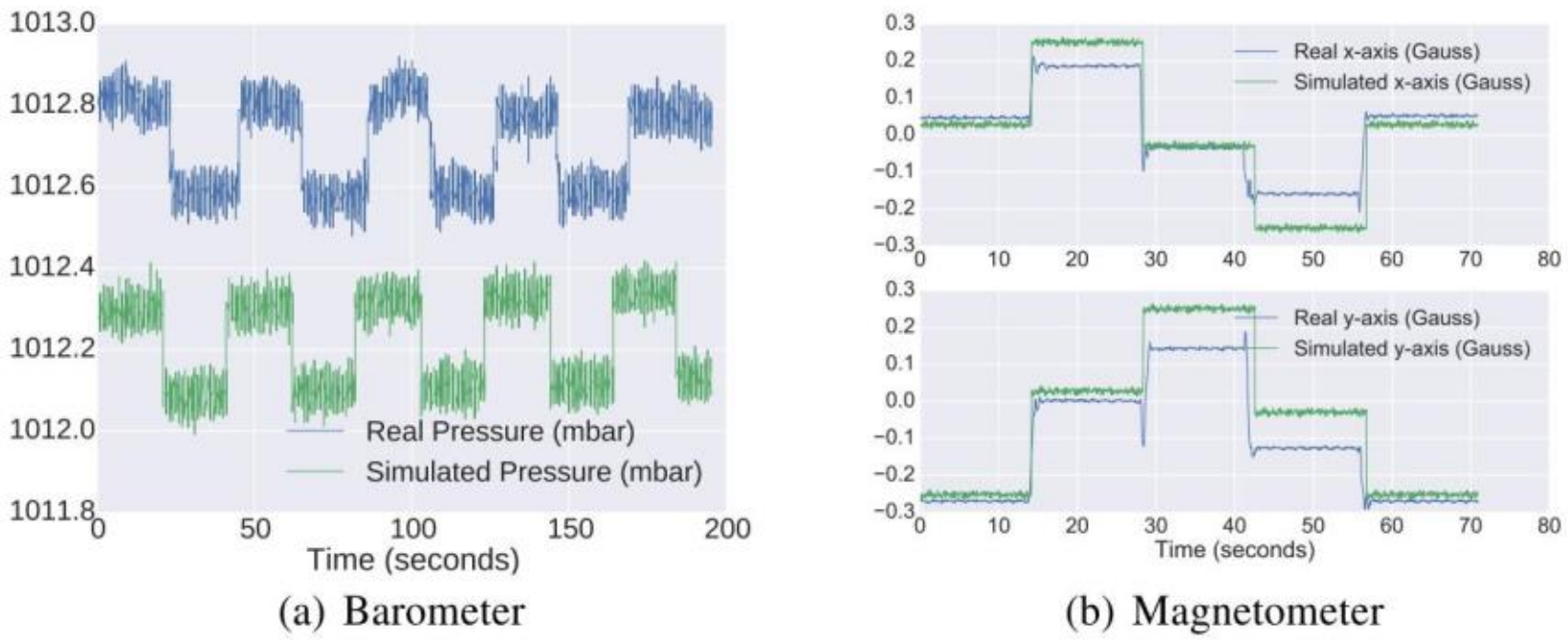

*Figure 19: Barometer and magnetometer simulated and real-world comparison, AirSim [40].*

The AirSim researchers performed the following tests for each sensor.

**IMU:** Measured readings from the accelerometers and gyroscope as the vehicle was stationary and flying, and compared the variance between simulation and the real hardware.

**Barometer:** Raised the sensor periodically between two fixed heights: ground level and then elevated to 178 cm (both in simulation and real-world).

**Magnetometer**: Placed the vehicle on the ground and then rotated it by 90◦ four times. [40]

To summarize, the previous sections, mathematical rigor is provided within the simulated models and care is taken to match performance between physical designs and simulation.

However, there is also an obvious tradeoff between computational efficiency, speed to market, and quality. While many aspects of the development cycle need to be rigorously developed, there are certainly options for modeling components with less specificity. In an interview with Waymo simulation and modeling developers, a reporter asked, “what about oil slicks on the road? Or blown tires, weird birds, sinkhole-sized potholes, general craziness. Did they simulate those? He [the Waymo engineer] said, sure, they could, but ‘how high do you push the fidelity of the simulator along that axis? Maybe some of those problems you get better value or you get confirmation of your simulator by running a bunch of tests in the physical world’ ” [29].

## Simulation Graphics Fidelity

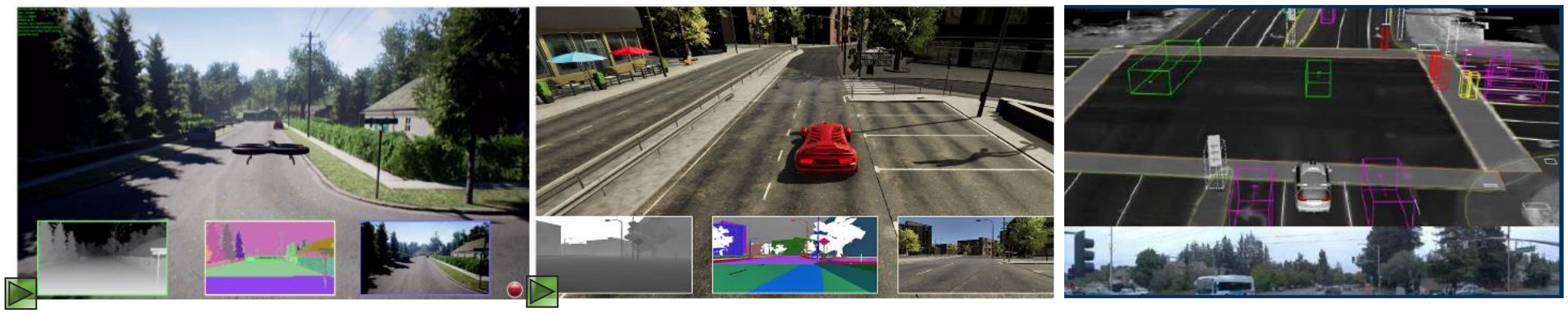

*Figure 20: Overview of different simulation graphics fidelity. (Left) A snapshot from AirSim shows an aerial vehicle flying in an urban environment. The inset shows depth, object segmentation and camera streams generated real time [31]. (Middle) Another snapshot from AirSim, showing a self-driving car in an urban environment. The inset shows depth, object segmentation and camera streams generatee real time [17]. (Right) Waymo simulation environment where vehicles are bounded by green and purple boxes, pedestrians with yellow, and cyclists with red [40].*

Two examples of graphical fidelity are provided for comparison. The left two images in Figure 20 are from Microsoft AirSim, while the image on the right is from Waymo’s Carcraft. Both simulation environments Microsoft AirSim and Waymo’s Carcraft provide a rich environment for the development of self-driving cars (an aerial vehicles). The main difference between the two is the inclusion or lack of photo-realism within the virtual environment. Neither is a ‘correct’ implementation, but a function of the intended use-cases. Photorealistic environments enable the testing of perception based algorithms. Perception based techniques could include machine learning methods, such as deep learning, which are responsible for appropriately classifying/identifying objects within the scene, such as a traffic light changing from red to green, a pedestrian crossing the street, an aggressive vehicle merging in ‘your’ lane. Data used by these machine learning techniques include various sensor feeds, which could be simulated and tested within such a visually realistic environment.

AirSim is developed using Unreal Engine or Unity which provides “cutting edge graphics features such as physically based materials, photometric lights, planar reflections, ray traced distance field shadows, lit translucency, etc. The screenshots from AirSim highlight near photo-realistic rendering capabilities. Further, Unreal’s large online Marketplace has various pre-made elaborate environments, many of which are created using photogrammetry techniques” [40]. Photorealistic simulation is an approach taken not only by Microsoft AirSim, but NVIDIA, RightHook, Cognata, CARLA sim, and ANSYS to name a few.

In stark contrast, companies such as Waymo and Uber do not utilize a photorealistic environment, and thus do not model the perception problem in simulation. After years of work developing machine learning models to detect and classify objects in a scene using real world data, Waymo skips that object-recognition step. Instead of feeding the car’s software raw simulated data which it then identifies as a pedestrian, the simulation tool simply tells the car: A pedestrian is here [29]. Uber also utilizes real world data to train algorithms for the perception problem, foregoing reliance on simulation [45].

Undoubtedly, there is a trade-off between the two differing strategies. Fidelity and realism of the graphics will incur a penalty on throughput/speed of the simulation framework to iterate quickly on different test scenarios. However, if the fidelity of photorealistic environments is truly comparable to the real world, this could potentially enhance the speed of developing machine learning based perception algorithms, as well as testing sensor capability. A visually realistic environment could provide ample training data (images, data feeds, etc.) for training machine learnt models, and enable a bootstrapping effect/transfer learning to the real world.

## Modeling Perception

This section briefly discusses the pipeline for tackling the perception problem, assuming the simulation environment utilizes photorealistic graphics.

The pipeline starts with simulated scenario data (visually rendered) being provided to a reduced order model (ROM) of the virtual camera (the ROM was generated within a physics-based sensor simulation capable of modeling any combination of sensors) [3]. The virtual camera generates an image, which is then adapted for an actual camera's electronic control unit (ECU) (if hardware in the loop simulation is performed). The camera ECU then provides hardware level data feeds of the perceived virtual environment which are filtered by Perception SW for detection and classification of objects (road lanes, pedestrians, vehicles and many others). From here, the filtered data is provided to the rest of the self-driving car's software stack for path planning, behavioral prediction, trajectory planning, and vehicle control. The new state of the vehicle is then calculated based on the physics of the environment and vehicles dynamics model.

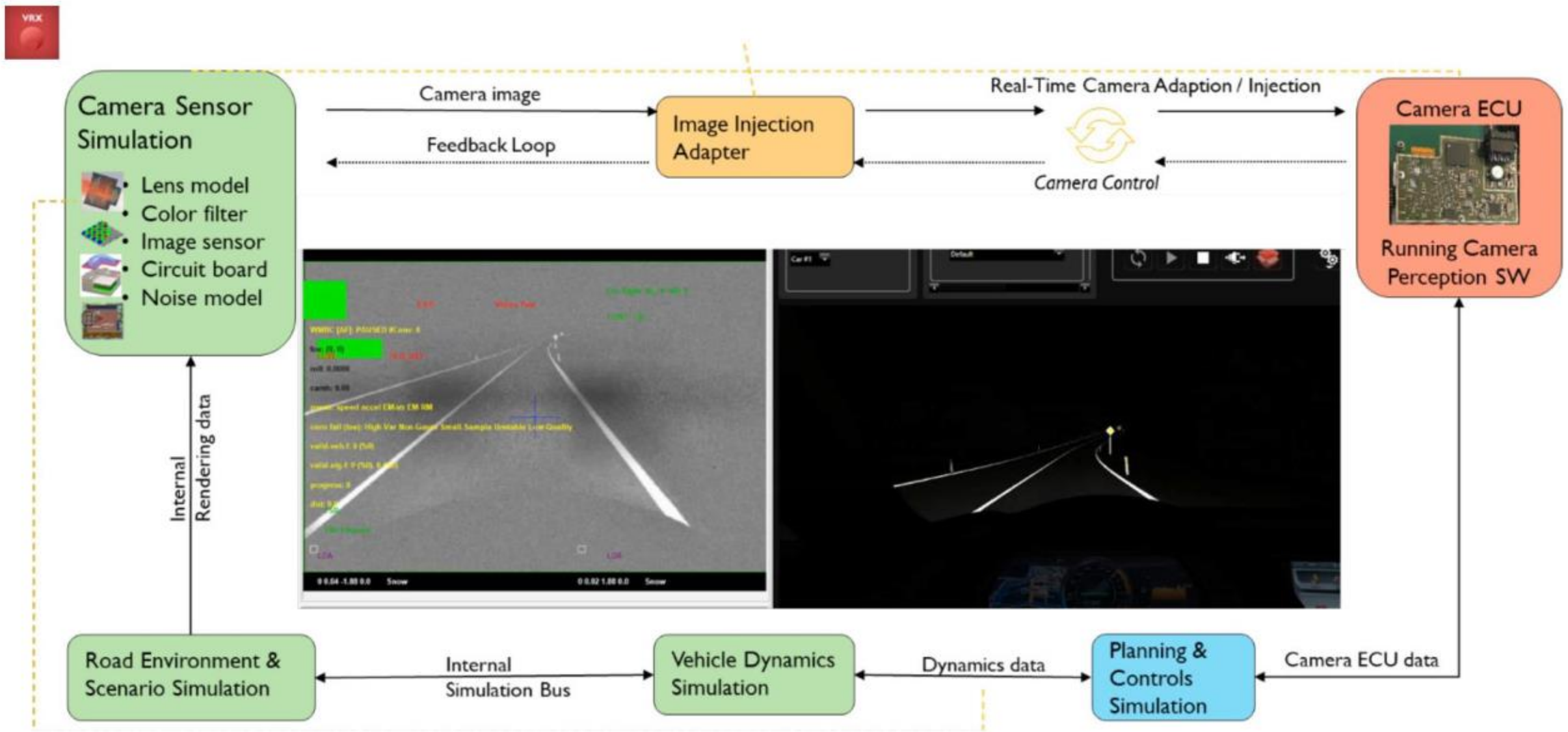

*Figure 21: Pipeline for simulating a camera sensor virtually, and utilizing the data feed to inform Perception SW and subsequent software layers in the stack. The image on the left is the simulated camera image. The image on the right is a virtual driver-side perception of the environment. These images can be provided to engineers real time [3].*

If the perception problem is not done within simulation (also in conjunction with real world training/testing) then the alternative method is to collect enough real world data (images, sensor feeds) which can/must be accurately labeled for training on machine learning algorithms. Solving perception via the real world data approach is considered within the section on Infrastructure.

## Simulation Characteristics

The following subsections discuss several characteristics and features of robust simulation toolsets. These characteristics include the ability to collect data from running simulations, control the simulation scene, extend capabilities, provide proper interfaces between internal and external components, facilitate parameterization of a scene (turn one scenario into a thousand), provide usability features for developers and scenario designers (ex. make scenario development easy), reusable designs, high throughput/performance, experience replay, and appropriate visualization of a scene (ex. vehicle/model behavior visualization).

### *Data Collection*

One of the primary purposes of simulation toolsets is to generate data: for testing, for validation, regarding performance metrics, etc. A nice feature of simulation toolsets is to enable gathering of simulation data for future use cases, and possible offline training/testing. For instance, photo-realistic and labeled data from simulation could be utilized for training and/or testing machine learning techniques, potentially minimizing the need for real world data collection.

Take Microsoft AirSim as an example. Within Airsim, there are two ways to capture training data for machine learning techniques, such as deep learning. Quoting the Microsoft developers, “the easiest way is to simply press the record button in the lower right corner. This will start writing pose and images for each frame. The data logging code is pretty simple and you can modify it [freely].” The second and ‘better way’ to generate training data with precision is by accessing the APIs. This allows full control of how, what, where and when you want to log data [31].

### *Controllability*

Controllability is the ability to specify the initial state of a scenario and workload of the system under test [28]. This is important for a number of reasons from development of algorithms, safety assessments, to testing and validation. In order to achieve controllability a deterministic simulation environment, or the capability to simulate a scenario and its constituent models deterministically, is necessary. This can be provided by specifying models with an internal pseudo-random number generator that can be set to predetermined seed values [28] . This enables the capture of real world randomness while enabling control of the simulation state.

Additionally software toolsets need to be built into the simulation to enable an engineer/scenario designer to specify or replay environment configuration from a selected time stamp. For instance, assume a self-driving car is specified to operate in a complex, multi-vehicle scenario complete with a number of sensors, pedestrians, scenery, and in a specific urban location. The autonomous vehicle maneuvers through this environment for a number of time steps, accelerating, braking and maneuvering as needed. At some point in the simulation, the autonomous vehicle approaches an on-coming vehicle and the test engineer notices an irregular behavior from the AV’s software stack. Instead of decelerating as expected, the trajectory planner causes the vehicle to accelerate.

In such a scenario, it is useful for the test engineer to be able to capture the current state of the environment and vehicle just prior to the unintended behavior for further testing of subsequent software iterations, or to manually test look-ahead scenarios of this environment. Providing a 'look-ahead' capability will allow a user to test functionality had the vehicle maneuvered left, or right, or properly decelerated, instead of speeding up. This approach is common in variations of Monte-Carlo tree search.

### *Extensibility*

Extensibility of a simulation environment and its constituent components is a necessary feature, especially in the modern age of crowd-sourced, and open-source development. Software in particular is a continually adapting, improving and changing toolset. In order to keep up with trends, development standards, best practices, and taking advantage of relevant externally created feature-sets, an extensible platform is a must. Extensibility is a loaded term which can include the ability to inherit from, or extend prior software functionality (i.e. in terms of software development, inheriting from a software class), leverage open source capability, and add new development features (software classes, models, etc.) that were not envisioned at the time of initial simulation design.

Specific examples are provided by AirSim. Due to the utilization of Unity (and Unreal Engine) as the foundational game engine for the AirSim simulation environment, Microsoft and AirSim developers can leverage many freely available, high quality models, and virtual environments from the Unity (Unreal Engine) asset stores. "The Unity Asset Store provides an expansive library of high-quality content that you can use to quickly and easily build complex virtual environments" [17]. Further, "Unity's own machine learning initiative ML-Agents can be integrated into AirSim's capabilities, allowing for even more experimentation" [17].

Similarly, AirSim was also developed as an Unreal plugin. This enables the AirSim toolset to be seamlessly dropped into any Unreal environment [31].

### *Component Level Interfaces (External and Internal)*

Appropriately scoped, well defined, and usable interfaces are necessary within any simulation toolsets. Adequate interfaces enable coupling between internal simulation modules as well as external modules, developed within the company or by a 3rd party. In general, interfaces need to specify a set of APIs (Application Program Interfaces) for developers, and the communication protocol utilized (HTTP, UDP, TCP, etc.).

"AirSim exposes APIs so you can interact with the vehicle in the simulation programmatically. You can use these APIs to retrieve images, get state, control the vehicle... The APIs are exposed through the RPC [Remote Procedure Call], and are accessible via a variety of languages, including C++, Python, C# and Java" [31]. RPC enables a service-client oriented architecture, where a client can communicate with a service providing some business intelligence located on another computer within the network.

Another example of interfacing between company developed modules and customer developed modules is more apparent through the work of RightHook, as shown in Figure 22.

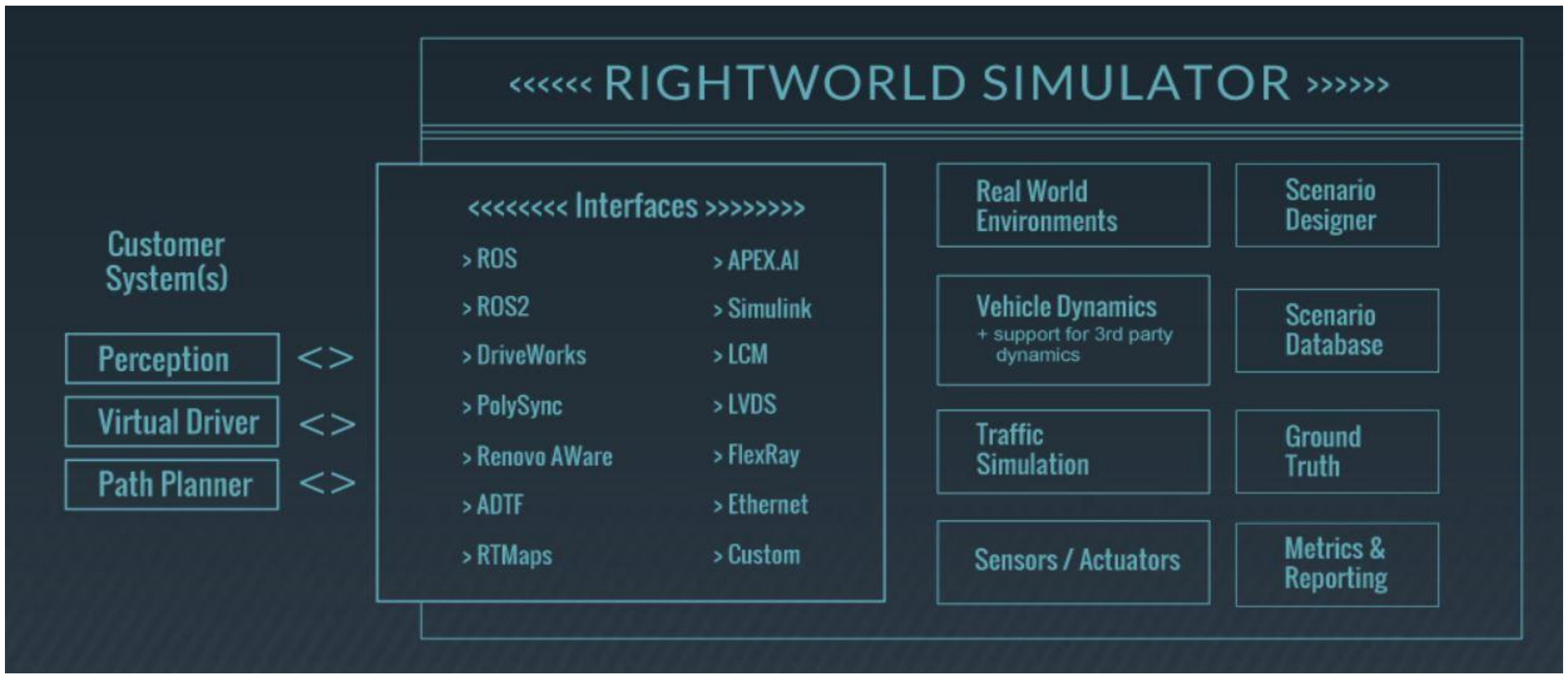

*Figure 22: RightHooks' RightWorld Simulator overview. The diagram breaks down some of the available interfaces [38].*

RightHook is a startup for consumer product simulations. Therefore they have to maintain a more robust set of interfaces between the various components and modular pieces of software to satisfy the various needs of their customers. Simulation toolsets utilized directly within a single company can provide a more common set of interfaces that are agreed upon within the requirements prior to development. However, it is important to note that adding additional interfaces will likely occur throughout development, even for non-commercial products. As new technologies are provided (for machine learning, vehicles dynamics, etc.) it is imperative that a sophisticated simulation toolset provide the ability to easily incorporate such features.

In the case of RightHook, they have provided a number of interfaces to common platforms so that users can incorporate their proprietary modules (such as perception algorithms, path planners, and a virtual driver). These interfaces include NVIDIA's DriveWorks, MathWorks Simulink, the Robot Operating System (ROS), and also custom interfaces.

### *Parameterization*

Parameterization as described here expresses the idea of turning a single scenario into a thousand scenarios incorporating slight variations of the original. Other terms that are commonly used, but have similar meanings are hyper-parameter sweeps and 'fuzzing'. Essentially, the scenario designer can provide various ranges for model parameters, or sampling strategies for selecting parameter values from a range, and sweep over these configurations to generate a number of unique scenario tests.

An appropriate illustration of utilizing parameter sweeps within scenario development and testing is provided by Google's Waymo. Waymo's Self Driving Vehicle Safety Report provides an enlightening use case. "For example: at the corner of South Longmore Street and West Southern Avenue in Mesa, Arizona, there's a flashing yellow arrow for left turns. This type of intersection can be tricky for humans and self-driving vehicles alike — drivers must move into a five-lane intersection and then find a gap in oncoming traffic. A left turn made too early may pose a hazard for oncoming traffic; a turn made too late may frustrate drivers behind" [46].

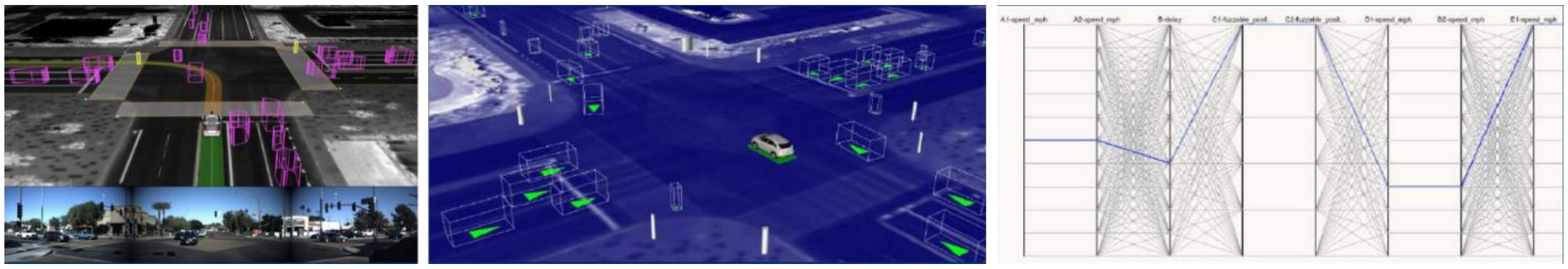

*Figure 23: (Left) Waymo self-driving vehicle encounters a flashing yellow left turn arrow in Mesa, Arizona. Simulated replication of real-world intersection, shown in bottom inset. (Middle) Simulated alteration of intersection scenario shown on left. Additional vehicles and modified vehicle arrangement has been provided. (Right) Waymo's visual depiction of 'fuzzing', which is generating a number of unique scenarios for testing by sweeping over model parameter ranges [46].*

Waymo engineers can "multiply this one tricky left turn to explore thousands of variable scenarios and 'what ifs?' Through a process called fuzzing, we [Waymo engineers] alter the speed of oncoming vehicles and the timing of traffic lights to make sure our vehicles can still find a safe gap in traffic. The scene can be made busier and more complex by adding simulated pedestrians, motorcycles 'splitting the lane,' or even joggers zig-zagging across the street" [46].

### *Observability*

*Observability* is the ability of the tester to observe the state of the system to determine whether a test passed or failed [28]. Assume that a developer has run a number of tests sweeping over all the variations/parameters within the scenario based on reasonable ranges of models contained: speeds and positions of vehicles, multiple variations on pedestrian behaviors, multiple cyclists, the list goes on. Once all of this data is collected, the "problem really becomes analyzing all these scenarios and simulations to find the interesting data that can guide engineers to be able to drive better. The first step might just be: Does the car get stuck?" [29]

Observability can be a difficult problem to address. For example, in a vehicle-level obstacle test the criteria for 'success' might be that the vehicle leaves sufficient clearance as it passes an obstacle. But, even if the system 'passes' a test by avoiding collision that could simply be "due to the system getting lucky in avoiding an obstacle". The vehicle might not have actually detected the obstacle. Further, the system "might hit the obstacle on the next test run – or perhaps hit it 2000 test runs later. This lack of observability is one facet of the robot legibility problem, which recognizes the difficulty of humans understanding the design, operation, and 'intent' of a robotic system" [28].

Simulation tools, as a possible way of mitigating this issue, could provide the capability to define heuristics or logical rules defining criteria for 'success' and 'failure' within simulation runs, which can provide engineers or analysts a sort of spotlight into problematic areas of automotive vehicle design.

### *Usability and Simulation Management*

Considering usability of a simulation environment is to consider the ability of engineers, designers, and quality assurance personnel to rapidly test, evaluate and iterate on virtual scenarios and the various models incorporated. Usability of a simulation environment can be provided in a number of different forms from quality User Interfaces to enable top level modifications, intuitive user controls such as drag and drop functionality of models, the ability to manage the simulation state: stopping, starting, restarting, saving an environment (also tied to Controllability), and easy customization of model parameters through high level interfaces or scripts. Several examples of simulation usability are provided by excerpts from Waymo and DSpace.

DSpace has several toolsets to facilitate scenario development, notably through their ModelDesk framework. ModelDesk is a graphical user interface for simulation, intuitive model parameterization and parameter set management. ModelDesk includes: parameter management, a road generator, traffic editor, maneuver editor, and simulation management among other capabilities [16]. ModelDesk's Road Generator and Traffic Editor Programs are shown in Figure 24. "The Road Generator supports the definition of intersections and complex road networks. The user interface provides a list of road segments (upper left in figure), an overview of the whole road network (middle), and a view of lane details (right). The Traffic Editor is the graphical user interface for defining the segment based definition of movements for fellow vehicles on a road network" [16]. Each of the ModelDesk programs facilitate the construction of simulated scenarios.

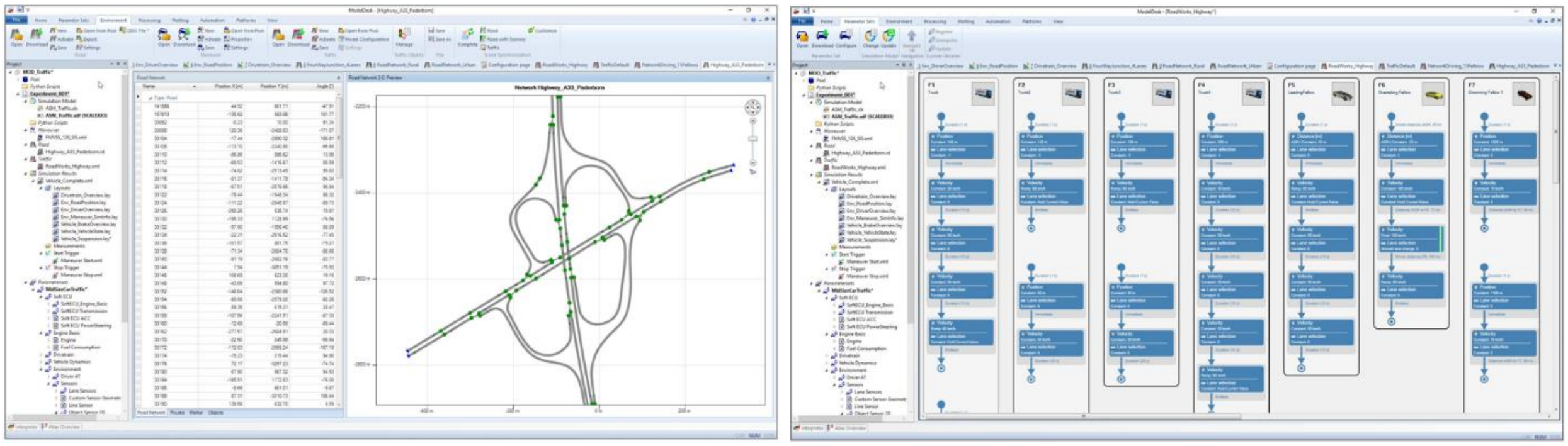

*Figure 24: DSpace Automotive Toolsuite. (Left) The DSpace Road Generator program, and (Right) the DSpace Traffic Editor program. Both programs are incorporated in the ModelDesk suite [16].*

Waymo's simulation 'Carcraft' provides users with a scenario builder program which can program model logic to configure precise movement in order to test specific behaviors. As a Waymo simulation engineer states, there is a "spectrum between having control of a scenario and just dropping stuff in and letting them [vehicle, pedestrian models] go" [29]. Waymo designers can easily drop in synthetic cars, cyclists, and pedestrians either through the UI dropdown, or through specified hot-keys. After configuring the scenario through the UI scenario builder, Waymo engineers can simply press a button and the "objects on the screen begin to move. Cars act like cars, driving in their lanes, turning. Cyclists act like cyclists. Their logic has been modeled from the millions of miles of public-road driving the team has done" [29].

### *Component/Software Reusability*

Through the simulation design and development process it is worth ensuring construction of an architecture that facilitates code reusability. This could be reuse of simulation features, software systems, sub-systems, components, and interfaces. Take Microsoft AirSim as an example.

"AirSim offers sensor models for accelerometer, gyroscope, barometer, magnetometer and GPS. All our sensor models are implemented as C++ header-only library and can be independently used outside of AirSim. Like other components, sensor models are expressed as abstract interfaces so it is easy to replace or add new sensors" [40].

Additionally, the "core components of AirSim including physics engine, vehicle models, environment models and sensor models are designed to be independently usable with minimal dependencies outside of AirSim" [40]. This is just one example of many possibilities to explore when considering software reuse.

### *Throughput*

The necessary throughput required of a simulation toolset is dependent on the intended functionality and scope/complexity of the environment simulated. There is undoubtedly a trade-off between complexity and speed. Therefore, specific numbers that are directly applicable/translatable from one company to another are difficult to quantify. Nevertheless this is a critical aspect of enabling useful insights from simulation. As explained by Waymo, "the iteration cycle is tremendously important to us and all the work we've done on simulation allows us to shrink it dramatically. The cycle that would take us weeks in the early days of the program now is on the order of minutes" [29].

To provide rational for throughput, some numbers/examples are specified below from various companies' simulation environments.

- BMW company boasts that it can perform "up to 2 million scenarios per day…[utilizing] a mix of driving simulation, hardware in the loop testing, proving ground testing and global fleet testing" [6].
- NVIDIA provides software in the loop (SIL) and hardware in the loop (HIL) based simulation testing. Regarding their HIL capabilities, "driving decisions from DRIVE AGX Pegasus [NVIDIA's Self-Driving Car computer/processor] are fed back to the simulator 30 times every second, enabling hardware-in-the-loop testing" [37].
- Microsoft AirSim designed software for its simulated physics that can "run its update loop at '1000 Hz' modeling linear and angular drag, acceleration of the vehicle factoring in the presence of drag forces and torques, as well as additional forces on the body" [40].

### *Experience Replay*

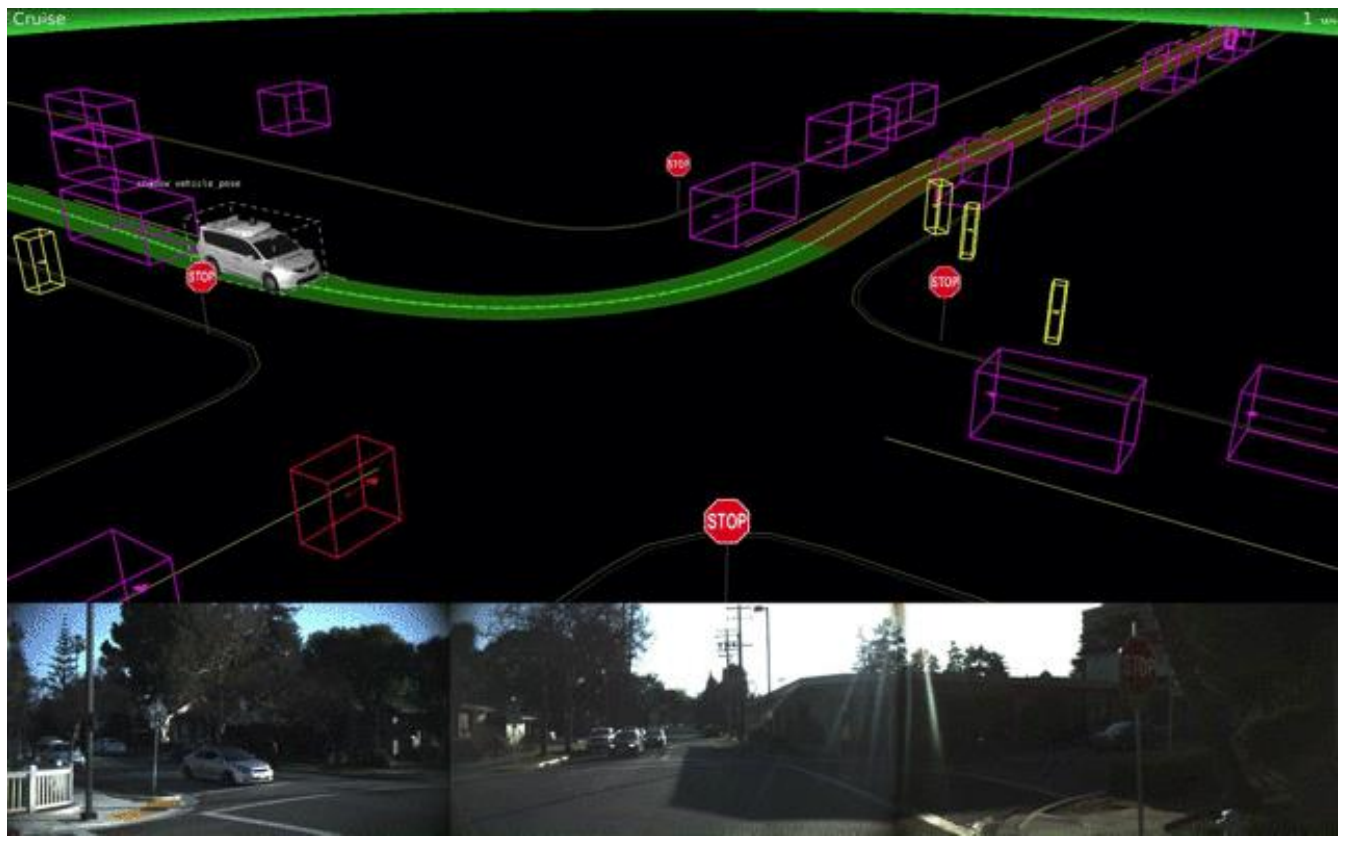

*Figure 25: Waymo showcasing the ability to replay real world experiences, as well as overlay the real world and updated software in the loop to compare performance [29].*

Waymo provides the ability to replay experiences derived from the real world, as well as overlay the real world experience with the updated vehicle software within simulation. Additionally, NVIDIA re-simulation enables real data from sensors placed on test vehicles driving through public roads to be fed into the simulation [37].

In the case of the animation provided in Figure 25, the simulation is showing a 4-way-stop intersection with both the new AV software overlaid with the shadow vehicle (dashed grey box) highlighting the original behavior of the vehicle. Below the simulated environment is a capture of the real world data from the experience in which the AV erroneously stopped for a pedestrian.

The ability to replay a scenario with new behaviors overlaid on the old data can highlight differences, improvements, failures, etc. This is beneficial to give developers insight into additional training [29].

### *Behavior Visualization*

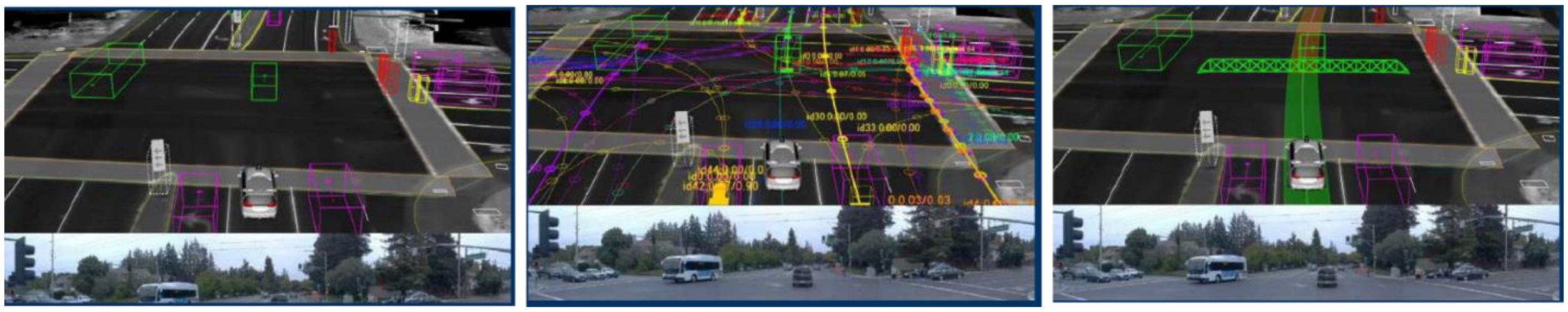

*Figure 26: (Left) Waymo vehicle perceives surroundings. (Middle) Waymo's software predicts the various behaviors of entities in the environment with probabilistic likelihoods. (Right) The vehicle software algorithms have selected a viable trajectory [46].*

It is important to visualize the various perceptions and behaviors calculated by the autonomous vehicle. The series of images in Figure 26 depict the pipeline of high level decisions an autonomous vehicles calculates. First the vehicle needs to understand, 'what are the entities around me'. This is provided by the bounding boxes. The vehicle then predicts the probability of each entities possible trajectories, as well as its own intended path. Once a viable path is determined, the autonomous vehicle selects and displays a trajectory to maneuver, as shown in the rightmost image.

Not only is this information useful for software developers, quality assurance engineers, and scenario designers/tester, but, depending on the quality of the visualization, it can also be utilized within the autonomous vehicles infotainment system. This would serve two-fold purposes: enabling the customer to understand the behavior of an autonomous vehicle and instil confidence in its behavior over time.

## Multi-Simulation Toolset Pipeline

The series of sections that follow provide examples of some additional simulation tools related to autonomous vehicle development, with the exception of mission level simulation tools previously discussed. There is an entire pipeline of simulation toolsets needed to test and validate the performance of all aspects of autonomous vehicles from multi-physics: structural properties of the chassis, vehicle dynamics, sensors, thermal properties, electronics, to mission level simulation - testing the full vehicle/system functionality within an environment, utilizing Reduced Order Models (ROMs) of its sub-system components.

The ANSYS autonomous vehicle open simulation platform is an example of such a multi-simulation capability, and provides a basis for some of the simulations of specific sub-systems/models addressed in the following sections. ANSYS's simulation platform addresses the challenge of modeling and simulating the vehicle development lifecycle by integrating physics, electronics, embedded systems and software to accurately simulate the complete autonomous driving stack. ANSYS capabilities "span the simulation of all sensors, including LiDARs, cameras, radars and ultrasonic sensors; the multi-physics simulation of physical and electronic components; the analysis of system functional safety; as well as the design and automatic code generation of safety-certified embedded software. Sensor simulation can be integrated into a closed-loop simulation environment that interacts with traffic simulators, enabling thousands of driving scenarios to be executed virtually" [3].

Further, the ANSYS Medina Safety environment is closely coupled to the aforementioned system. ANSYS Medini Analyze "helps to manage the safety validation process by implementing key safety analysis methods such as failure modes and effects analysis (FMEA) while supporting safety analysis and design according to ISO 26262 for electrical/ electronic systems and software-controlled safety-related functions" [3].

The key take away here is that multiple levels of simulation tools are required to test and validate performance, and ensure safety. As the rigor on systems engineering, and model-based systems engineering continues to evolve, an ideal tool chain will incorporate all aspects of development from system and safety level requirements, to software and hardware level requirements, to the actual implementation details of each sub-system and sub-system component. Along the design chain and through development, analysis and metrics obtained via simulations tools (as well as other means of testing) will feed into verifying that requirements for system, safety, software, and hardware are achieved.

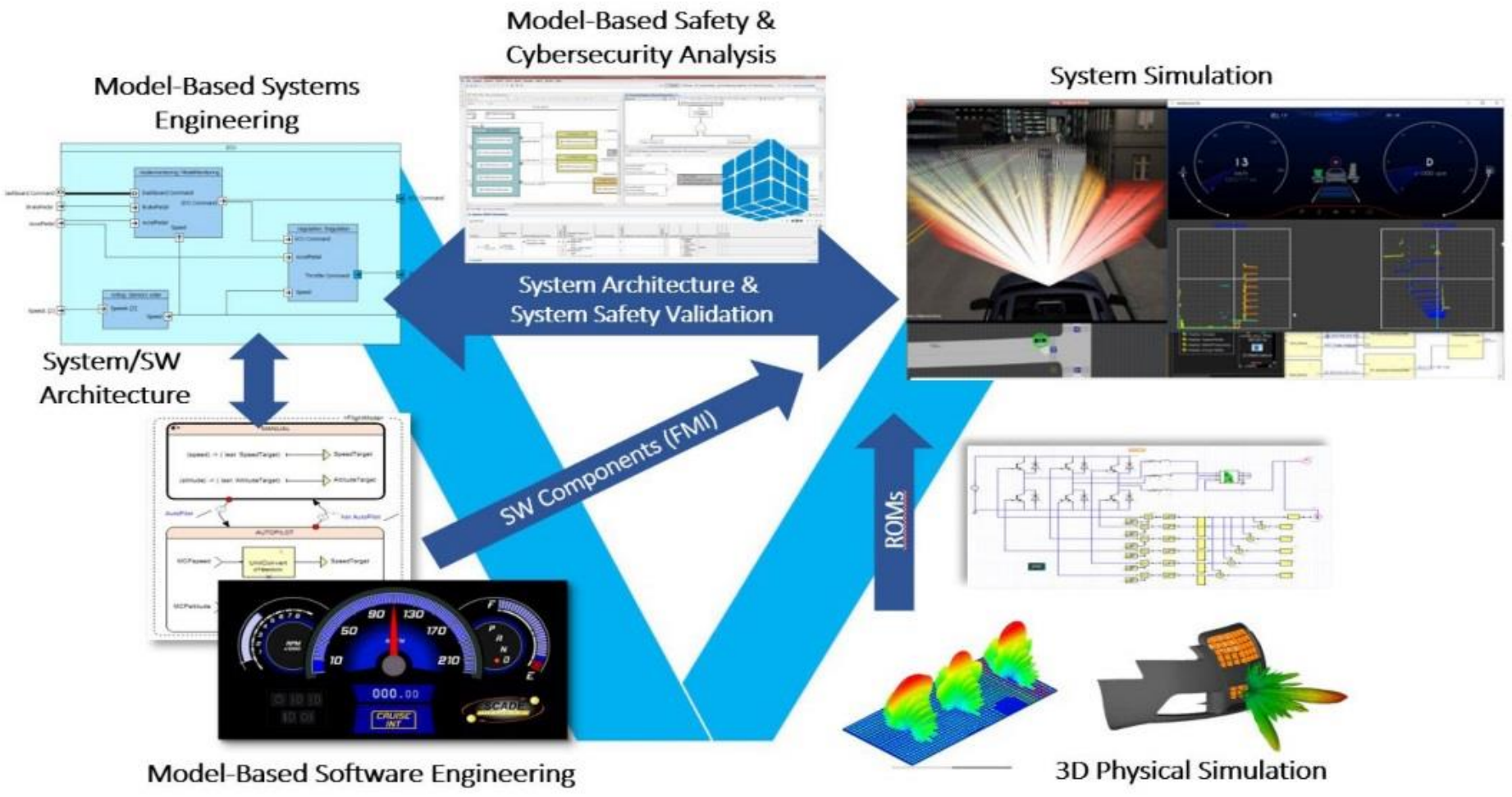

*Figure 27: ANSYS integrated pipeline of simulation toolsets covering the full lifecycle of autonomous vehicle development [3].*

## Multiple Autonomous Vehicle Components to Simulate

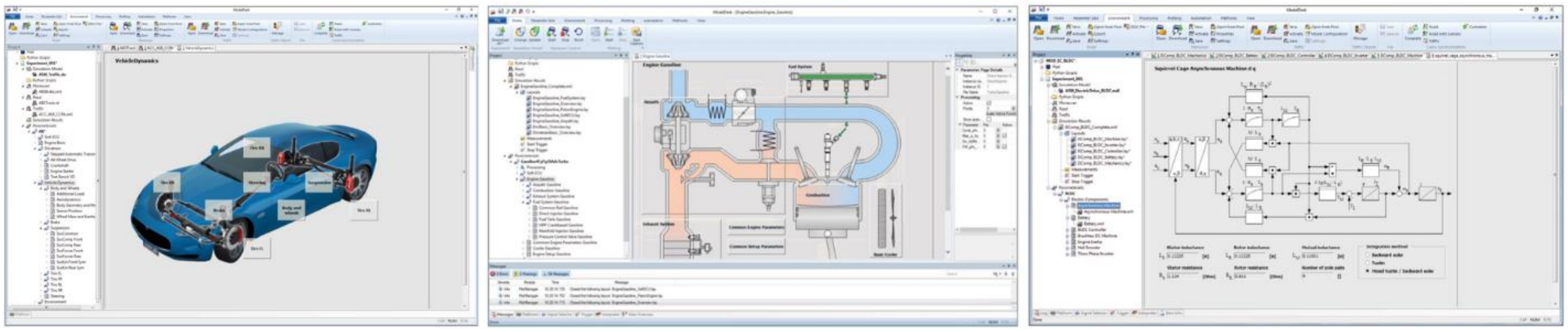

*Figure 28: Images of various DSpace simulation environments for several facets of vehicle development. (Right) Vehicle dynamics, (Middle) engine simulation, (Right) electronic components simulation.*

DSpace provides a number of simulation environments and models covering a multitude of automotive vehicle systems. Multiple sub-system level components can be tested within simulation from the diesel and gasoline engine design, and truck, car and trailer vehicle dynamics. Within DSpace, these vehicles dynamics incorporate a multi-body system consisting of car body and 4 wheels, 13 degrees of freedom (DoF) drive train, table based suspension kinematics, and 3 DoF steering model, to name a few. The electronics simulation could provide processor-based and FPGA-based plant model components, including motors, power electronics, batteries, and position sensors [16].

Through simulation such models can be modified for the desired performance and then the Reduced Order Models (ROMs) can be incorporated into higher order simulations, such as mission level simulations.

## Multi-Physics Simulation

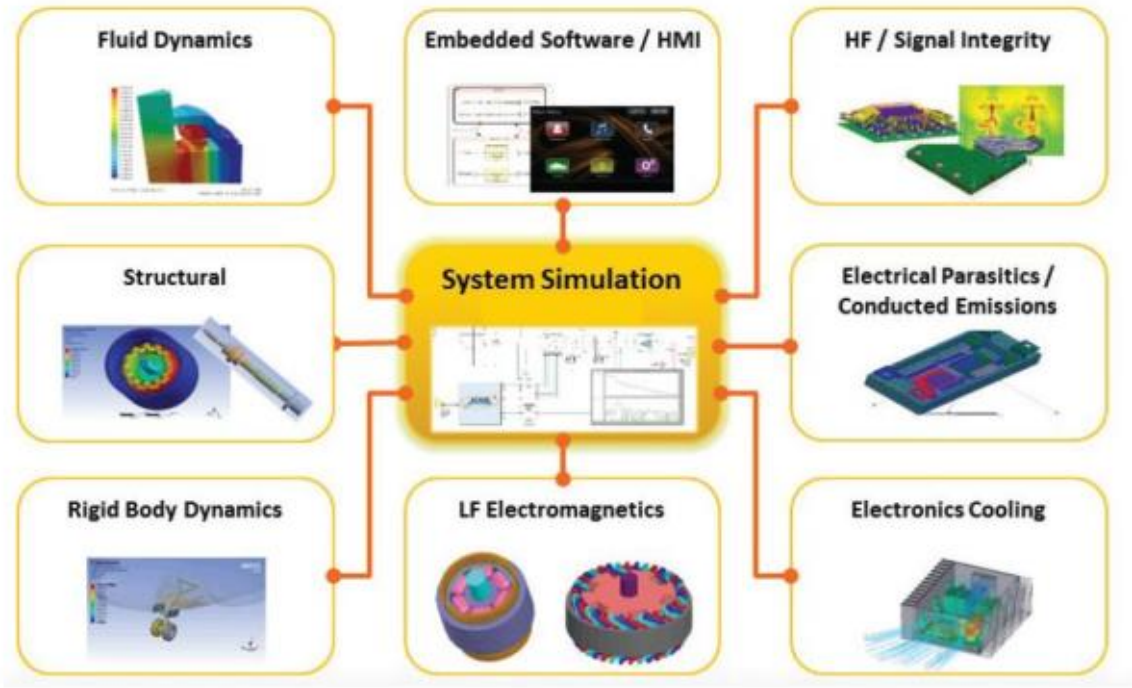

*Figure 29: ANSYS multi-physics simulation suite for systems [4].*

Multi-physics simulation tools provide a number of different functionalities used throughout the development lifecycle of an autonomous vehicle from structural analysis due to forces, heat, and electromagnetic effects, electromagnetics simulation predicting signal integrity, power integrity and thermal integrity of products, to name a few. Additionally, simulation can be used to monitor the thermal performance of electronic control units (ECUs). [4]

## Sensor Validation via Simulation

Sensors enable self-driving cars to perceive their environment detecting and classifying obstacles, predicting velocities, and assisting in localizing the vehicle precisely within its environment. The following subsections discuss the simulation tools and process for designing and constructing radar and optical sensors.

### *Radar*

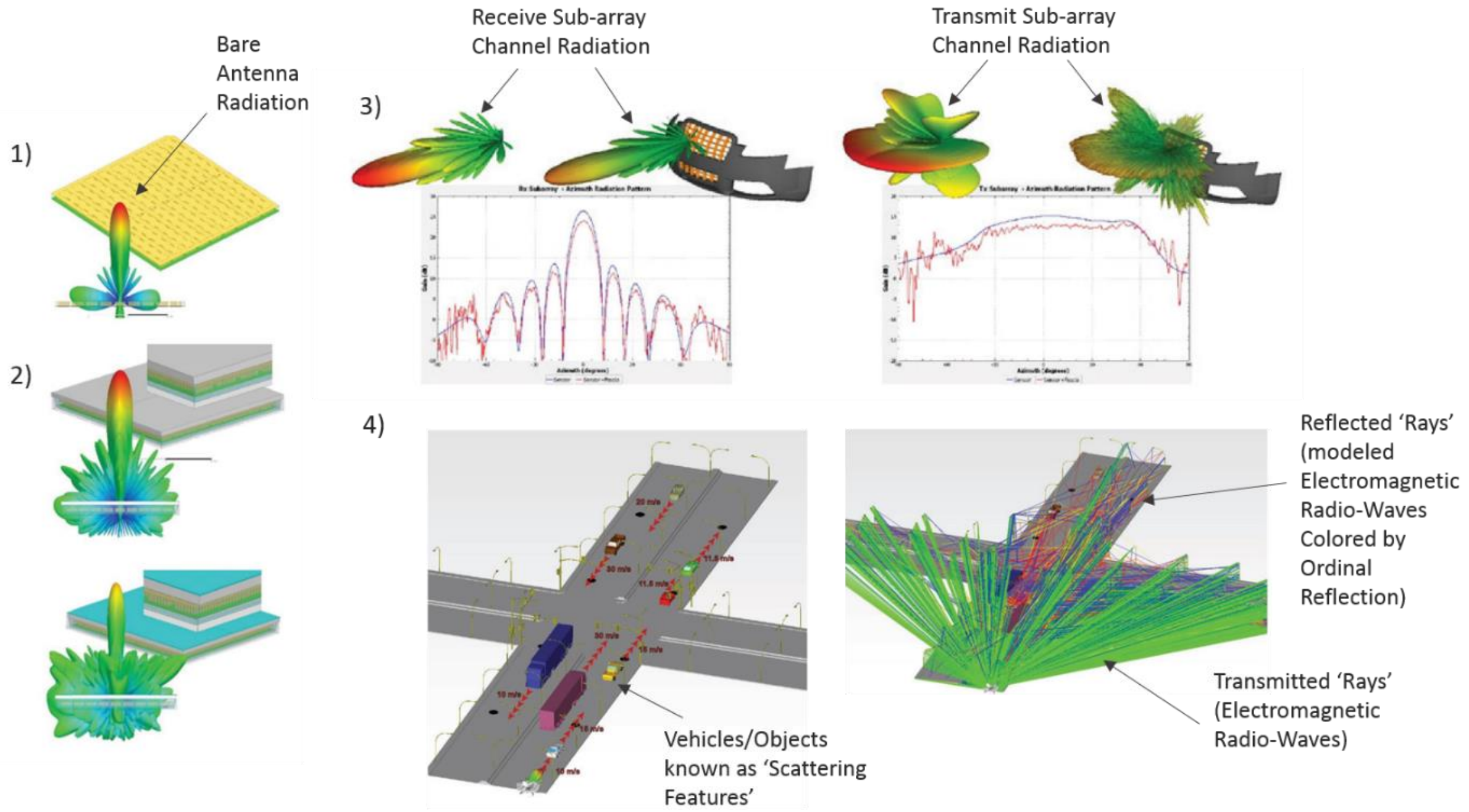

*Figure 30: Sensor (radar) model validation via simulation provided by ANSYS. The collection of images illustrate the series of steps for sensor design and validation. Bare antenna array design and simulated radiation pattern (1); antenna array model integrated with radome and packaging. Shows simulated radiation pattern (2-top); antenna package with a thin layer of water over the radome and the simulated radiation performance of the array (2-bottom). Receive channel sub-array radiation pattern (3-left) and transmit channel radiation pattern (3-right) showing radiation patterns for the module in isolation, and including fascia and bumper interaction due to installation. Shooting and bouncing rays traced from a radar transmit channel throughout the environment. Multiple colors correspond to ordinal reflection for each ray track pictured (4) [8].*

Prototyping, testing and validating radars via simulation is a multistep process as outlined by ANSYS [8]. An overview of the series of steps is provided below and illustrated in Figure 30.

1) Prototype and "tune" antenna topologies which form the interface of a radar system. The antenna both radiates energy from a transmitter in the form of electro-magnetic radio waves and intercepts reflected radio-waves.
2) Test antenna variations (and their resulting radiation pattern) within their structural packaging called radome (top). Additional tests could include environmental conditions, such as a thin layer of water over the radome (bottom).
3) Test the selected antenna's radiation pattern (both for transmit and receive) due to external interactions such as connection to the vehicle's front bumper.
4) Generating a synthetic model of the sensor for use in simulation using the shooting and ray bouncing (SBR) technique. The sensor model is installed on a virtual vehicle in a simulated environment interacting with other vehicles to test performance and coverage.

Previous approaches to radar sensor modeling have used simple point-source statistical scattering models that assume all radar scattering originates from a single point. Appropriate application of the SBR technique can provide full-physics simulation with reasonable efficiency [8].

An example of the SBR technique used by the simulation model is shown in step 4 of Figure 30. The images illustrate how the simulator spreads energy to the objects in the environment (scattering features). Each ray carries an amount of energy that is weighted by the transmit sub-array's far-field radiation pattern (noted in Figure 30). The rays simulate the high frequency radio waves transmitted by the antenna. In a second processing step, the solver integrates the radiation of the reflected electromagnetic waves back at the receive subarray, weighting the reception by the receive subarray's radiation pattern [8].

Within the process outlined above the sensor design is optimized. The antenna designs can be automatically tuned to achieve the best possible combination of performance criteria. For instance, a "specific radiation pattern beam width may be required, with minimal elevation side lobes, while presenting a good electrical match to the transmit power amplifier". Additionally, an antenna may be combined with many antenna to form a multichannel array to achieve specific performance criteria [8].

Modeling and simulation reduces the physical prototyping, installation and testing process from periods originally as long as 9 months to days. "A digital twin of the radar sensor can be quickly evaluated on a digital twin of the vehicle design, providing a complete virtual prototype of the integrated sensor and vehicle front-end. No sheet metal is cut, painted or primed, no plastic injection is done and no prototype sensor is physically built. The cooperation required between vehicle manufacturer and sensor developer is reduced to an efficient exchange of CAD models" [8].

### *Optical*

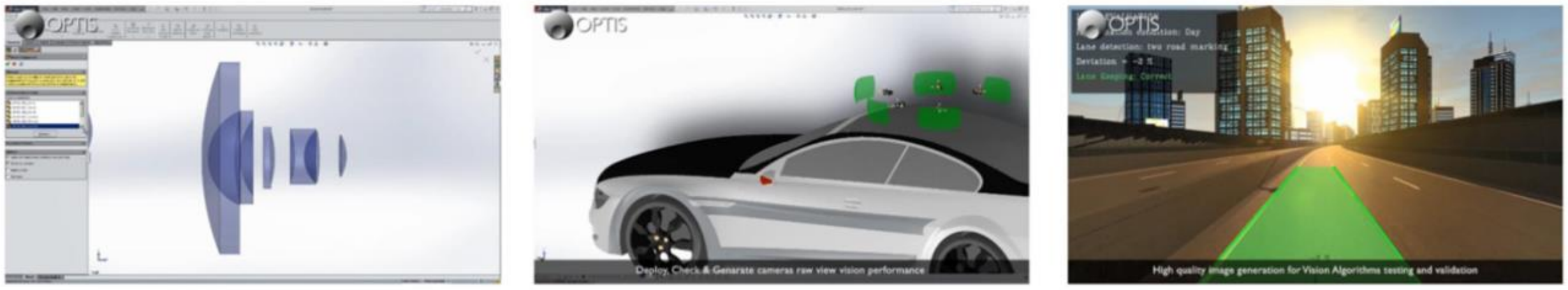

*Figure 31: (Left) Detailed component simulation with ANSYS SPEOS. (Middle) Simulation of in-car installation with SPEOS. (Right) Simulation of sensor performance in driving with VRXPERIENCE, ANSYS's mission level simulation environment [3].*

A similar approach for designing and verifying radar through simulation is utilized for optical sensors.

As an example, ANSYS SPEOS optical simulation software "validates the performance of the optical sensors through physics based simulation that accurately represents their real-world performance. SPEOS enables engineers to perform a detailed physics simulation of optical cameras and LiDARs, taking optical lenses, mechanics, sensors, materials and light properties into account and merging images obtained by multiple cameras" [3].

After component level simulation to generate an initially acceptable bare antenna radiation pattern, "SPEOS simulates the in-vehicle installation to determine the impact of the installation on images produced by the cameras and LiDARs. Finally, a reduced order model (ROM) of the camera and LiDAR models is integrated into a driving scenario that provides the performance needed for real-time simulation. The simulation accurately duplicates the images generated by the cameras and LiDAR, tracing rays of light as they enter the sensors to identify factors such as sun glare and reflections on the road or glass buildings, while validating the perception of the sensors" [3].

## Digital Twin

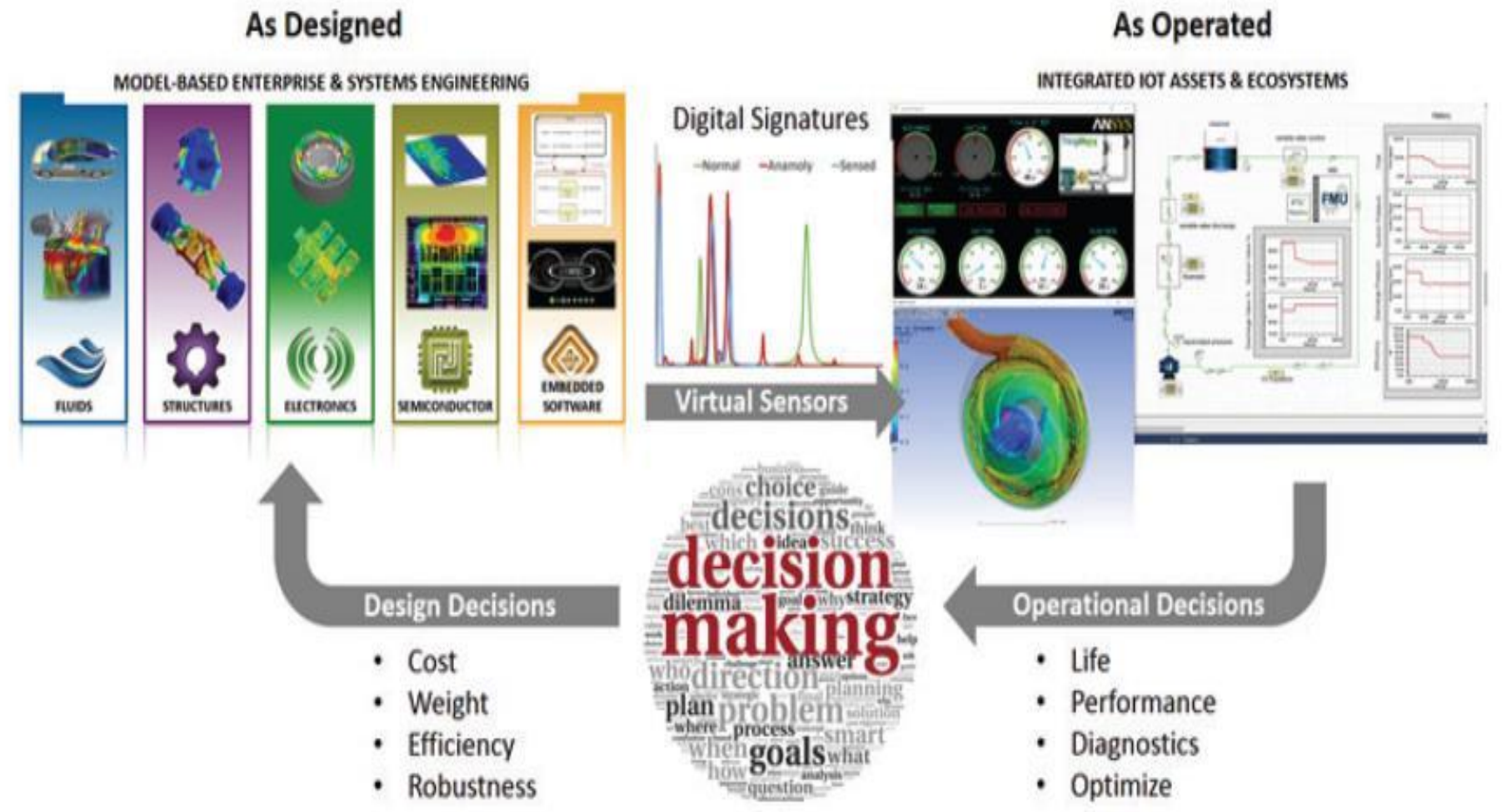

*Figure 32: Internet of Things enabled digital twin pipeline between the simulated model and real world design [4]*

With the increasing feasibility and proliferation of the Internet of Things (IoT) in everyday electronics, the connection between a simulated digital twin and physical implementation of a product is strengthened. Through IoT enabled devices and communication channels, operational logistics of the physical product can be provided to the simulated model in order to drive business value and insight into performance and usability optimization. This is highlighted in Figure 32, which illustrates that operational performance can inform current and future design decisions modeled within simulation.

The IoT connects simulation to the product or process in near real-time, just-in-time or in replay mode to aid in operating and maintaining the product or process. Digital Twins can be utilized to improve real-time performance of products in operation, or provide business and engineering intelligence for improved next-generation design [4].

## High-Fidelity Mapping

High definition mapping of a physical environment are necessary for the current success of self-driving vehicles. HD Maps enable more accurate perception: detection, classification, entity behavior prediction, as well as localization of a vehicle within its environment. Additionally, high definition mapping forms the foundation of many high fidelity simulation environments: Waymo's Carcraft, Uber's AV simulation, NVIDIA DRIVE Sim. The construction of HD Maps is being undertaken not only by suppliers/mapping companies such as HERE, but also the OEMs and self-driving car companies such as Waymo, GM, Ford, and Uber to name a few. The following sub-sections discuss the purpose of HD Maps and their differences between traditional maps, the data sources used to enable mapping, HD Maps as a service, and mapping layers/structure.

## HD-Mapping Purpose

There is a difference between traditional maps utilized by human drivers and the maps intended for autonomous vehicle navigation. Some companies, such as Mapper.ai refer to the distinction as 'human maps' versus 'machine maps'. Both kinds of maps are utilized for location-based decisions. However the end users are distinct and "as a result the types and quantities of information each contains are remarkably different". HD Maps or 'machine maps' contain a much higher level of resolution, greater detail regarding semantic information of the environment: road types, road lanes, curbs, direction of traffic flow, position and relative distances of infrastructure, etc. Further, HD maps can be utilized for navigation of an autonomous vehicle even without a GPS signal, by providing centimeter level accuracy versus meter level accuracy within traditional mapping schemes [19].

There are a number of use cases for HD maps, from driving business intelligence, understanding real time logistics within an environment, and enabling autonomous vehicles to navigate urban and sub-urban environments. The purposes focused on here is HD maps utilization to localize a vehicle within its environment, and to provide enhanced situational awareness for an autonomous vehicle.

HD maps allows the car to localize. Any autonomous car using LiDAR will generate a point cloud in real-time. By overlaying the point cloud against a pre-existing map, the car's on board software can utilize simultaneous localization and mapping (SLAM) algorithms to determine its location by comparing the semantic values within both the 3D point cloud and on-board HD maps [19].

The second major use case for autonomous vehicles discussed here is enhancing situational awareness, as shown in Figure 33.

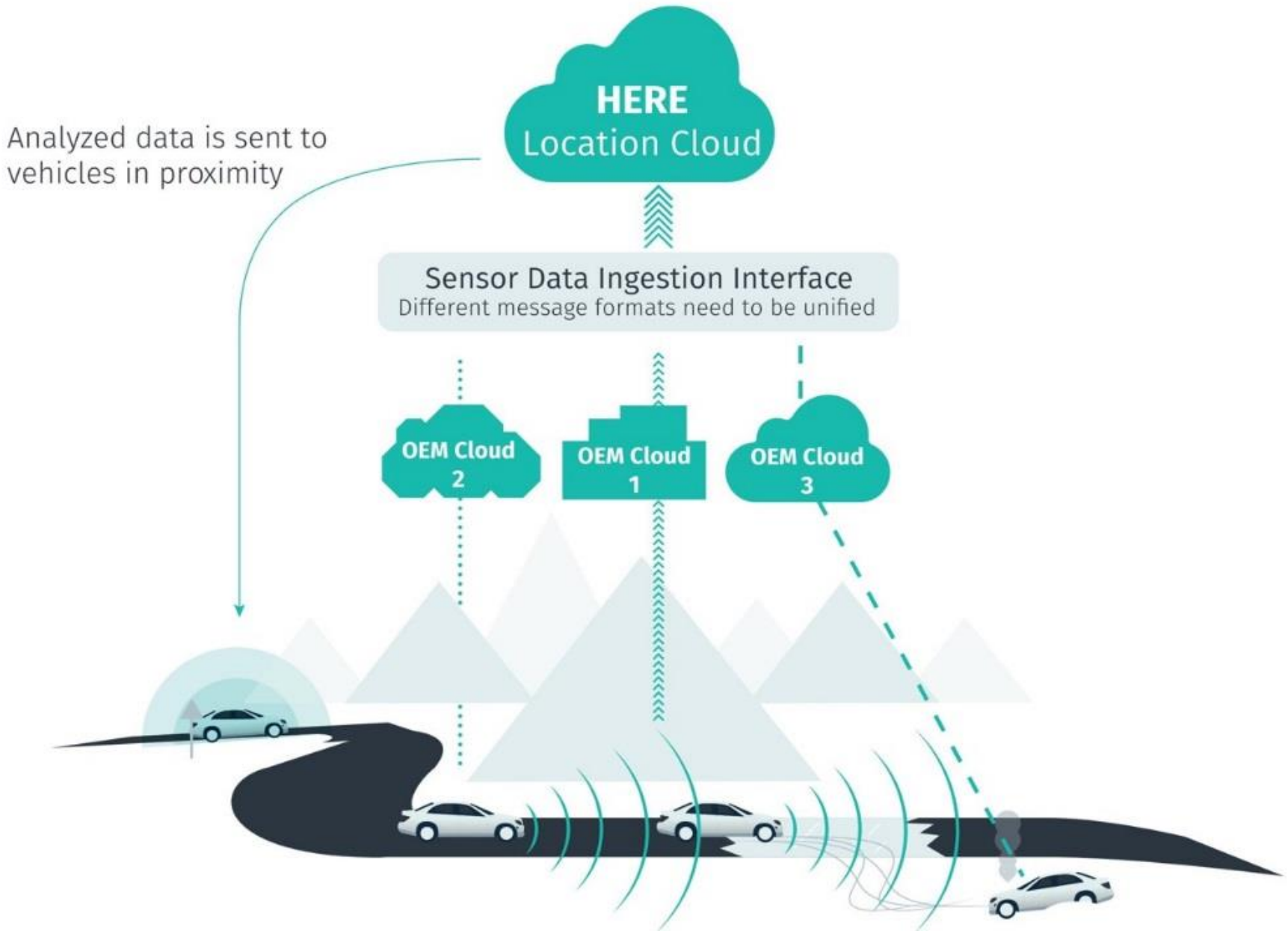

*Figure 33: Example use case of Real Time HD Maps, such as those provided by HERE [25]. In this case, multiple OEM vehicles are comparing their on board maps with live sensor feeds. Discrepancies between the current version of a map and new information, such as real time sensor data, is sent to the HERE cloud database. The real time updates are then sent to other vehicles within close proximity. In this example, the updates provide an alert to nearby vehicles of an accident and icy roads ahead.*

The real-time LiDAR and other sensors on the platform are constantly working to provide data feeds for object detection and avoidance. An accurate point cloud map serves as a baseline view of the environment. With more of an environment pre-mapped, the on-board computer is able to spend less resources on differentiating static from dynamic elements.

By installing maps on board their vehicles, Waymo's self-driving system can focus on the aspects of an environment changing dynamically: such as other vehicles, pedestrians, cyclists. As an example, Waymo's system can utilize the HD maps to "detect when a road has changed by cross-referencing the real-time sensor data with its on-board 3D map. If a change in the roadway (e.g., a collision up ahead that closes an intersection) is detected, our vehicle can re-route itself within the system's operational design domain and alert our operations center so that other vehicles in the fleet can avoid the area" [46].

High definition maps are essential for current implementations of self-driving vehicles.

## Sources used to Construct HD Maps

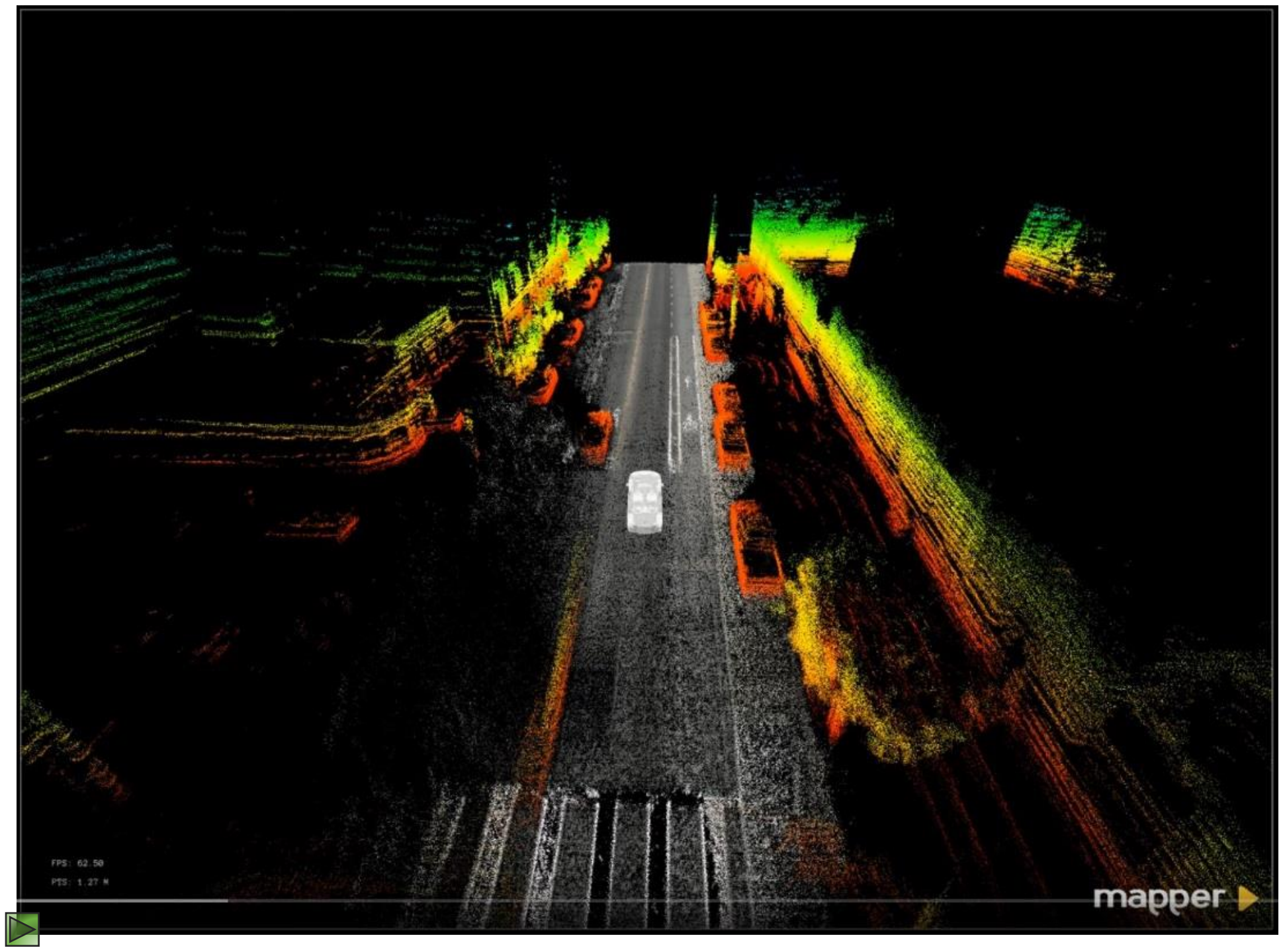

*Figure 34: 3D point cloud from LiDAR sensor of the Mission neighborhood in San Francisco near Mapper.ai office [19]. LiDAR 3D Point Clouds are used as a base for HD Maps.*

An insightful explanation of data feeds utilized to construct HD Maps is provided by HERE technologies, specifically concerning their HERE HD Live Map which is intended to be updated real time as environmental conditions change. Note, other companies utilize a similar process.

The foundation of HERE HD Live Map is sourced from high-end industrial-capture vehicles. These vehicles include four cameras with 96-megapixel resolution, a 32-beam spinning Velodyne LiDAR camera, and an IMU inertial sensor unit that capture precise road data. HERE vehicles collect 700,000 3D data points at a time, which accumulates over 140 gigabytes of location data a day [25].

In addition to industrial capture vehicles, HERE uses other sources - such as satellite imagery, aerial imagery, government data, crowd-sourced vehicle sensor data, and mobile probes. The combination of sources provides high-quality environmental data - such as road and lane information, and roadside objects like signs and barriers - to ensure vehicles have precise positioning for lateral and longitudinal control, as well as inform proactive driving decisions [25].

## HD Mapping Process

There are roughly four primary steps in constructing an HD Map, such as those provided by Mapper.ai and HERE.

1. Construct the base map utilizing sensor (LiDAR) equipped capture vehicles.
2. Collect data provided by partners or crowd-sourced vehicles.
3. Aggregate and fuse the various data sources in the cloud, resolving discrepancies and differing vehicle perspectives.
4. Publish the new map information, intelligently. I.e. to ensure efficient data transfer, only the updates that occur within the specific tile for the specific layer - the Road Model, HD Lane Model, and HD Localization Model - get sent to the OEM's cloud and the vehicle [25].

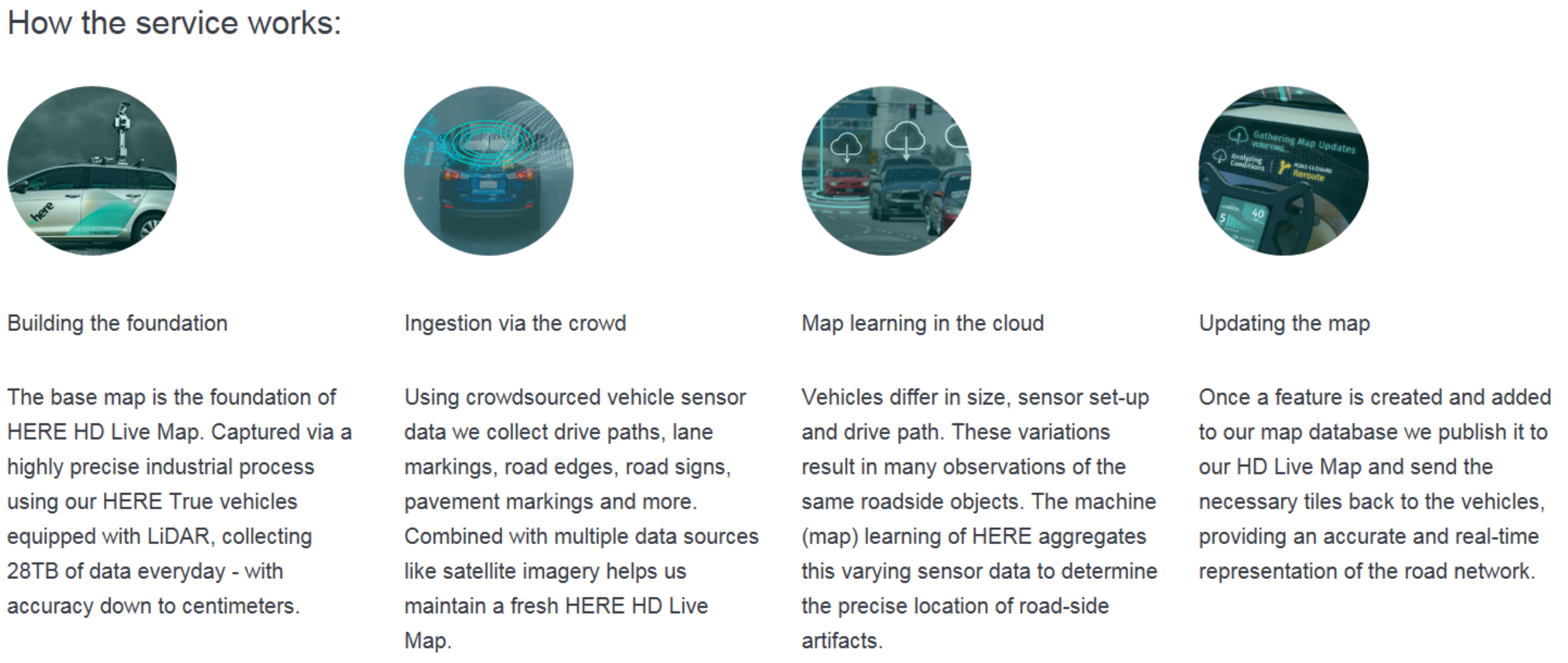

*Figure 35: Breakdown of the process for building, enhancing, aggregating and updating HERE HD Maps [26].*

Additionally, published data has a 'Quality Index' that informs vehicles and users the predicted quality/accuracy of the data provided. For instance, if the speed displayed on a street sign has not been updated within several years, there is a lower probability of correctness, and the map needs to be refreshed.

### HD Map Structure/Layering

HERE’s HD Map is organized as a series of tiles (grids within a particular region/area) and several layers containing different types of information: HD Localization Model, Road Model, and HD Lane Model. An illustration of the three layers are provided in Figure 36 (left). Additionally, the image on the right provides another organization of HD Map semantics, provided by Mapper.ai.

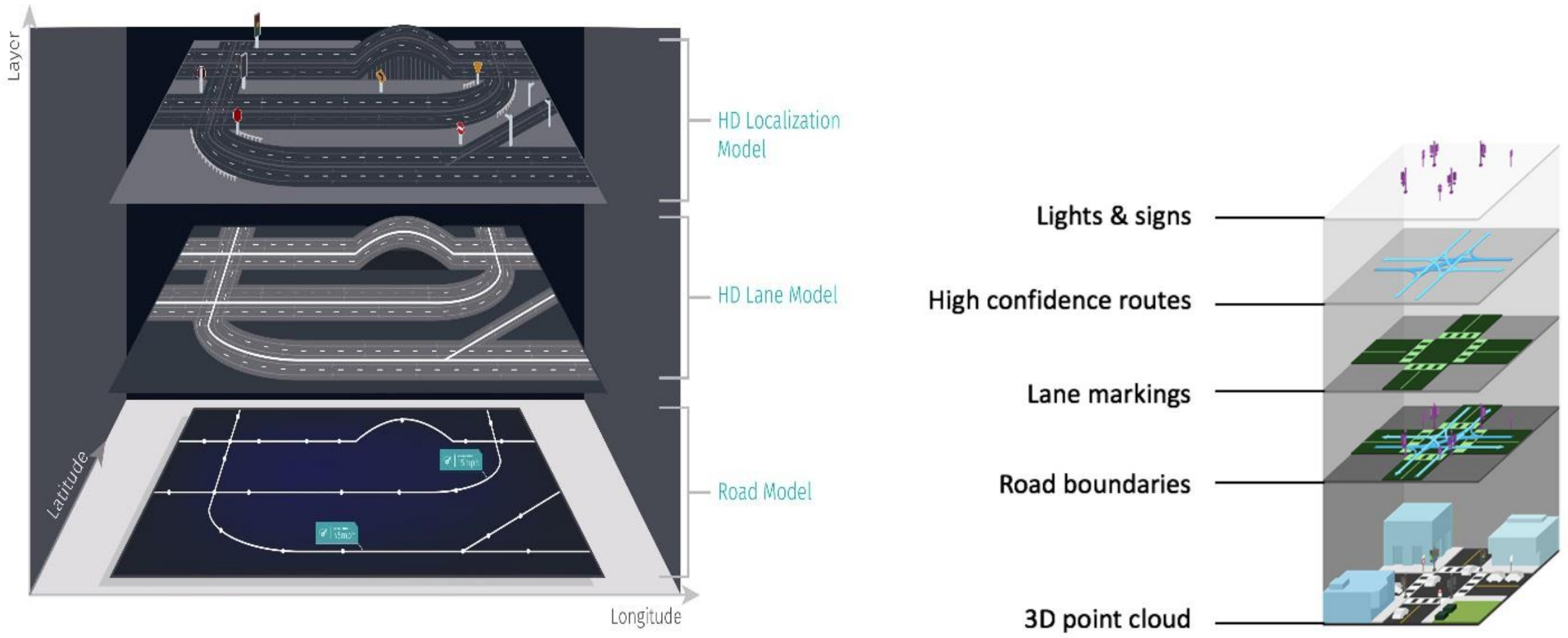

*Figure 36: (Left) HERE HD Live Map Layers which organize the relevant map data into distinct categories [25]. (Right) Mapper.ai semantic layers organization [19].*

A synopsis of the different layers is provided by HERE [25]:

**HD Localization Model:** The top layer helps the vehicle localize itself in the world by using roadside objects like guard rails, walls, signs and pole like objects. The vehicle identifies an object, then uses the object’s location to measure backwards and calculate exactly where the vehicle is located.

**HD Lane Model**: The next layer – the HD Lane Model – provides more precise, lane-level detail such as lane direction of travel, lane type, lane boundary, and lane marking types, to help self-driving vehicles make safer and more comfortable driving decisions.

**Road Model:** The Road Model offers global coverage for vehicles to understand local insights beyond the range of its onboard sensors such as high-occupancy vehicle lanes, or country-specific road classification.

## Simulation to Real World Pipeline

The simulation to real world process is multi-facetted. While aspects of this pipeline are obviously not performed directly within a simulated environment, the information obtained by real world testing: either through closed course and city, suburban, and rural road tests provide data and insights into new, relevant scenarios that can directly inform the simulation environment.

The typical pipeline adhered to by most autonomous vehicle companies is to collect real world data to inform/construct the simulation environment and train any relevant machine learning techniques on labeled versions of the data. From here, hardware in the loop testing, closed course testing, and/or high fidelity simulators provide a more realistic intermediary step, enabling developers to validate their software and hardware for safety and functionality before deploying to vehicles driving in public areas. Each of these development stages: simulation, closed course testing and real-world testing form a multi-directional feedback loop, where each stage helps inform the others.

The following sections discuss the various methods and processes that add realism and credibility to simulation.

## Operational Design Domain

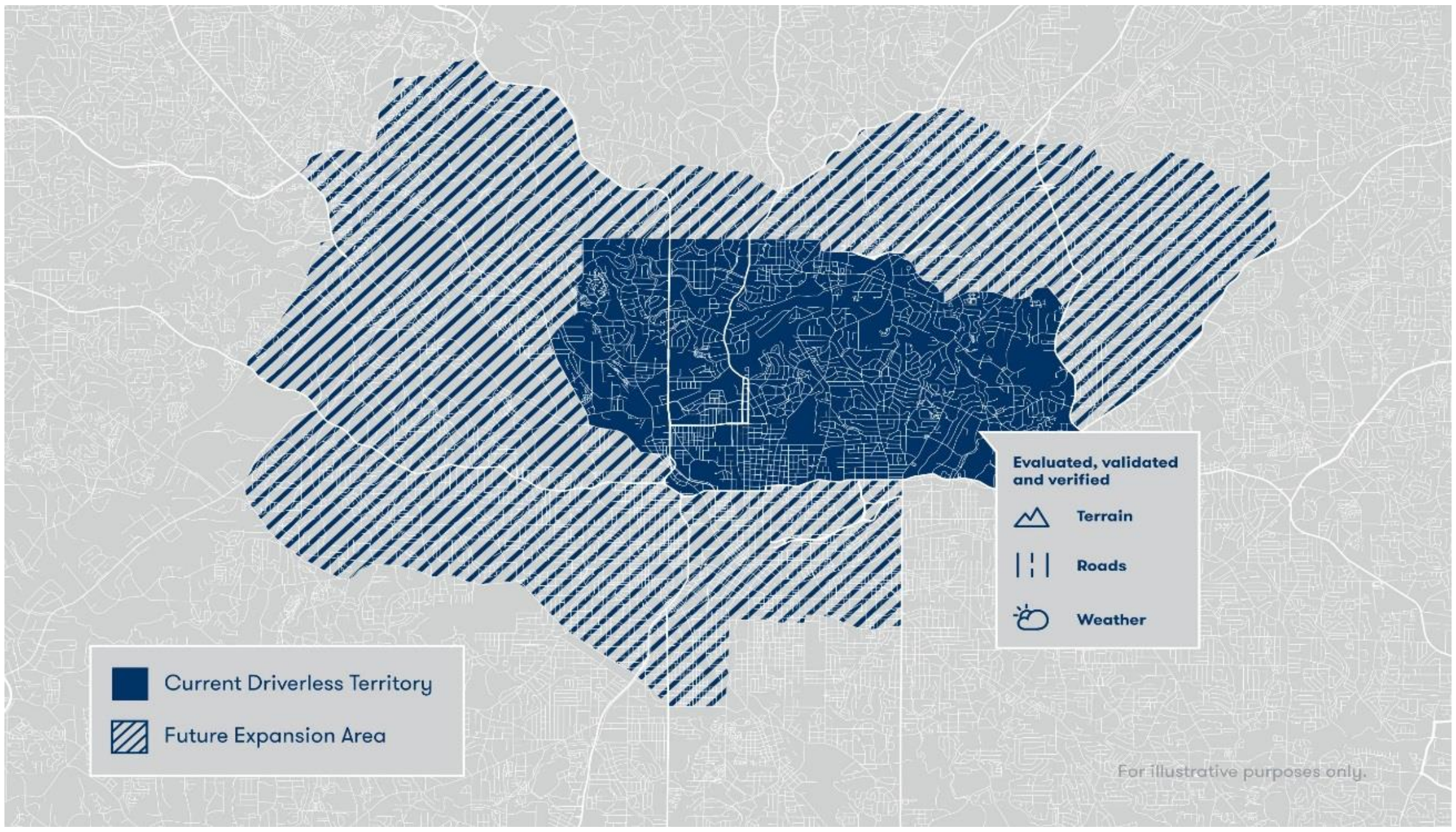

*Figure 37: Operational Design Domain graphic [46].*

The operational design domain (ODD) describes the conditions under which a self-driving vehicle is intended to operate. The ODD could specify any number of conditions. Waymo's domain includes geographies, roadway types, speed range, weather, time of day, and state and local traffic laws and regulations [46].

ODD is an important concept to consider not only from a safety perspective but also as an informant to the types of simulation environments companies (like Waymo and Uber) address. The simulation environment primarily contains regions that have been fully mapped and incorporated into their ODD [46].

According to Uber the ODD characterization process includes, quote: [45]

• **Driving the area manually** to collect detailed data and logs on the scenarios and actors that exist within the ODD.

• **Adding data tags** to camera and LiDAR footage collected from manually-driven logs, highlighting potentially relevant attributes of actors in and around the road as well as attributes of road design (e.g. road geometry or curvature, traffic control measures).

• **Synthesizing the tagged data** to identify and break down information on all scenarios and subsequent system behavior requirements for each scenario.

• **Creating representative simulation and track tests** to evaluate current and future software releases.

The operational design domain is the methodology by which self-driving car companies iteratively increase the scope of their vehicles capabilities and operating region.

## Software in the Loop (SIL)

The premise behind software in the loop is to incorporate the software actually executing on a real platform, within simulation. In the case of NVIDIA the output of their DRIVE Sim simulator could be fed into a server containing the complete AV software stack normally running on the DRIVE AGX Pegasus car computer. The software could process the simulated sensor data and output vehicle controls [37]. As an another example, an aerial vehicle running in AirSim could operate using the typical flight controller firmware such as PX4, ROSFlight, or Hackflight, which takes desired state and sensor data as inputs, computes an estimate of current state, and subsequently outputs the actuator control signals to achieve the desired state [40].

Software in the loop is addressed throughout the document.

## Hardware in the Loop (HIL)

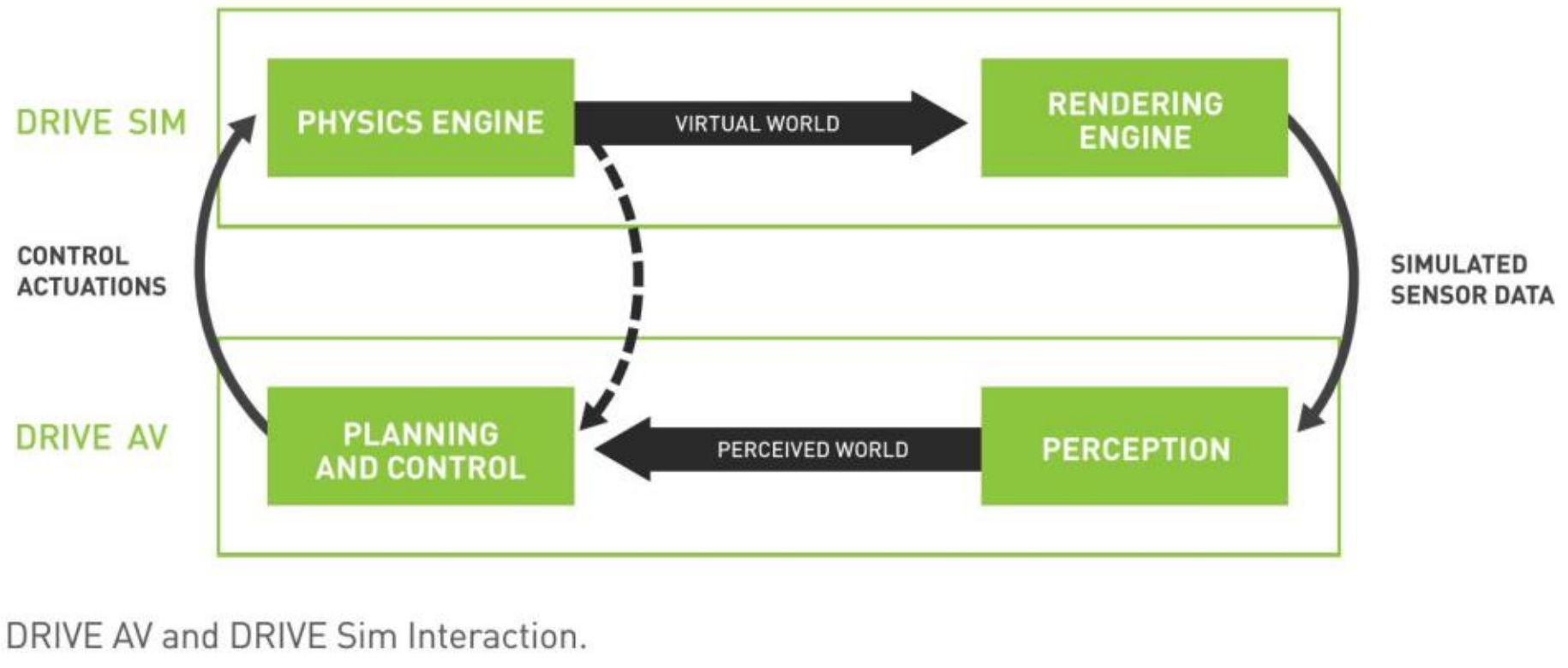

*Figure 38: NVIDIA DRIVE AV and DRIVE Sim interaction [37]. DRIVE Sim is the autonomous vehicle simulation environment and DRIVE AV contains the hardware in the loop and autonomous vehicle software stack.*

Hardware in the loop capabilities enable the autonomous vehicles hardware processor to run while interfacing with the simulation environment. This feature adds another aspect of realism and rigor to the testing capabilities provided via simulation. A good example of HIL is provided by NVIDIA's autonomous vehicle tool-suite, DRIVE Constellation.

In short, DRIVE Constellation is a closed-loop simulation plus processor platform running two servers. One server uses simulation to generate photorealistic imaging to create the sensor output of an autonomous vehicle. The simulation server then sends the sensor data to the DRIVE AGX Pegasus platform running on a second server to perform the various software algorithms on the hardware - handling perception, planning, and virtually navigating the vehicle. Specifically, the DRIVE AGX Pegasus vehicle computer runs the complete, unmodified binary autonomous vehicle software stack (DRIVE AV) that operates inside an autonomous vehicle. DRIVE AGX processes the simulated data as if it were coming from the sensors of a vehicle actually driving on the road, and sends actuation commands back to the simulator. The control output updates the virtual model of the vehicle within simulation. This happens at 30 frames a second [37].

## High-Fidelity Simulators

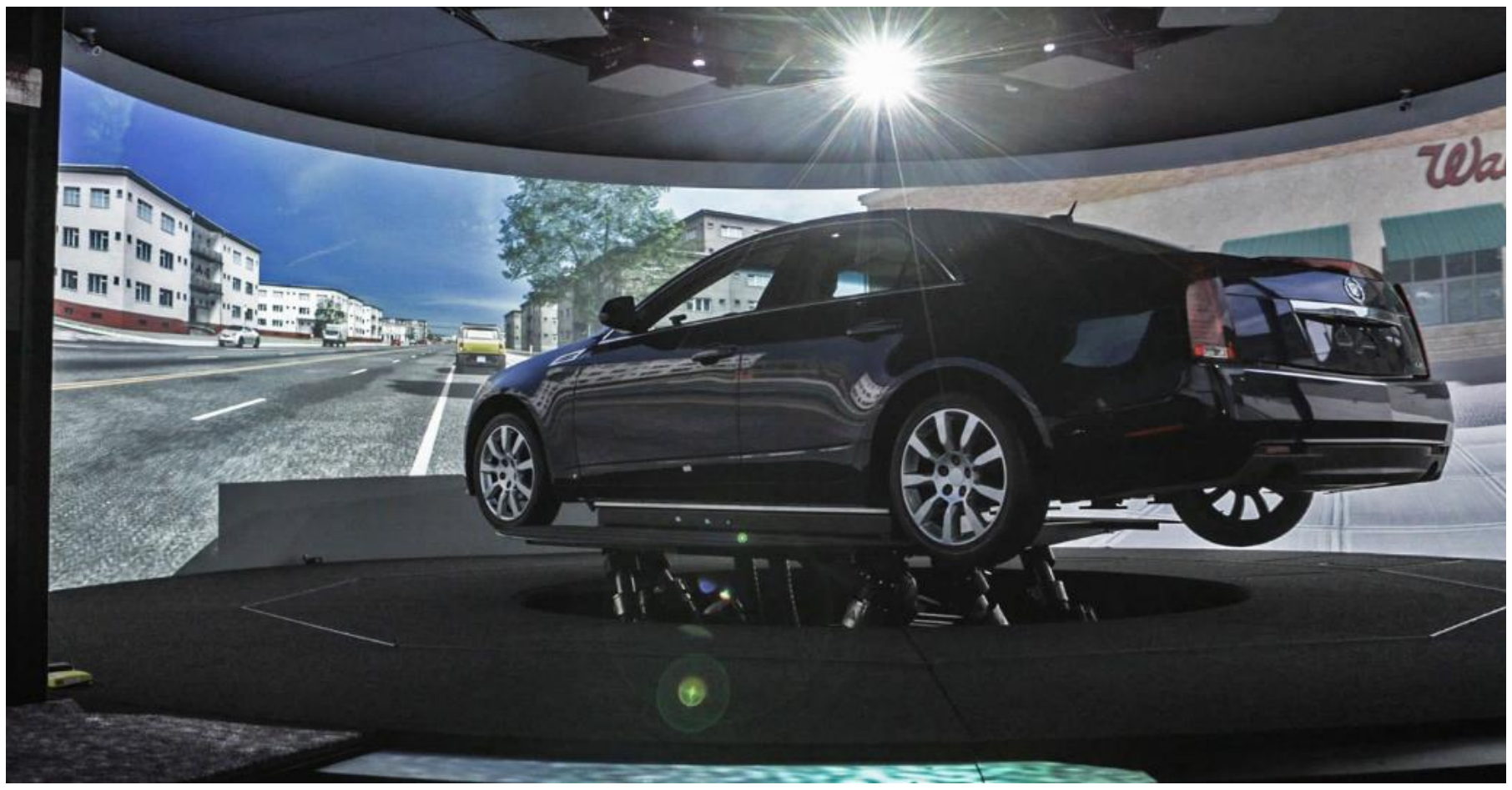

*Figure 39: GM's 360 degree, high-fidelity vehicle simulator [47].*

Both BMW and GM provide examples of high fidelity vehicle simulators, which enable not only software and hardware testing in the loop but can mimic physical environment and road conditions through multi-axis actuated platforms, synced vehicle dynamics and visuals, as well as sophisticated technologies for monitoring driver behaviors.

BMW's high fidelity simulator connects systems under test to a vehicle mock-up attached to a platform within an encapsulating dome. The dome can be moved via an electric drive to emulate longitudinal, transverse and rotational movements of a vehicle. Longitudinal and transverse acceleration forces of up to 1.0 g can be generated. The controllable platform is used to test new systems and functions by replicating highly dynamic evading maneuvers, full braking and hard acceleration. What's more, the simulator provides an extremely detailed rendering of the world while synchronizing the visuals with the multiple degrees of freedom movements of the vehicle. [7]

GM provides a similar functionality with its 360 degree high fidelity simulators. The simulator features a movable platform capable of maneuvering a full-size car to simulate roll, pitch and yaw. High fidelity visuals are provided with a 5-terabyte-per-second image generator supporting 4K resolution. This allows response to steering and pedal force inputs to take place within 50 milliseconds. Additionally, engineers can monitor the facial expressions and biometrics to determine how passengers and drivers feel about the experience [47].

## Closed-Course Testing

After testing and validating software and hardware functionality within simulation (software in the loop and hardware in the loop), closed course testing is used as an intermediary step to refine software and develop new test scenarios, before driving autonomous vehicles in public areas. Most autonomous vehicle companies mentioned in this paper utilize closed course testing from Waymo, Uber, GM, Ford, and BMW.

Ford and Waymo are two such examples discussed.

On the University of Michigan's 32-acre M-city simulated urban environment, Ford conducts a number of closed course tests within various conditions from sun to snow [34]. Ford uses closed course environments to subject the vehicles to edge cases and difficult situations such as deliberately injecting faults into fail-functional components - braking and sensors. Additionally, they test the system's ability to transition to a Minimal Risk Condition during malfunctions with safety operators close at hand [18].

Waymo has set up a private, 91-acre, closed-course testing facility in California specially designed and built for unique testing needs. This private facility is known as 'Castle', and is set up like a mock city, including everything from high-speed roads, suburban driveways, railroad crossing, to various props like cones, toys, bicycles, etc. The Waymo team uses this and other closed-course facilities to validate new software before it's released to the fleet of vehicles on the road. Closed course testing is also useful to stage challenging or rare scenarios enabling vehicles to gain experience handling unusual situations within a controlled environment [46]. These difficult scenarios can then be reproduced within simulation from the data collected while navigating the closed course. Waymo has developed more than 20,000 simulation scenarios at Castle, which "might take hundreds of thousands of driving miles to encounter on public roads" [46].

## Scalability and Infrastructure

Simulation tools are not enough, on their own, to test and validate autonomous systems. The simulation environments, machine learning algorithms which support the vehicle software stack, hardware, and various models and sensors are computationally expensive, generating an enormous amount of data.

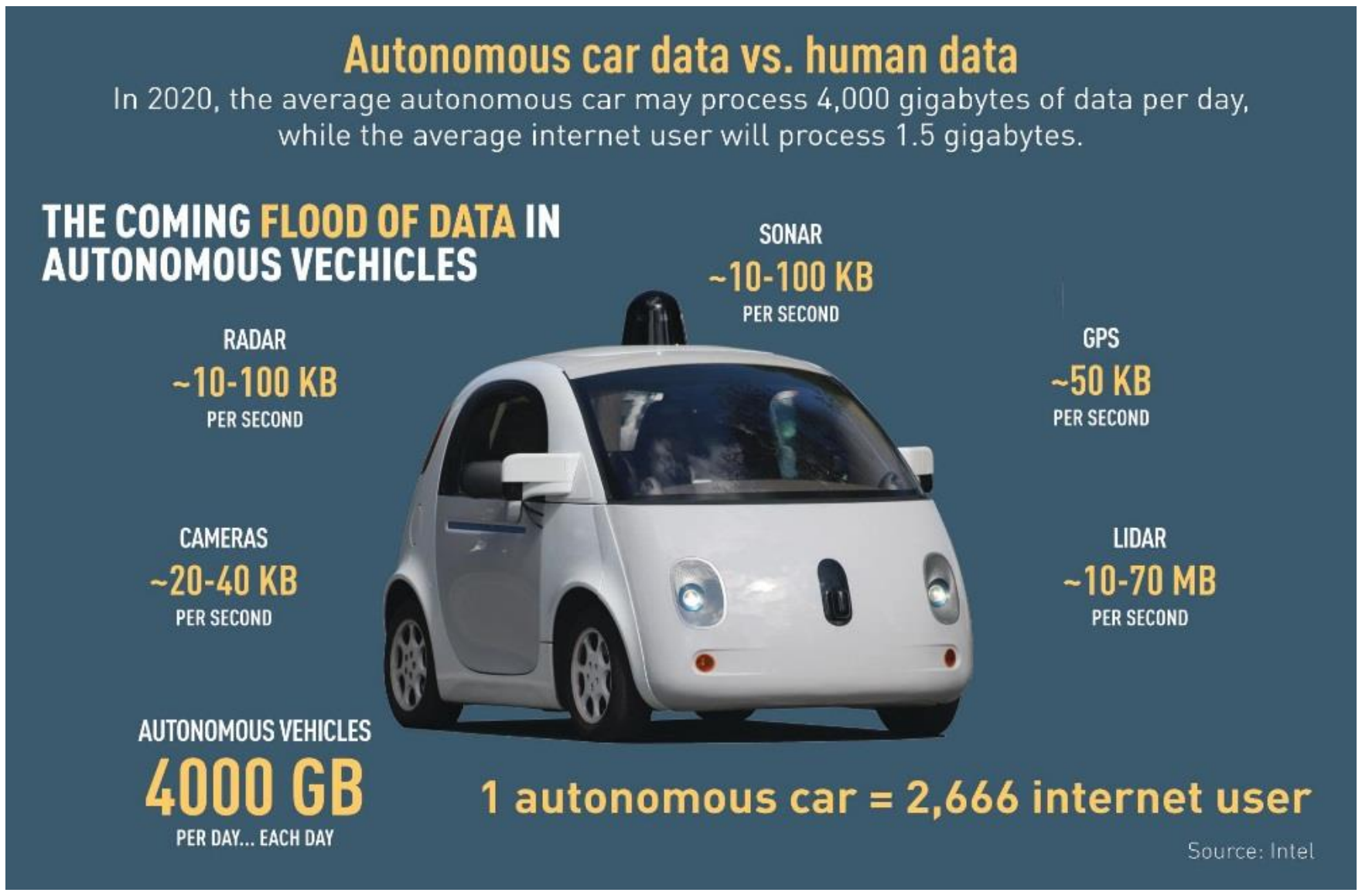

*Figure 40: Autonomous vehicle configured with various sensors versus a human in terms of data processing. A single autonomous vehicle is comparable to 2667 internet users [21].*

As shown in Figure 40, an impressive amount of data is generated from a single autonomous vehicle, particularly due to the advanced sensor systems: radar, LiDAR, GPS, cameras, etc. on board. Approximately 4 TB (terabytes) of data are generated by each autonomous vehicle daily. This is 2667 times greater than the amount of data generated by a single internet user [21]. Another way of looking at the amount of data generated by an AV is through comparing the storage in a common flash drive. Assuming the flash drive has a capacity of 16 GB, it would take 250 flash drives to store the data of a single autonomous car. As an NVIDIA representative further elaborated, the "computational requirements of fully autonomous driving are enormous—easily up to 100 times higher than the most advanced vehicles in production today" [37].

In order to leverage the insights provided by simulation techniques and data generated from virtual testing (hardware and software in the loop), closed course testing, and public tests, a powerful collection of supplementary machinery such as large scale compute clusters, cloud architectures, and machine learning pipelines are required. The following sections shed light on some of these supplementary, but necessary requirements, to assist in simulating an autonomous vehicle.

## Compute Infrastructure

The compute infrastructure can be utilized for the simulation-to-real-world pipeline, as well as to train and test machine learning algorithms comprising the autonomous vehicles on-board software. Concretely the compute resources utilized by the various companies mentioned assists in the deployment of simulation directly, as well as the development of supplementary programs and modules utilized by simulation.

What follows are multiple examples of several companies compute infrastructures utilized for developing various aspects of autonomous systems. Similar to the section on simulation throughput, it is difficult to quantify the scale of infrastructure needed as this greatly depends on the specific simulation toolset and intended use case, as well as the amount of data collected, rate of collection, expected lifespan of the data, and again the intended use case of the data. By way of example (and as we will see in later sections), machine learning algorithms utilized within autonomous vehicles require an abundance of accurately labeled training and test data. The size of the model and desired performance, and speed to convergence affect the infrastructure requirements.

In order to help provide a common comparison to the different compute infrastructures, both on and off-board the vehicle, the performance of various Apple laptops and desktops is provided in Figure 41.

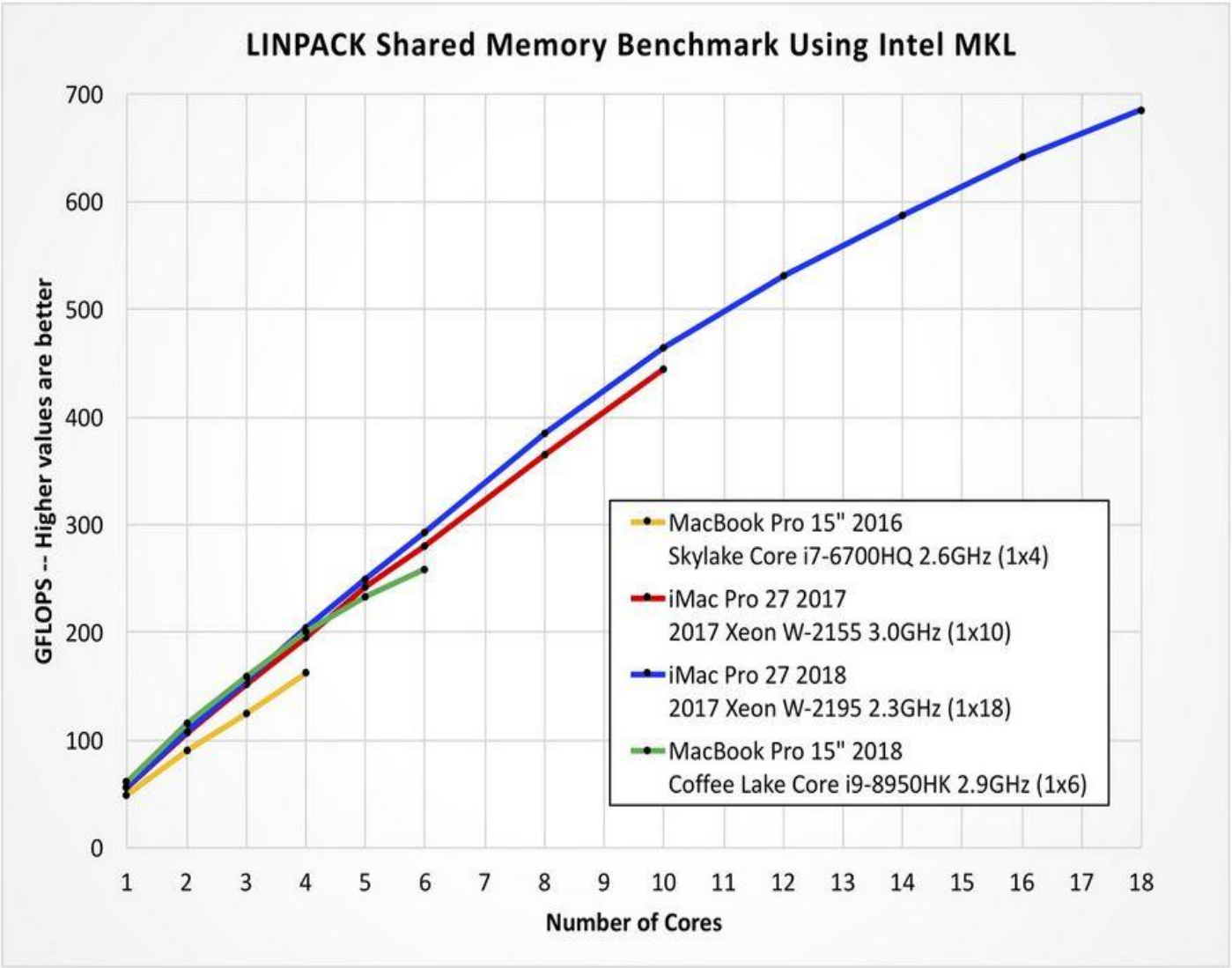

*Figure 41: Computational performance of various Apple computer models in GFLOPS (floating point operations per second). The performance of each Apple computer was determined by computing the LINPACK benchmark: solving a dense system of linear equations [27].*

The computational performance of various Apple computer models is shown in GFLOPS (floating point operations per second). The performance of each Apple computer was measured by computing the LINPACK benchmark: solving a dense system of linear equations by LU decomposition. The benchmark solves a system of 15,000 equations using Intel's Math Kernel Library [27]. As shown in the figure, the 2018 iMac Pro 27, a consumer desktop computer, has the highest performance of the compared models, using 18 cores to achieve 690 GFLOPS. This metric will be used to help shed light on the relative capabilities of autonomous vehicle infrastructures.

## On Board Vehicle Compute

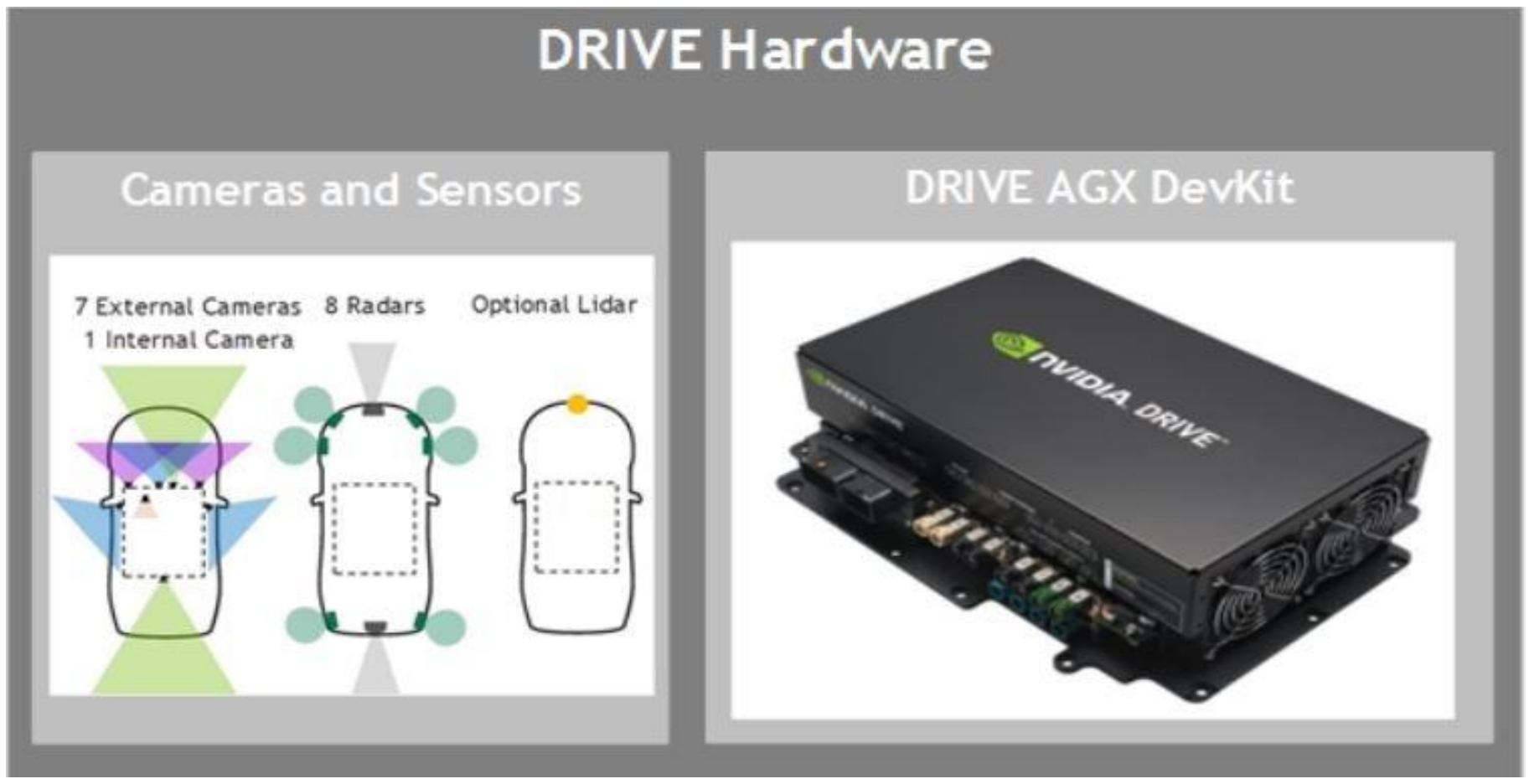

*Figure 42: NVIDIA DRIVE Pegasus AGX development kit providing SAE Level 5 Autonomous Vehicle processing capabilities [36].*

This subsection highlights some of the compute resources available for processing that occurs within the autonomous vehicle. NVIDIA is known for high performance compute, and are providing a number of products marketed toward autonomous vehicle development. DRIVE Pegasus AGX is the most recent on-board vehicle processing unit offered by NVIDIA. DRIVE AGX has a number of compute interfaces to support self-driving vehicles, as shown in Figure 42: 7 external cameras, 1 internal camera, 8 radars, and an optional LiDAR [36].

While it is not possible to find specifications on the throughput (given as FLOPS), NVIDIA representatives have provided details. The DRIVE AGX platform is "the size of a laptop [and] delivers the performance of more than 60 laptop computers. DRIVE AGX Pegasus development platform can scale up to 320 TOPS [integer, not floating point operations] of peak performance" [43]. Further information is available on DRIVE AGX's predecessors, however.

| NVIDIA DRIVE PX Specification Comparison | | |
|---|---|---|
| | DRIVE PX | DRIVE PX 2 |
| **SoCs** | 2x Tegra X1 | 2x Tegra "Parker" |
| **Discrete GPUs** | N/A | 2x Unknown Pascal |
| **CPU Cores** | 8x ARM Cortex-A57 + 8x ARM Cortex-53 | 4x NVIDIA Denver + 8x ARM Cortex-A57 |
| **GPU Cores** | 2x Tegra X1 (Maxwell) | 2x Tegra "Parker" (Pascal) + 2x Unknown Pascal |
| **FP32 TFLOPS** | > 1 TFLOPS | 8 TFLOPS |
| **FP16 TFLOPS** | > 2 TFLOPS | 16 TFLOPS? |
| **TDP** | N/A | 250W |

*Figure 43: NVIDIA DRIVE PX and DRIVE PX 2 specifications. [41]*

DRIVE AGX is utilizing new hardware components: 2 Xavier Systems on a Chip (SoC), and will include 2 next generation discrete GPUs. Therefore it should provide additional compute to the 8 TFLOPS (32-bit) offered in DRIVE PX 2, which is roughly equivalent to 11.6 2018 iMac Pros (assuming details provided in Figure 41).

## Off Board Vehicle Compute

| SYSTEM SPECIFICATIONS | |
|---|---|
| GPUs | 16X NVIDIA® Tesla V100 |
| GPU Memory | 512GB total |
| Performance | 2.1 petaFLOPS |
| NVIDIA CUDA® Cores | 81920 |
| NVIDIA Tensor Cores | 10240 |
| NVSwitches | 12 |
| Maximum Power Usage | 12kW |
| CPU | Intel® Xeon® Platinum 8174 CPU @3.10GHz, 24 cores per CPU |
| System Memory | 1.5TB |
| Network | 8X 100Gb/sec Infiniband/100GigE Dual 10/25/40/50/100GbE |
| Storage | OS: 2X 960GB NVME SSDs Internal Storage: 30TB (8X 3.84TB) NVME SSDs |
| Software | Ubuntu Linux OS See Software stack for details |
| System Weight | 360lbs (163.29kgs) |
| Packaged System Weight | 400lbs (181.44kgs) |
| System Dimensions | Height: 17.3 in (440.0 mm) Width: 19.0 in (482.3 mm) Length: 31.3 in (795.4 mm) - No Front Bezel 32.8 in (834.0 mm) - With Front Bezel |
| Operating Temperature Range | 5°C to 25°C (41°F to 77°F) |

| SYSTEM SPECIFICATIONS | |
|---|---|
| GPUs | 8X Tesla V100 |
| Performance (Mixed Precision) | 1 petaFLOPS |
| GPU Memory | 256 GB total system |
| CPU | Dual 20-Core Intel Xeon E5-2698 v4 2.2 GHz |
| NVIDIA CUDA® Cores | 40,960 |
| NVIDIA Tensor Cores (on V100 based systems) | 5,120 |
| Power Requirements | 3,500 W |
| System Memory | 512 GB 2,133 MHz DDR4 RDIMM |
| Storage | 4X 1.92 TB SSD RAID 0 |
| Network | Dual 10 GbE, 4 IB EDR |
| Operating System | Canonical Ubuntu, Red Hat Enterprise Linux |
| System Weight | 134 lbs |
| System Dimensions | 866 D x 444 W x 131 H (mm) |
| Packing Dimensions | 1,180 D x 730 W x 284 H (mm) |
| Operating Temperature Range | 5–35 °C |

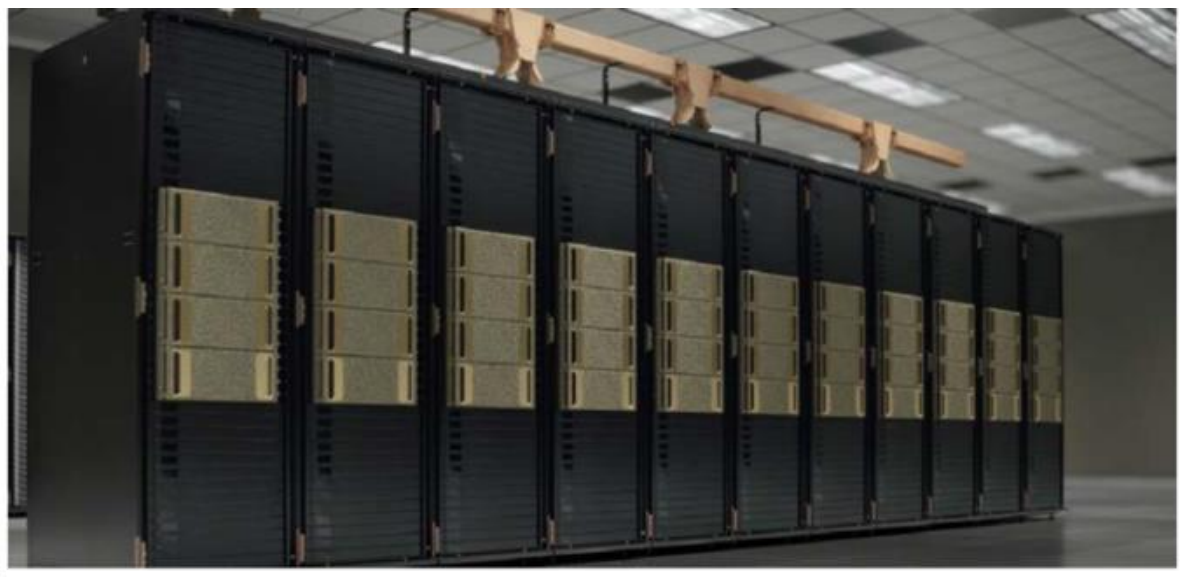

*Figure 44: (Left) Specs for NVIDIA DGX-2H supercomputer data center, successor to DGX-1 [14]. (Middle) Specs for NVIDIA DGX-1 supercomputer [13]. (Right) NVIDIA DRIVE Perception Infrastructure supporting data collection and post processing provided by information gathered from large autonomous fleets. The infrastructure runs on NVIDIA DGX SaturnV - an AI supercomputer [37].*

Additionally, capable compute clusters - interconnected hardware and processors - are needed for data collection, data analysis, machine learning development, running simulations, and more. NVIDIA and BMW provide examples.

NVIDIA offers several supercomputers: DGX-1 and DGX-2H, as well as large scale infrastructure, SaturnV, utilizing many of these supercomputing datacenter's in a box. The specifications for DGX-2H and DGX-1 are provided in Figure 44. DGX-1 and DGX-2H provide 1.0 petaFLOPS and 2.1 petaFLOPS, respectively. Comparatively, this performance is 1450 times and 3043 times higher than a 2018 iMac Pro. According to NVIDIA, “DGX-2 has the deep learning processing power of 300 servers occupying 15 racks of datacenter space, while being 60x smaller and 18x more power efficient” [9]. Taking compute capability a step further, multiple DGX data-centers can be interconnected.

“NVIDIA DRIVE Perception Infrastructure delivers and supports massive data collection, deep learning development, and traceability to support large autonomous fleets. It runs on the NVIDIA DGX SaturnV—our AI supercomputer comprised of 660 NVIDIA DGX-1™ systems with 5280 GPUs—and is capable of 660 petaFLOPS for AI model development and training” [37].

While compute is obviously important, storage is likewise necessary for collecting large datasets. BMW intended to provide 70 PB of data storage beginning 2018 with a target capacity of 200 PB in 2019. This roughly compares to 50,000 days-worth of autonomous vehicle data (assuming 200 PB storage and 4 TB/vehicle/day). Additionally, there is roughly 85 PB storage capacity through their partnership with Intel [6].

In order to leverage the benefits offered within simulation, large scale datasets, and machine learning techniques robust and powerful compute is needed.

## Cloud Architectures

Instead of managing one’s own data center with compute and storage devices, cloud computing and cloud enabled architectures allow companies to leverage infrastructure as a service. There are a number of suppliers for infrastructure as a service from Microsoft Azure, Amazon Web Services, and Google Cloud Platform to name a few popular options. Maintaining infrastructure: updating software packages, cybersecurity, hardware maintenance, troubleshooting, etc. is a non-trivial task. For these reasons, companies can choose to leverage the cloud.

Cognata, an AI based simulation startup, partnered with Nvidia and Microsoft Azure to launch a cloud based autonomous vehicle simulation platform to enable utilization of additional compute on the fly, and help mitigate the hardware dependency barriers. Further, Cognata’s simulation engine cloud-based services allow “auto makers to avoid the expense of substantial upgrades to their infrastructure that the validation process would otherwise require. ‘We are enabling vehicle OEMs, tier-one and tier-two companies in the autonomous market to drive millions of miles that would normally require multi-million-dollar initial investments in hardware’”, a Cognata representative noted [39].

By moving to the cloud, the Cognata simulation platform is not hosted on local servers, therefore users are not limited by local server capacity.

## Data Utilization - Data Management Infrastructures

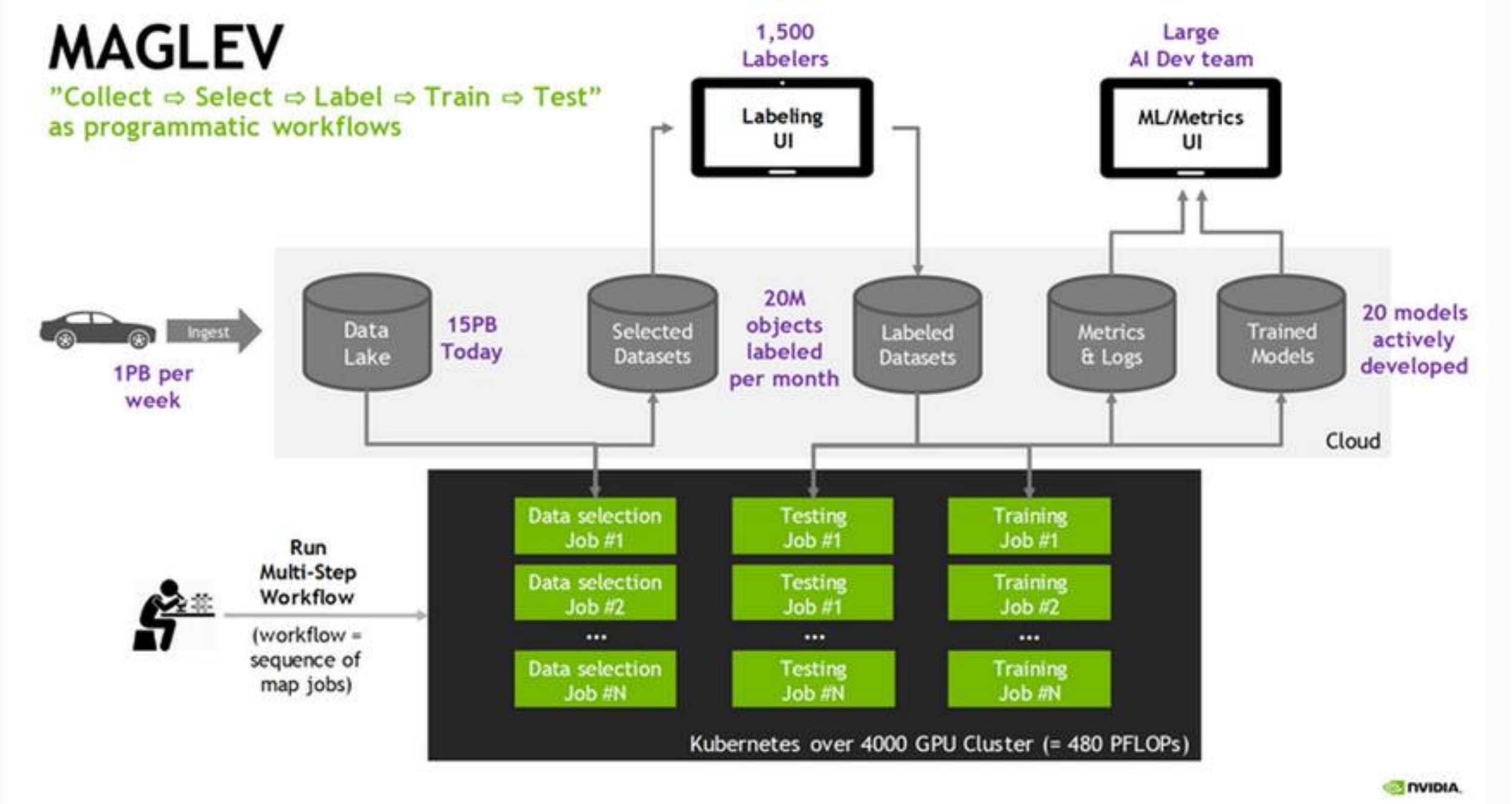

*Figure 45: NVIDIA's Project "MagLev" is a platform to develop AI deep learning models, and to manage the workflow as the model is developed, tested, and optimized. The platform is supported by a dedicated 4000-GPU cluster providing 480 petaFLOPS [33].*

Quality data is highly useful, not only for replay in simulation, informing new virtual scenarios to test an autonomous vehicle, but also to train machine learning models - commonly associated with vehicle perception but could also include aspects of the control algorithm as well (any type of model can be trained assuming relevant data is provided).

NVIDIA's Project MagLev, provides customers and partners with an end-to-end platform for automating the development of models learning through data. The following outlines the series of processes contributing to Project MagLev [33], represented in Figure 45:

1) **Data collection and labeling**: The process starts by collecting and accurately labeling a large set data. NVIDIA started with a small fleet of 30 vehicles, each equipped with 12 sensors (cameras, radar and LiDAR). These vehicles actively collect 1 Petabyte of road data every week. As of now this has generated a 15-petabyte (PB) dataset.

2) **Training:** The dataset can then be utilized for training neural network models (as an example) that run on the vehicle's onboard DRIVE AGX system. Training is performed within the 4000 GPU cluster as part of MagLev.

3) **Simulation testing and validation:** Once training is complete, the models are deployed within simulation or on hardware in the loop, and tested through interaction with the simulation. The NVIDIA DRIVE Sim virtual environment is deployed on the 4,000 GPU cluster which interfaces with 100 DRIVE AGX Pegasus Systems (vehicle computers), sending simulated sensor data and receiving DRIVE AGX output controls.

In order to support the development of supervised learning models, raw data must be labeled. To address this, NVIDIA employs 1,500 people to label the objects in the database, at the rate of 20 million objects every month [33].

## Scalability and Parallelization

Where possible, it is desirable to optimize the development cycle by parallelizing self-contained processes simulating various autonomous vehicle scenarios.

Using Waymo as an example. “Each day, as many as 25,000 virtual Waymo self-driving vehicles drive up to eight million miles in simulation, refining old skills and testing out new maneuvers that help them navigate the real world safely” [46].

Many of the 25,000 virtual vehicles should be capable of running in parallel, interacting with their own self-contained environments, in order to expedite testing and validation.

## Compute Interfaces

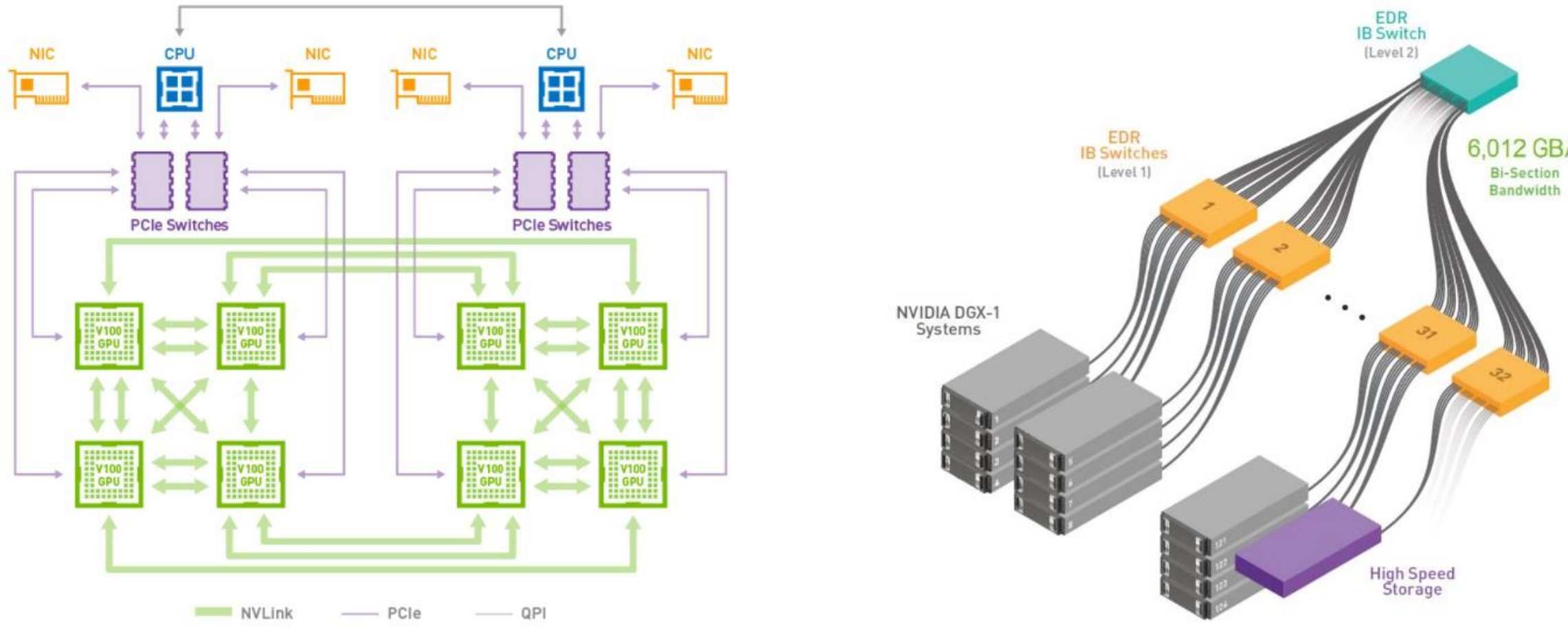

*Figure 46: (Left) NVLink topology of DGX-1 with Volta GPUs. (Right) Example network topology for infrastructure requiring high performance computing and data throughput. Topology uses enhanced data rate (EDR) InfiniBand (IB) network structures to facilitate fast data transfer [23].*

Not only is having fast, and efficient computers within a data center desirable, it is also necessary to have high data rate connections between compute resources within a cluster (collection of computers on the same network, likely working toward a common solution). Consider both the enormous data required to train machine learning models (100 exaFLOPS for Google’s near human level machine learnt translation model [23]) that perform on an autonomous vehicle, as well as the considerable data generated by real and simulated sensor feeds (autonomous vehicle generates 4 TB of data per day largely due to its sensors [21]).

Thus, the efficient transfer of data between compute resources, storage and other devices within a network is necessary to avoid a bottleneck within the pipeline, wherein one to many devices are waiting for data from another device. When such an event occurs, the compute resource might become idle/wait until such data is provided. If the compute resources can perform calculations faster than data can be exchanged within the network, then throughput is limited by the data exchange rate.

It turns out that in the overwhelming majority of cases, the data rate required in order to avoid a communications bottleneck exceeds the capability of the PCIe network. To address this problem within the DGX-1 supercomputer, NVIDIA developed a dedicated GPU interconnect called NVLink, which in its second release provides over 300GB/s bidirectional bandwidth per Tesla V100 GPU [23]. However, high transfer rates are also needed between multiple computational and storage devices (multi-node communication), not just internally to a particular device. Therefore, it is recommended to structure a High Performance Computing (HPC) data center topology with enhanced data rate connections such as InfiniBand, which enable up to 200Gb/s data speeds (as shown in Figure 46) [30].

## Safety and Validation

Autonomous vehicles are needed to perform well within an incalculable number of scenarios including different weather and atmospheric conditions: sun, rain, snow, twilight, night, backlit, etc., while avoiding and detecting a variety of  impediments: bikes, motorcycles, trains, pedestrians, stop signs, potholes, black ice, traffic circles, etc.

In order to validate an autonomous vehicle, driving needs to occur for billions of miles. Thus it is not remotely possible to generate this amount of data, nor test the variety of scenarios, with real vehicles [33].

Simulation can greatly enhance the safety and validity of vehicle performance by enabling testing of difficult to replicate or inherently unsafe scenarios, and generate a vast quantity of data to inform vehicle improvements. Further simulation can help avoid software reversion via regression testing.

### Safety through Edge Case Testing

There are a number of scenarios which are difficult to test within the real world. Some might call these edge-case scenarios. However, many of the so called edge cases are not truly edge cases in the sense that they are uncommon.  Particularly for autonomous driving, the reality is that 'edge case' scenarios are daily occurrences which are dangerous, non-ideal situations such as a crash, erratic pedestrian behavior, unexpected objects, not following rules of the road, etc.

These scenarios happen with frequency every day but are difficult or undesirable to recreate in the real world environment. Simulation allows for a low risk path to test and enhance vehicle performance within these scenarios, without danger to reality.

In the case of Waymo's safety program, 28+ core competencies [provided by U.S Department of Transportation] are tested within simulation, including thousands of scenario variants to ensure their system can safely handle the challenges of real-world environments [46].

In addition to testing core behavioral competencies, Waymo engineers also conduct crash avoidance testing across a variety of scenarios. In 2015, NHTSA published data showing the distribution of the most common pre-crash scenarios. For example, just four crash categories accounted for 84% of all crashes: rear end crashes, vehicles turning or crossing at an intersection, vehicles running off the edge of the road, and vehicles changing lanes. Therefore, avoiding or mitigating those kinds of crashes is an important use case for simulation. Waymo utilizes simulation to test operation of self-driving cars on NHTSA's 37 pre-crash scenarios [46].

### Safety through Data Generation

Simulation enables the generation of one scenario into a thousand, testing all the slight variations in an environment: behaviors, speeds, trajectories, etc. without relying on real world data capture.

Simulation is cheap, flexible, and reproducible, with high throughput compared to constructing real world scenarios. Leveraging this availability of quality data enables the autonomous vehicles to train on a plethora of diverse scenarios and situations, which can provide a more general and less biased end system as it's exposed to a wider variety of experiences.

### Safety via Regression Tests

Simulation regression tests, such as those used by Uber, simulate a set of representative on-road scenarios against which all software releases are tested for regression. Regression tests flag self-driving behaviors that fail a scenario that it had previously passed [45]. In doing so faulty modification to software are potentially caught early in the development cycle.

## Conclusion

As shown throughout this paper, appropriate simulation and modeling for autonomous vehicles, and the utilization thereof, incorporate a number of factors from software design, machine learning, high fidelity mapping, infrastructure development and the utilization of datasets. Additionally, there is a strong interconnection and iterative series of development tools between simulation and the real world from software in the loop (SIL), hardware in the loop (HIL), high fidelity simulators, and closed course testing.

The paper started with an introduction and overview for autonomous systems and their value proposition, as well as the increasing necessity for appropriate simulation tools within this domain. A synopsis of key players discussed throughout the paper was provided.

The following section provided a rational for autonomous systems as a use case for understanding state of the art simulation toolsets. This section additionally looked at how other simulations and domains do not provide adequate scope or capability for understanding state of the art simulation technologies.

Next an overview of the purpose of simulation and methodologies for programming/interacting with simulation tools was provided.

Subsequent sections offer the main details and focus of the paper. An entire section was devoted to simulation toolsets within autonomous systems: both aerial and land (with a focus on land based vehicles). The section on autonomous vehicle simulations provided insight into the pipeline of simulation toolsets from multi-physics simulation, sensor specific simulation, to mission level simulation, and the possibility for managing these separate simulation environments cohesively. Primary focus was given to mission level simulations as this enables modeling of the entire software stack as well as the dynamic and static entities within an environment. Mission level simulation environments can be constructed via high definition maps combining LiDAR point clouds with additional semantic information, through photo-realistic game engines, or through a combination of the two.

Within a simulation environment for autonomous systems, there are a number of different models to define from static elements, weather and atmospheric conditions, vehicles and pedestrians, communication channels within a single object as well as between one entity and another. Additionally, sensor noise (and model imperfections in general), failure modes and entity collisions likewise need to be appropriately modeled. These models are typically defined mathematically, through programmatic logic or probabilistically, and can be compared to real world behaviors for added realism. Visual fidelity and modeling perception within a higher fidelity graphics environment was discussed. Then several important characteristics of autonomous vehicle simulation (and simulation in general) was provided. These characteristics include: data collection, controllability, extensibility, component-sim interfacing, parameterization of scenarios, observability of entity state, usability and management of the simulation toolsets, component/software reusability, throughput, experience replay and behavior visualization.

The section on autonomous system simulation tools concluded with a discussion on high fidelity mapping which are not only critical for the development of autonomous vehicles, but also enable the construction of highly precise simulation environments.

From here, the simulation to real world pipeline was discussed: operational design domain, software in the loop, hardware in the loop, closed course testing, high fidelity simulators and in general how all of these components (as well as simulation) are interconnected. As an example, data from closed course testing helps inform simulation, and vice versa.

In order to run large scale simulation tests, and provide meaningful utilization of collected data, a robust compute infrastructure is needed; whether this comes from an infrastructure as a service provider (cloud based infrastructures), or is managed and monitored locally.

Finally, simulation will have an increasing role in ensuring safety of not only autonomous vehicles but products developed via simulation in general. The ability to demonstrate various scenarios, some of which are difficult or dangerous to reproduce in the real world is a key benefit.

Simulation and modeling will continue to help engineers and product owners design higher quality systems, especially within autonomous vehicles.

# Appendix

The following sections provide supplementary materials.

## Visualization Tools

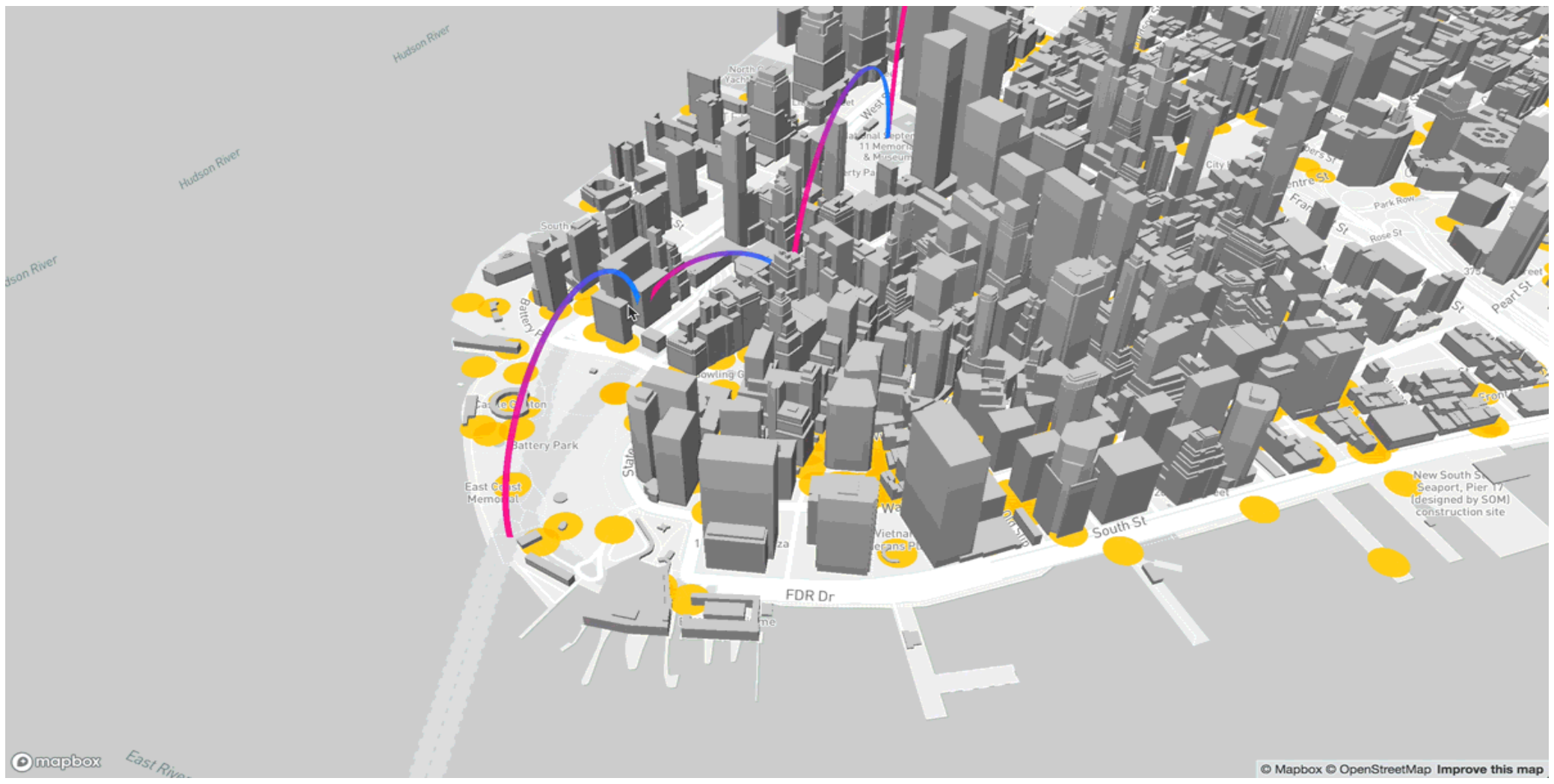

*Figure 47: Mixing deck.gl and mapbox layers. Uber open source software [11].*

Through the research effort, various visualization tools were noted, particularly those from Uber. While these toolsets did not have a direct relation to autonomous vehicles, they are related to modeling and visualizing mobility within geography. Thus, these open source toolsets might warrant future investigation.

The toolsets will not be explained, but representative images are displayed to help gauge intrigue. The location of Uber's open source toolsets are located at: [https://uber.github.io/#data_visualization](https://uber.github.io/#data_visualization). A snapshot of these toolsets is provided in Figure 48.

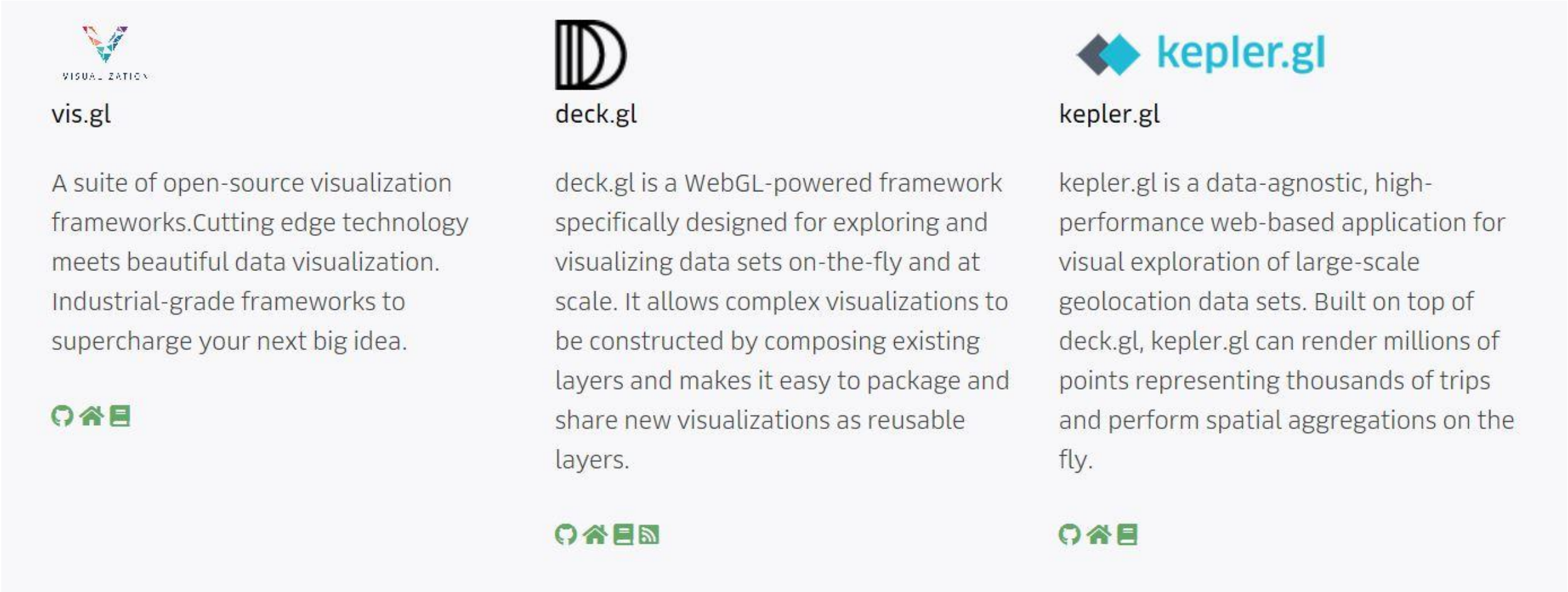

*Figure 48: Uber open source data visualization packages and frameworks. [44]*

The following are intriguing images from Uber's open source toolsets.

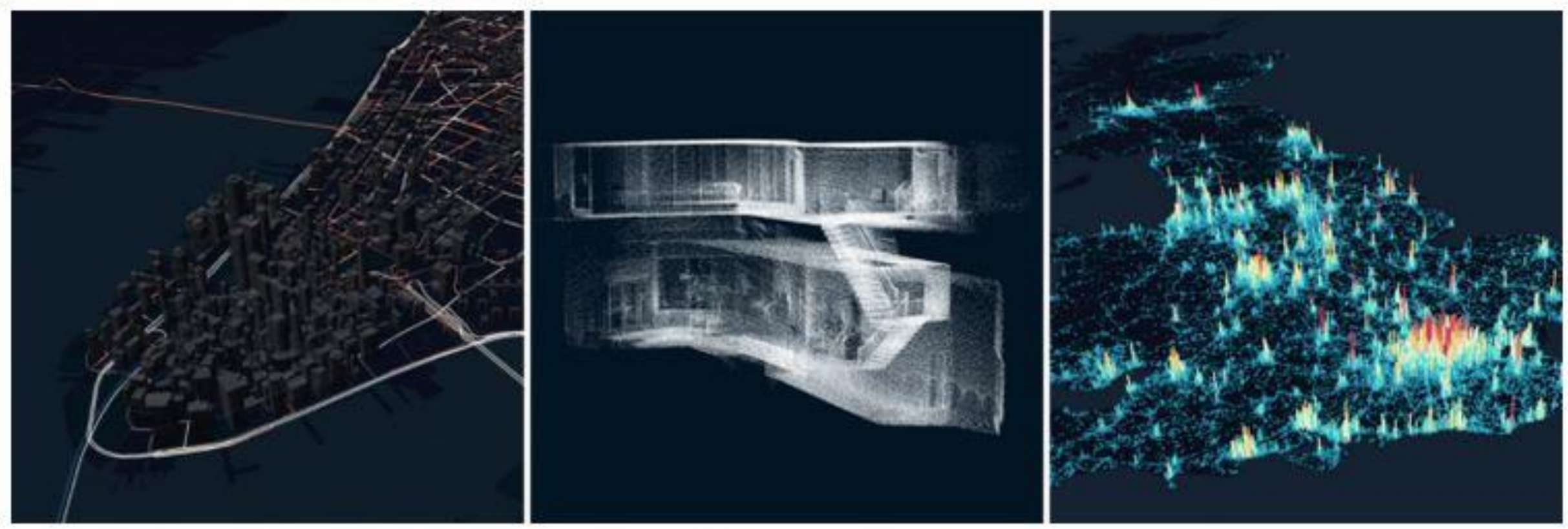

*Figure 49: Uber's deck.gl layers render GeoJSON, point cloud, and grid visualizations [10].*

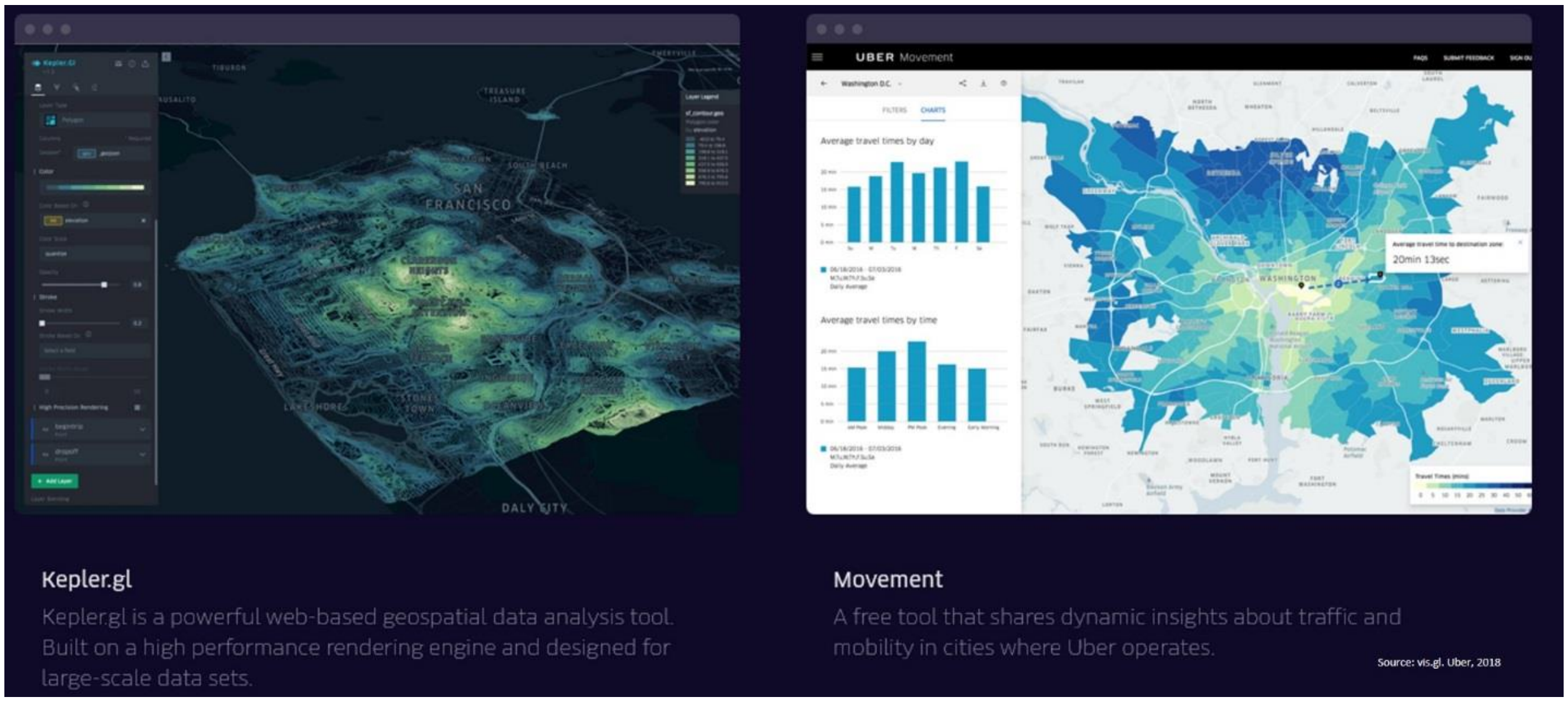

*Figure 50: Uber visualization tools*

## Guiding Questions

As the paper was brainstormed and researched, the following were questions that guided the direction of the paper. Some questions are answered directly within the paper while other questions are provided over multiple sections. These are some of the questions:

- How does simulation affect safety?
- How are simulation tools constructed?
- How does simulation address the perception problem, and how is this tied to simulation's visual fidelity?
- What is the right level of fidelity?
- How are the different aspects of a vehicle tested in simulation?
- What aspects/characteristics define a good simulation?
- How does machine learning play into simulation?
- How is the real world tied to simulation? What is the feedback loop?

- What kind of simulation architecture is needed? How is it architected?
- What kind of compute infrastructure is needed to support simulation?
- What kind of data is ingested by simulation and what type of data is generated by simulation?
- What are supporting toolsets to autonomous vehicle simulation?

These questions were answered by researching how current companies within the autonomous vehicles industry utilize simulation, supporting toolsets, infrastructure, and software programs, as well as data.

THE BOEING COMPANY

# Simulation, and Data Utilization within Industry – Paper Synopsis

## An Autonomous Vehicles Perspective

**Fadaie, Joshua G**

**12/4/2018**

## High Level Simulation Tool Requirements/Synopsis

According to research by the RAND Corporation, 11 billion miles are needed to be driven in order to conclude with 95% confidence that autonomous vehicles decrease failure rates (1.09 fatalities for every 100 human miles driven) by 20%. Real world fleet vehicle testing alone is infeasible to meet this metric.

Simulation will be increasingly more important to ensure system functionality and safety by providing high throughput testing and validation, ability to test edge cases and dangerous scenarios, feasibly generate data, and avoid software regression.

What follows is a brief synopsis of the information contained in the main document. (For the information provided below, references are cited within the paper).

**Simulation Characteristics:**

- **Data generation**: Ability to utilize the simulation tool to generate data for offline training/testing.
- **Controllability**: Ability to control the initial state and configuration of scenario and entities within the environment. Ability to 'fork' a simulation scenario from a currently running scenario for further testing.
- **Extensibility**: Extensibility is a loaded term which can include the ability to inherit from, or extend prior software functionality (i.e. in terms of software development, inheriting from a software class), leverage open source capability, and add new development features (software classes, simulation models) that were not envisioned at the time of initial simulation design.
- **Component Level Interfacing (Internal and External)**: Well defined, usable interfaces enable coupling between internal simulation modules as well as external modules, developed within the company or by a 3rd party. In general, interfaces need to specify a set of APIs (Application Program Interfaces) for developers, and the communication protocol utilized (HTTP, UDP, TCP, etc.).
- **Parameterization:** Parameterization as described here expresses the idea of turning a single scenario into a thousand scenarios incorporating slight variations of the original parameters. This usually entails providing ranges for a particular sets of variables which are then run in parallel or run in batches of simulation jobs.
- **Observability**: Observability is the ability of the tester to observe the state of the system to determine whether a test passed or failed. Observability is easier said than done (refer to section on Observability). Simulation tools could provide the capability to define heuristics or logical rules defining criteria for 'success' and 'failure' within simulation runs.
- **Usability and Sim Management:** Usability considers the ability of engineers, designers, and quality assurance personnel to rapidly test, evaluate and iterate on virtual scenarios and the various models incorporated. Usability of a simulation environment can be provided in a number of different forms from quality UI to enable top level modifications, intuitive user controls such as drag and drop functionality of models, the ability to manage the simulation state: stopping, starting, restarting, saving an environment (also tied to Controllability), and easy customization of model parameters through high level interfaces or scripts.
- **Component/Software Reusability:** This could be reuse of simulation features, software systems, sub-systems, components, and interfaces.  Good Object Oriented Programming (OOP), Model Based Development, and appropriate interfaces facilitate this capability.

- **Throughput:** As explained by Waymo, "the iteration cycle is tremendously important to us and all the work we've done on simulation allows us to shrink it dramatically. The cycle that would take us weeks in the early days of the program now is on the order of minutes".
- **Experience Replay:** Providing the ability to replay experiences derived from the real world, as well as overlay the real world experience with the updated vehicle software within simulation.
- **Behavior Visualization:** Visualize the various perceptions and behaviors calculated by the autonomous vehicle through its software stack from Perception, Localization, Path Planning, Behavior Prediction, and Trajectory Planning.

**Thoughts on Simulation High Level Requirements:**

Simulation toolsets in addition to embodying the above characteristics (characteristic requirements) and qualities must include these capabilities:

- **Software development via**:
  - Document based (object oriented design and functional) programming
  - Model Based Development
- **Connection of safety features and safety/system requirements to software development and to other simulations tools in the toolsets**
  - An ability to manage all simulations, their components and gathered metrics within a common location
- **Support the pipeline of simulation toolsets** needed to test and validate the development of all aspects of autonomous vehicle development: multi-physics, system level design, and mission level simulation.
- **Support for various vehicle and environmental models**:
  - Static environment models: buildings, vegetation, traffic signs, infrastructure, road types, lane markings, etc.
  - Dynamic environment models: pedestrians, vehicles, animals, etc. and their dynamical behaviors
  - Weather and atmospheric conditions: gravity, air-density, air pressure, magnetic field, characteristics of light, rain, fog, snow, sunshine
  - V2X communication: vehicle to vehicle, vehicle to pedestrian, vehicle to network, vehicle to infrastructure communication
  - Other: sensor noise, environment imperfections/noise, collision detection, failure mode simulation
- **Real world support**:
  - Software in the loop (SIL): test software running on vehicle within simulation
  - Hardware in the loop (HIL): test software and hardware running on vehicle within simulation. See sections on HIL and Compute Infrastructure for details (equivalent of 11.6 2018 iMac Pros provided on board vehicle through NVIDIA DRIVE AGX hardware [assuming accurate FLOPS provided])
  - Feedback from real world to simulated world
  - High fidelity simulators with passenger in the loop feedback, electro-mechanical actuation and visual fidelity and synchronization

- Closed course support facility
- **Provide sufficient rigor to the simulated environment**
    - Mathematically or statistically accurate models
    - Simulation models verifiably match the real world performance
    - HD Mapping as a means of constructing the simulation environment arena
- **Visualization/Graphics fidelity**
    - If modeling the perception problem in simulation, the tool must include photo-realistic rendering of a scene (photo-realism provided by game engines such as Unity and Unreal Engine)
- **Ability to deploy simulation environment to a cluster of computers**
    - This entails the ability to parameterize the simulated scenario and deploy each parametric/individual scenario to any computer in the network and then 'consolidate' the results

**Requirements for Support Equipment**:

Simulation toolsets by themselves are insufficient to provide an adequate solution to the autonomous vehicle problem domain. Sophisticated support equipment is necessary:

- **Sufficient infrastructure** to deploy and scale the simulation toolset
    - NVIDIA DGX SaturnX AI Supercomputer as example: equivalent to 956k 2018 iMac Pros (assuming iMac Pro offers 690 GFLOPS and SaturnX has 660 petaFLOPS)
    - One AV generates 4 TB data per day, largely due to sensors
- **Solving the perception problem** (and in general data curation), regardless of the simulation tools ability to provide photo-realistic visualization, is a massive endeavor.
    - NVIDIA employs 1,500 people to label objects in a database at rate of 20 million objects per month.
- **Requires not only compute but robust compute interconnection**: large data transfers between compute resources (100 exaFLOPS to train Google's near human level machine learnt translation model). Think 200 Gb/s supported by enhanced data rate connections such as InfiniBand.
- **Consider cloud architecture** as a way of mitigating the non-trivial responsibility of maintaining infrastructure: updating software packages, cybersecurity, hardware maintenance, troubleshooting, etc.
    - Look at simulation startup Cognata's partnership with NVIDIA and Microsoft Azure as an example. Infrastructure as a Service (IaaS) providers: Amazon Web Services, Google Cloud Platform, Microsoft Azure to name a few.

**Some Open Questions:**

- Exactly how a centralized simulation management tool manages and organizes the SW architecture, safety and system requirements (and lower level requirements), various simulation tools (simulation pipeline), metrics, etc. and ties all of these disparate sources together.
- What is the equivalent of an HD map for air vehicles? What are the associated traffic lanes, formal and informal rules of flight, what additional features need to be included, etc.?